\documentclass[useAMS,usenatbib, usegraphicx]{mn2e}

\usepackage{verbatim} 
\usepackage[usenames,dvipsnames]{color}
\usepackage{amssymb, amsmath}
\usepackage{multicol}
\usepackage{longtable}
\usepackage{rotating}
\usepackage{pdflscape}
\usepackage{url}
\usepackage{times}
\usepackage{epstopdf}

\makeatletter
\def\url@leostyle{%
  \@ifundefined{selectfont}{\def\UrlFont{\sf}}{\def\UrlFont{\small\ttfamily}}}
\makeatother
\DeclareRobustCommand{\ion}[2]{%
\relax\ifmmode
\ifx\testbx\f@series
{\mathbf{#1\,\mathsc{#2}}}\else
{\mathrm{#1\,\mathsc{#2}}}\fi
\else\textup{#1\,{\mdseries\textsc{#2}}}%
\fi}

\title[Evolution of the SFRD in the local universe]{Galaxy And Mass Assembly: Evolution of the H$\alpha$ luminosity function and star formation rate density up to $z<0.35$}

\author[Gunawardhana \it{et.\,al}]{M.\,L.\,P.\,Gunawardhana$^{1,2}$\thanks{E-mail: mlpg@phyiscs.usyd.edu.au}, A.\,M.\,Hopkins$^{2}$\thanks{ahopkins@aao.gov.au}, 
J.\,Bland--Hawthorn$^1$,  S.\,Brough$^2$
\newauthor
R.\,Sharp$^3$, J.\,Loveday$^4$, E.\,Taylor$^{1,5}$, D.\,H.\,Jones$^{6}$,  M.\,A.\,Lara-L\'opez$^2$, A.\,E.\,Bauer$^2$
\newauthor
M.\,Colless$^{2,3}$, M.\,Owers$^2$, I.K.\,Baldry$^7$,  A.\,R.\,L\'opez--S\'anchez$^{2,12}$, C.\,Foster$^{13}$, S. Bamford$^{8}$
\newauthor
M.J.I.\,Brown$^{5}$, S.P.\,Driver$^{9,10}$, M.\,J.\,Drinkwater$^{11}$, J.\,Liske$^{13}$, M. Meyer$^{9}$, P.\,Norberg$^{14}$
\newauthor
A.S.G.\,Robotham$^{9}$, J.H.Y.\,Ching$^1$, M.E.\,Cluver$^2$, S.\,Croom$^1$, L.\,Kelvin$^{9,10}$ 
\newauthor
M.\,Prescott$^{15}$, O.\,Steele$^{16}$, D.\,Thomas$^{16}$, L.\,Wang$^{14}$ \\
$^{1}$Sydney Institute for Astronomy (SIfA), School of Physics, University of Sydney, NSW 2006, Australia\\
$^{2}$Australian Astronomical Observatory, PO Box 915, North Ryde, NSW 1670, Australia\\
$^{3}$Research School of Astronomy $\&$ Astrophysics, Australian National University, Cotter Road, Weston Creek, ACT 2611, Australia\\
$^4$Astronomy Centre, University of Sussex, Falmer, Brighton, BN1 9QH\\
$^{5}$School of Physics, The University of Melbourne, Parkville, VIC 3010, Australia\\ 
$^6$School of Physics, Monash University, Clayton, Victoria 3800, Australia\\
$^7$Astrophysics Research Institute, Liverpool John Moores University,Twelve Quays House, Egerton Wharf, Birkenhead, CH41 1LD, UK\\
$^{8}$School of Physics \& Astronomy, University of Nottingham, University Park, Nottingham NG7 2RD, UK\\
$^{9}$International Centre for Radio Astronomy Research (ICRAR), University of Western Australia, Crawley, WA 6009, Australia \\
$^{10}$Scottish Universities' Physics Alliance (SUPA), School of Physics and Astronomy, University of St Andrews, North Haugh, St Andrews, KY16 9SS, UK\\
$^{11}$School of Mathematics and Physics, University of Queensland, QLD 4072, Australia \\
$^{12}$Department of Physics and Astronomy, Macquarie University, NSW 2109, Australia\\
$^{13}$European Southern Observatory, Karl-Schwarzschild-Str.~2, 85748 Garching, Germany\\
$^{14}$Institute for Computational Cosmology, Department of Physics, Durham University, South Road, Durham, DH1 3LE, UK\\
$^{15}$Department of Physics, University of the Western Cape, Private Bag X17, Bellville 7535, South Africa \\
$^{16}$Institute of Cosmology and Gravitation, University of Portsmouth, Dennis Sciama Building, Portsmouth, PO1 3FX \\
}

\begin{document}

\date{Accepted 2013 May 20}

\pagerange{\pageref{firstpage}--\pageref{lastpage}} \pubyear{2002}

\maketitle

\label{firstpage}

\begin{abstract}
{Measurements of the low--$z$ H$\alpha$ luminosity function, $\Phi$, have a large dispersion in the local number density of sources ($\sim0.5-1$ Mpc$^{-3}$dex$^{-1}$), and correspondingly in the SFR density. The possible causes for these discrepancies include limited volume sampling, biases arising from survey sample selection, different methods of correcting for dust obscuration and AGN contamination. The Galaxy And Mass Assembly (GAMA) survey and Sloan Digital Sky Survey (SDSS) provide  deep spectroscopic observations over a wide sky area enabling detection of a large sample of star--forming galaxies spanning  $0.001<SFR_{H\alpha}\, (M_{\odot}yr^{-1})<100$ with which to robustly measure the evolution of the SFR density in the low--$z$ universe. The large number of high SFR galaxies present in our sample allow an improved measurement of the bright end of the luminosity function, indicating that the decrease in $\Phi$ at bright luminosities is best described by a Saunders functional form rather than the traditional Schechter function. This result is consistent with other published luminosity functions in the FIR and radio. 
For GAMA and SDSS we find the $r$--band apparent magnitude limit, combined with the subsequent requirement for H$\alpha$ detection leads to an incompleteness due to missing bright H$\alpha$ sources with faint $r$--band magnitudes.} 

\end{abstract}

\begin{keywords}
surveys -- galaxies: luminosity function -- galaxies: evolution 
\end{keywords}

\section{Introduction}

The evolution of the global star formation rate (SFR) density is now traced out to $z\sim10$ using star formation indicators across a broad wavelength range, from x--ray/gamma rays to radio emission. Direct information on the star formation rate has been collected from nebular emission lines such as  [\ion{O}{ii}]~$\lambda$3727,  [\ion{O}{iii}]~$\lambda$5007, H$\alpha$, H$\beta$ \citep[][and references therein]{Glazebrook04, Westra10} tracing massive stars, ultraviolet, far and mid infrared emission \citep{Schiminovich05, Perez05, Reddy07, Reddy09, LS10, Bouwens10, Bouwens11} revealing young star--forming regions, radio emission produced in supernova remnants \citep{Haarsma00, Seymour08}, X--ray emission produced from high mass X--ray binaries \citep{Georgakakis03, Fabbiano05}, and gamma ray bursts produced from massive stellar explosions. \citep{Woosley06, Yuksel08, Kistler09}.

The cosmic star formation history (SFH) indicates a global increase in star formation activity since the formation of the first galaxies, reaching a peak at $z\sim2-3$. This is followed by a rapid decline in average star formation of approximately a factor of ten \citep[e.g.][]{Lilly96, Madau96, Hopkins04, HB06, Perez08}. This is typically interpreted in the model of mass dependence \citep{Cowie99}, that states that high--mass galaxies formed their stars early and rapidly, with lower--mass systems forming more slowly and at later times. Evidence supporting this idea in the context of the mass--dependence of the SFH has accumulated over recent years \citep[e.g.][]{Feulner05, Juneau05, Zheng07, Mobasher09}.

Numerous studies have been conducted to determine the physical processes contributing to the shape of the cosmic SFH, particularly the substantial decline in star formation activity since $z\sim2$. Hydrodynamic simulations examining hot and cold mode accretion indicate a close relationship between the global gas infall rate and the cosmic SFR \citep{vande11, Keres05}. A mechanism associated directly with star formation itself has been proposed that moderates the relation between neutral gas and SFR in galaxies \citep{Hopkins08}. Through the analysis of rest--frame $u$--band luminosities \cite{Prescott09} find evidence of a decline in characteristic luminosity (L$^*$) over $0<z<1.2$, coinciding with the decline in global star formation. A change in the rate and mode of star formation since $z\sim1$ is assumed to be responsible for this. The strong decrease in the fraction of galaxies undergoing starbursts \citep{Dressler09} and the decline in galaxy interactions such as tidal encounters and mergers \citep{Lefevre00, PCL07, Lotz08, Lotz11} are given as possible explanations. Recent work \citep{Nakamura03, Sobral09, Westra10, LS10} points out a link between star formation and galaxy morphology, indicating that the merger--induced star formation tends to dominate in galaxies with L $>$\,L$^*$, with a more quiescent mode dominating in fainter galaxies.

While the SFH based on different SFR tracers gives a broadly consistent picture of the evolution of the global star formation, the dispersion between individual measurements at a given redshift is striking and can span more than $0.5$ dex \citep[e.g.][]{HB06}. Ideally, studying the cosmic SFH using SFR indicators covering a large spectral range would provide a robustly consistent picture of galaxy formation and evolution. In reality, different SFR indicators suffer from different selection and calibration biases (e.g.\, the sensitivity to the stellar metallicity abundance and the ionisation state in the case of [\ion{O}{ii}]~$\lambda$3727), and are affected by and treated for dust obscuration differently \citep{Gilbank10, WAS10}, introducing systematic uncertainties to measurements. H$\alpha$ emission, as a direct tracer of instantaneous star formation in a galaxy, is a good candidate for providing an accurate view of the evolution of SFR density. It is however, currently restricted to low--to--moderate redshift, and even with the use of a common SFR indicator, a compilation of local H$\alpha$ SFR densities \citep{Sobral09, Westra10} still shows large discrepancies between measurements. The possible causes for this dispersion include cosmic (sample) variance, the differences in selection criteria between surveys, and the uncertainties coming from the measurements, corrections and assumptions that go into the final estimate of SFR densities. We aim to understand and interpret the observed evolution of cosmic SFR density paying special attention to the advantages, and drawbacks of survey and sample selection.

An additional complication is that different SFR indicators probe different stellar mass ranges (e.g.\,H$\alpha$ emission traces stars with masses $\geq 10$\,M$_{\odot}$). In order to infer a SFR density therefore requires the assumption of a stellar initial mass function (IMF). The stellar IMF is widely accepted to have a universal form regardless of environment and time \citep[e.g.][]{Bastian10}. There are a number of recent studies, however, that suggest variations in the stellar IMF with respect to redshift \citep{Wilkins08a, Wilkins08b, Chary08, Dave10}, surface brightness \citep{M09, HG08} and SFR or SFR surface density \citep{Gun11} or colour \citep{Dutton11}. An environment dependent and/or evolving IMF directly impacts the derived cosmic SFR densities. Although not explicitly explored in this paper, the incorporation of such an IMF can be a potential solution towards reconciling the observed discrepancies in the evolution of the cosmic SFR and stellar mass densities \citep{Wilkins08b}.   

Many local SFR densities come from narrowband filter surveys \citep[e.g.][]{Jones01, Pascual01, Fujita03, Glazebrook04, Ly07, Ly11, Shioya08, Dale08} complementing those from spectroscopic surveys \citep[e.g.][]{Tresse98, Sullivan00, Tresse02, Perez03, Shim09, Westra10, Gilbank10}. In contrast to spectroscopic surveys, narrowband surveys at optical wavelengths provide deep imaging over a narrow redshift slice, yielding relatively large volume--limited samples of galaxies. Also, the target selection is done through emission--lines. The two main advantages with narrowband surveys are that they are most effective at detecting faint emission--line sources, and the galaxies are selected using a quantity they aim to measure, which scales with SFR \citep{Jones01, Westra08}. 

There are however, a number of drawbacks to narrowband surveys. The main disadvantages are the need to assume common corrections for stellar absorption, dust obscuration, contamination by AGN, and insensitivity to low equivalent widths. These assumptions introduce large uncertainties and can lead to a systematic underestimate of the final SFR density \citep{Spector11, Massarotti01, James04}. In contrast, spectroscopy allows the determination of such corrections individually for each galaxy. Moreover, a survey with a large sky coverage is generally preferred in order to overcome cosmic (sample) variance and small number statistics. Despite being deep, the current generation of narrowband surveys only cover a relatively limited sky area. Even for spectroscopic surveys, only the Sloan Digital Sky Survey (SDSS): Stripe 82 \citep[][area $\sim$ 275\,deg$^2$ and $z \lesssim 0.2$\footnote{Redshift ranges given here do not necessarily denote the redshift coverage of the survey, but rather the redshift coverage corresponding to a particular emission line (e.g.\,H$\alpha$).}]{Gilbank10}, Universidad Complutense de Madrid (UCM) survey \citep[][area $\sim$ 472\,deg$^2$ and $z \lesssim 0.045$]{Gallego95, Perez03} and now the Galaxy And Mass Assembly (GAMA\footnote{\url{http://www.gama-survey.org}}) survey \citep[][area $\sim$ 144\,deg$^2$ and $z \lesssim 0.35$]{Driver09, Driver11} provide substantial sky coverage. 

The layout of this paper is as follows. We describe sample selection in \S\,\ref{Data} and provide a brief introduction to the GAMA and SDSS surveys. \S\,\ref{PhysicalProp} details the derivation of physical properties such as H$\alpha$ SFRs for the two samples. In \S\,\ref{LF}, we describe the technical details of the derivation of the luminosity functions (LFs), taking into account different survey selection criteria. This section also presents the resulting GAMA and SDSS LFs. \S\,\ref{functions} describes the details of the functional types used to fit the LFs. In \S\,\ref{CSFH}, we infer SFR densities for our GAMA and SDSS LFs, and in the Appendix we explore the potential biases influencing our estimates of SFR densities. 

The assumed cosmological parameters are H$_0 = 70$ km\,s$^{-1}$\,Mpc$^{-1}$, $\Omega_M = 0.3$, and $\Omega_{\Lambda} = 0.7$. All magnitudes are presented in the AB system. 

\section{Data}\label{Data}

In this study, we utilise the GAMA phase--I survey , which covers three equatorial fields of 48 deg$^2$ each, with two fields reaching a depth of $r_{AB}<19.4$ magnitude and the third extending to $r_{AB}<19.8$ magnitude. There are $\sim136\,000$ galaxies with measured spectra available from GAMA observations \citep{Driver09, Driver11}. The availability of such a large galaxy sample with deep spectroscopic observations ($\sim$ 2 magnitudes fainter than SDSS) over a wide sky area, covering a modest redshift range allows the determination of the evolution of the SFH in the local universe in a consistent manner with reduced systematic and sampling biases. 

We also use the SDSS Data Release 7 (DR7) spectroscopic galaxy sample \citep{Abazajian09} in this study. SDSS--DR7 covers an sky area of $>8000$\,deg$^2$, with $0<z<0.38$ and $r_{AB}<17.77$, providing the largest galaxy sample to date.

\subsection {GAMA survey and data}\label{GAMA}

GAMA is a spectroscopic survey undertaken at the Anglo--Australian Telescope (AAT). 
GAMA spectroscopic targets were selected from the SDSS Data Release 6 \cite[DR6,][]{Adelman08} to limiting Petrosian magnitudes of $r<19.4$ in two fields, and $r<19.8$ in the third field. \cite{Baldry10} provides a detailed discussion of the GAMA input catalogue, and the tiling of the sources is described in \cite{Robotham10}. 

For this paper, we use GAMA I data consisting of GAMA, SDSS, 2--degree field Galaxy Redshift Survey (2dFGRS) and Millennium Galaxy Catalogue (MGC) sources. The GAMA spectra are obtained from the AAT with  the 2--degree Field (2dF) fibre feed and AAOmega multi--object spectrograph. AAOmega provides a resolution of 3.2\,\AA\, full width at half maximum (FWHM) with complete spectral coverage from 3700--8900\,\AA\, \citep{Sharp06, Driver11}. The spectra are sky subtracted following \cite{Sharp10}, and redshifts are assigned with RUNZ \citep{Saunders04}, a \textsc{fortran} program for measuring redshifts from reduced spectra. Spectra were given a redshift quality (nQ), with nQ$>2$ regarded as a secure redshift \citep{Driver11}. GAMA does not re--observe the majority of SDSS, 2dFGRS and MGC galaxies in the three GAMA regions.

GAMA I spectroscopic data set is over $98\%$ complete in spectroscopic followup \citep{Driver11}, the small spectroscopic incompleteness likely due to low--luminosity, low surface--brightness galaxies. In addition, GAMA, like all spectroscopic surveys, suffers from several other sources of incompleteness; imaging incompleteness, and redshift measurement failures, i.e.\,spectra with nQ$\leqslant2$ \citep{Loveday11}. The LFs presented in this paper are corrected for these sources of incompleteness, see  \S\,\ref{LF} and \S\,\ref{unccorr}.

All GAMA spectra are flux calibrated following the detailed discussion given in \cite{Hopkins13} and Liske et al.\,(in prep.). Briefly, the GAMA flux calibration process is essentially a two--step process. In the first instance, an initial flux calibration is achieved for each 2dF plate to correct for the wavelength--dependence of the system throughput. This is then supplemented by an absolute flux correction. 

Three fibres on each 2dF plate are assigned to standard stars. For each star a flux correction vector is derived by taking the ratio of the observed to its best fit model, the average between the three provides an unique wavelength--dependent correction for a given plate. Any lower--order shape in the continuum is removed by dividing the standard stellar spectrum by the unique correction vector. A fit to the residuals achieves an initial curvature correction that accounts for the poor CCD response at blue and red extremes of the spectrum. An absolute flux calibration is obtained by tying the spectrophotometry directly to the $r$--band petrosian magnitudes from the SDSS photometry.

The standard strong optical emission lines are measured from each curvature corrected and flux calibrated spectrum assuming a single Gaussian approximation and a common redshift and line--width within an adjacent set of lines (e.g.\, H$\alpha$ and the [\ion{N}{ii}]~$\lambda\lambda$6548, 6583 doublet), and simultaneously fitting the continuum local to the set of lines \citep{Hopkins13, Brough11}.  Corrections for the underlying Balmer stellar absorption, dust obscuration and fibre aperture effects, detailed below, are applied to these measurements. The GAMA sample consists of a relatively large number of low--$z$ galaxies. The observed recessional velocities of the nearest galaxies ($z<0.02$) are influenced by peculiar motions. For these objects the redshift--distances will be systematically under-- or over-- estimated if peculiar velocities are ignored. Parametric multi--attractor models provide directional--dependent prescriptions to estimate the effects of peculiar velocities. For this sample, the flow--corrections have been made using the approach of \cite{Tonry00}, as described in \cite{Baldry12}. The derived physical properties of galaxies, such as luminosities, are based on these flow--corrected redshifts (DistancesFramesv06). 

SDSS photometry in {\em u,g,r,i,z\/} filters is available for each GAMA galaxy. The intrinsic galaxy luminosities are measured in $r$--band defined elliptical Kron apertures \citep{Hill10, Taylor11}. $k$-corrections to $z=0$ \citep[\textsc{kcorrect V4\_2,}][]{Blanton07} are applied and all photometry is corrected for foreground (Milky Way) dust-extinction \citep{Schlegel98}. 

\subsubsection{This sample}\label{gama_sample}

Our sample is drawn from the $136\,000$ spectra (AATSpecAllv08) available at December 2011, and is comprised of  72880 galaxies with GAMA redshifts, measured H$\alpha$ emission, nQ$>$2 and H$\alpha$ emission signal--to--noise above 3. The H$\alpha$ signal--to--noise is defined as the ratio of the observed H$\alpha$ flux to the RMS noise over a $153$\AA\,window $12$\AA\,blue-wards of the redshifted wavelength of the [\ion{N}{ii}]~$\lambda$6548 feature. Furthermore, a selection of nQ$=$2 sources obeying the constraints detailed in \cite{Baldry12} are also included in the sample. 

The redshift source of the brightest galaxies in GAMA is SDSS as GAMA does not re--observe most of these galaxies, (Table\,\ref{table:stats}). The emission--line measurements for the SDSS galaxies are from the MPA--JHU DR7 database\footnote{\url{http://www.mpa-garching.mpg.de/SDSS/DR7/}}. There are $11675$ SDSS sources with detected H$\alpha$ emission included in the sample. The emission measurements for MGC sources are not currently available, and while the emission measurements for the 2dFGRS sources are available, the spectra from which these measures estimated are not flux calibrated. Therefore, these galaxies are excluded from our sample. The sample incompleteness introduced by the lack of 2dFGRS and MGC galaxies can be corrected for since the missing fractions are known, (see \S\,\ref{LF} and \S\,\ref{unccorr}).

Galaxies dominated by emission from active galactic nuclei (AGN) are excluded from the sample based on standard optical emission--line ([\ion{N}{ii}]~$\lambda$6584/H$\alpha$ and [\ion{O}{iii}]~$\lambda$5007/H$\beta$) diagnostics \citep[BPT;][]{BPT} using the discrimination line of \cite{Kewley01}.  In the case of galaxies for which only some of these four emission lines are measurable, AGNs can still be excluded using the diagnostics $\log$ ([\ion{N}{ii}]~$\lambda$6584/H$\alpha$)$\geq$0.2 and $\log$ ([\ion{O}{iii}]~$\lambda$5007/H$\beta$)$\geq$1. Overall $\sim9\%$ of GAMA galaxies are classified as AGNs and excluded from our sample. For the galaxies still unable to be classified in this fashion, we flag them as `unclassified', and retain them in the sample of star forming galaxies. Of the star forming sample 30\% are `unclassified' for this reason. We default to this solution rather than excluding them from the sample, as a galaxy with measured H$\alpha$ but without an [\ion{N}{ii}]~$\lambda$6584 or [\ion{O}{iii}]~$\lambda$5007 measurement is more likely to be star forming than an AGN \citep{Fernandes10}.  \cite{Robotham13} investigated the potential pitfalls of automated BPT classifications by visually examining a small sample of low--$z$ GAMA galaxies. They found that the majority of the BPT classified AGNs are low-powered LINER--like systems with weak H$\alpha$, H$\beta$ and [\ion{O}{iii}]~$\lambda$5007. Furthermore, their results indicate that majority of the automated spectral classifications ($\sim75\%$) agree with the visual classifications. The impact of erroneously including a small fraction of AGNs is in any case very small, and does not change any of the conclusions below.  

Furthermore, we exclude all galaxies with H$\alpha$ emission measurements affected by the presence of strong sky lines, and all galaxies with H$\alpha$ emission below a minimum flux limit of \mbox{$25\times10^{-20}$W\,m$^{-2}$},  hereafter called the detection limit. This detection limit is obtained from examining the spectra of a sample of low H$\alpha$ luminosity galaxies.

The GAMA emission--line sample spans $0<z\leq0.35$, and a large range in stellar mass \citep[$7\leq \log\, ($M/M$_{\odot}) \leq 12$;][]{Taylor11} and $0.001\leqslant$\,SFR (M$_{\odot}$yr$^{-1}$)$\,\leqslant100$. 

\subsection{SDSS and data release 7}

In addition to the GAMA H$\alpha$ LFs, we also construct the SDSS--DR7\footnote{\url{http://www.sdss.org/dr7/}} \citep{Abazajian09} H$\alpha$ LFs. SDSS \citep{York00} has imaged $\sim 10\,000$\,deg$^2$ in five optical broad--band filters, using a wide--field imager with a mosaic CCD camera on a $2.5$\,m telescope, and covered the sky in a drift--scan mode in five filters \citep{Gunn98}. Photometric catalogues are then used to identify the spectroscopic targets on the same telescope, using a 640--fibre--fed pair of multiobject double spectrographs. The wavelength coverage is from $\lambda\lambda$ 3800--9200\AA\, with a spectral resolution of $\lambda/\Delta \lambda \approx 2000$ (FWHM $\sim$ 2.4\AA\, at $\lambda$5000) \citep{Abazajian09}. The SDSS--DR7 release presents the spectra for $\sim 10^6$ objects over a total sky area of $9380$\,deg$^2$. The main galaxy sample \citep[MGC,][]{Strauss02} used in this study is complete to a Petrosian $r$--band magnitude limit of 17.77.

\subsubsection{This sample}\label{SDSSsample}

{As for the SDSS sources in the GAMA fields, the emission--line measurements of the SDSS galaxies are from the MPA--JHU DR7 database}, and the derivation of these measurements is detailed in \cite{Brinchmann04} and \cite{Tremonti04}. Briefly, each Galactic extinction corrected galaxy spectrum is compared with a library of single stellar population models generated using the \cite{BC03} population synthesis code to fit the continuum shape. This accounts for weak features, and Balmer stellar absorption. Once the best--fit stellar population synthesis model to the continuum is subtracted and any remaining residuals are removed, Gaussian profiles are fitted simultaneously to all the emission lines, requiring that all the lines belonging to Balmer and forbidden--line series have the same width, and velocity offset. This requirement on line widths, and velocity offsets, allow stronger/multiple lines to be used to constrain the weaker lines. {The main difference between GAMA and SDSS emission--line samples is that the latter includes an implicit correction for stellar absorption effects. A constant correction for stellar absorption is incorporated when deriving H$\alpha$ luminosities for GAMA galaxies (see \S\,\ref{PhysicalProp}). The assumption of a single value can introduce some uncertainty, and should be restricted to the examination of gross characteristics of large samples of galaxies \citep{Hopkins03}, as is the case here. This assumption was shown by \cite{Gun11} to have a minimal impact on all but the lowest SFR systems in the GAMA sample, and we explore this further in the context of the H$\alpha$ LF in \S\,\ref{constEWc}.}

Similarly to GAMA, the redshifts of all the nearby galaxies in SDSS--DR7 are corrected for peculiar motions using \cite{Tonry00}.
The photometric measurements are {from the New York University value added catalogue \citep{Blanton05}\footnote{\url{http://sdss.physics.nyu.edu/vagc/}}}, with k--corrections to $z=0$ and the maximum redshift ($z_{\rm max}$) for each object derived using \textsc{kcorrect\_v4\_2} \citep{Blanton07} and the spectroscopic and flow--corrected redshifts of each object. Strictly speaking, heliocentric redshifts should be used in the estimation of k--corrections, although the difference in k--correction when using heliocentric or flow--corrected redshifts is negligible \citep{Loveday11}. In summary, aside from the differences in emission--line and photometric measurements, other aspects such as the derivation of k-corrections and flow corrections are the same between the two samples.

The same flux selection in H$\alpha$ used to select the GAMA star forming sample is also applied to SDSS emission--line galaxies, and redshift warnings and standard flags given by the aforementioned databases are used to remove artefacts/sources near stars. The final SDSS emission--line sample consists of 491\,501 galaxies from which 14\% are classified as AGNs and excluded from our sample.

\section{Measuring luminosities and star formation rates}\label{PhysicalProp}
\subsection{Measuring H$\alpha$ luminosities}

As outlined in \cite{Gun11} and \cite{Hopkins03}, measuring H$\alpha$ luminosities, and star formation rates, from fibre spectroscopy requires not only corrections for stellar absorption and obscuration, but also a correction for the aperture sampled by the fibre. Corrections for these effects are applied to all GAMA and SDSS galaxies as described below.

Following \cite{Hopkins03}, we derive an aperture, obscuration and Balmer stellar absorption corrected luminosity ($L_{H\alpha, int}$ in the units of Watts) for the whole galaxy using their $k$--corrected absolute magnitudes (M$_r$), and emission--line equivalent widths (EW). A correction for the missing flux due to aperture effects is applied to each galaxy, using M$_r$ to estimate the continuum at the wavelength of H$\alpha$. This approach of applying aperture corrections to individual galaxies, described in detail in \cite{Hopkins03}, yields similar results to the more complex colour gradient--based method described in \cite{Brinchmann04}. This type of aperture correction can underestimate emission--line luminosity \citep{Gerssen12}, however such effects are likely to be minimal in this analysis as we are using a large sample of galaxies. The relation from \cite{Hopkins03} is:
\begin{eqnarray}
	L_{H\alpha, int} &=& (EW_{H\alpha} + EW_c) \times 10^{-0.4(M_r - 34.10)} \nonumber \\
	&& \times \frac {3\times 10^{18}}{[6564.61(1+z)]^2} \Big (\frac{F_{H\alpha}/F_{H\beta}}{2.86} \Big)^{2.36}.
	\label{maths:apertureObsCor_lum}
\end{eqnarray}
A constant correction for stellar absorption (EW$_c$=2.5\AA) in Balmer emission line EWs is assumed for the calculation of luminosities for the GAMA galaxies \citep{Hopkins13}.  This value is chosen by comparing a sample of line fluxes used in this study against a robust sub--sample of line measurements using GANDALF \citep{Sarzi06}. The choice of the stellar absorption correction, however, does not significantly affect the resulting LFs as shown in Figure\,\ref{fig:GAMAlowLF_stellar}.

Stellar absorption corrected emission line fluxes are used in the determination of Balmer decrements (ratio of H$\alpha$ to H$\beta$ fluxes, $F_{H\alpha}/F_{H\beta}$) for each object in the two galaxy samples. 
\begin{equation}
	{\frac{F_{H\alpha}}{F_{H\beta}} = \frac{\frac{(H\alpha EW + EW_c)}{H\alpha EW}\times f_{H\alpha}}{\frac{(H\beta EW + EW_c)}{H\beta EW}\times f_{H\beta}}},
\end{equation}
where, $ f_{H\alpha}$ and  $f_{H\beta}$ denote the measured emission line fluxes.
The dust obscuration in the Balmer lines H$\alpha$ and H$\beta$ can be determined from the comparison of measured Balmer decrements with the Case B recombination theoretical value of 2.86 at an electron temperature of \mbox{$10^4$~K} and an electron density of \mbox{100~cm$^{-2}$} \citep{Osterbrock89}. The departure of the Balmer decrement from 2.86 can be used to correct for the dust extinction intrinsic to the galaxy. The exponent of the Balmer Decrement in Eq.\,\ref{maths:apertureObsCor_lum} is defined to be $k(\lambda_{H\alpha})/[k(\lambda_{H\beta}) - k(\lambda_{H\alpha})]$, where $k(\lambda)$ is determined from the \cite{Cardelli89} Galactic dust extinction curve.

A small subset of galaxies in the GAMA and SDSS samples (13\% and 4\% respectively) have Balmer decrements $<2.86$. Balmer decrements less than the theoretical Case B value can result from an intrinsically low reddening combined with uncertainty in stellar absorption, and also from errors in the line flux calibrations, and measurements \citep{Kewley06}. Although, some of these low values are probably a result of galaxies hosting \ion{H}{II} regions with high electron temperature, for which the theoretical H$\alpha$/H$\beta$ ratio is lower than 2.86 \citep{LE09}. These galaxies are included in the final GAMA and SDSS samples, assuming no obscuration (i.e.\,Balmer decrement is set to 2.86).
\begin{figure}
\begin{center}
	\includegraphics[scale=0.42]{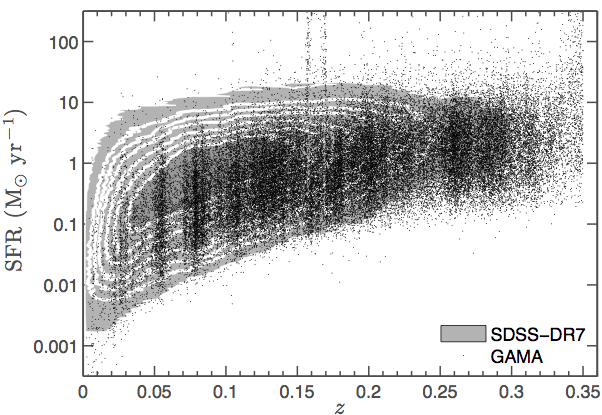}
	\caption{ SFR versus redshift distributions of GAMA (black data points) and SDSS--DR7 (grey contours) surveys. {While GAMA spectra are telluric absorption corrected, given the deep magnitude limits of the survey many of our sources are at or close to the S/N limit. The applied telluric absorption correction can therefore be unreliable over the wavelength ranges of strong atmospheric absorption bands as can be seen by the slight drop in GAMA SFRs centred at $z\sim0.16$, corresponding to the $z$ range where redshifted H$\alpha$ emission line overlaps with the O$_2$ atmospheric (A) absorption band. These galaxies are removed from our sample. The drop in SFRs evident at $z\sim0.14$ is due to atmospheric absorption effects on H$\beta$. For these galaxies, we estimate Balmer decrements empirically, see \S\,\ref{BD}. The SDSS sample is not limited by these constraints as the majority are bright sources with comparatively higher S/N than GAMA.}}
\label{fig:SFRs}
\end{center}
\end{figure}

The SFRs in units of M$_{\odot}$yr$^{-1}$ are derived using the calibration \citep{WAS10},
\begin{equation}
	SFR = \frac{L_{H\alpha, int}}{3.43\times10^{34} W},
	\label{eqn:SFR_calib}
\end{equation}
which assumes the IMF definition of \cite{BG03}.

The majority of the SFR measurements reported in the literature use the \cite{Kennicutt98} calibration based on the \cite{Salpeter55} IMF. The SFR densities reported in this paper assume a slightly flatter than Salpeter IMF, taken from \cite{BG03}. The motivation here is the observed GAMA SFR--IMF relationship \citep{Gun11}, where moderate--to--high SFR galaxies are characterised with flatter than Salpeter IMFs. The ratio of the calibration given in Eq.\,\ref{eqn:SFR_calib} to  the \cite{Kennicutt98} calibration is $\sim 2.4$ with our derived SFRs being lower than if the \cite{Kennicutt98} calibration had been used.  We use the SFR calibration based on the \cite{BG03} IMF throughout this paper, unless otherwise stated. The distributions of SFRs of the GAMA and SDSS samples are shown in Figure\,\ref{fig:SFRs}. 

\subsection{Estimating Balmer decrements}\label{BD}

The large number of weak emission line galaxies observed in GAMA gives the opportunity to investigate nearby low--SFR systems \citep{Brough11}, and the low--$z$ evolution of the SFR density. As H$\beta$ is a considerably weaker emission feature than H$\alpha$, not all weak H$\alpha$ sources in our final GAMA/SDSS samples have measured H$\beta$ fluxes (Table\,\ref{table:stats}). 
\begin{table*}
\caption{The total number of GAMA galaxies with H$\alpha$ fluxes above the detection limit (i.e.\,\mbox{$25\times10^{-20}$W\,m$^{-2}$}) in four different redshift bins up to $z\sim0.34$ and the approximate percentage of objects without measured Balmer decrements are given in the first part of the table. The SDSS galaxy numbers given are the SDSS galaxies in the three GAMA regions. Note that we have imposed a flux limit of  \mbox{$1\times10^{-18}$W\,m$^{-2}$} (see \S\,\ref{HaFluxlimit_main}) to construct the GAMA LFs presented in this paper. The second part of the table (the last two entries) indicates the total number of SDSS--DR7 galaxies with H$\alpha$ fluxes above the flux limit (i.e.\,\mbox{$1\times10^{-18}$W\,m$^{-2}$}) in two different redshift bins up to $z\sim0.2$ and the approximate percentage of objects without measured Balmer decrements. }
\begin{tabular}{| ccccccc |}
    \hline
    $z$ 									& 	Total No. of galaxies          	&	No. from		&	No. from		 	 &	No. from 		&   $\%$ without BDs	 & $\%$ with BD$<2.86$\\
    										&       GAMA + SDSS in GAMA		&	GAMA--9h	&	GAMA--12h		 & 	GAMA--15h	&										\\
    \hline
    \hline   
    $0.001<z<0.1$ 							
    & 	6928    + 3153				
    &	2936			
    &	4025 			
    &	3120			
    &  4				
    &	24 	\\
    
    $0.1<z<0.15$							
    & 	10700    + 2080
    &	2714			
    &	5179				
    &	4887		
    &  9			 
    &	16 \\
    
    $0.17<z<0.24$							
    & 	13287  + 462 				
    &	4284			
    &	5618				
    &	3847		
    & 17			 
    & 12 \\
    
    $0.24<z<0.34$  							
    & 12262  + 126   				
    &	3384			
    &	5962			
    &  3042		
    &  18			 
    &	 9	\\
    
    \hline
    $z$ 									& 	Total No. of galaxies          	&				&				 	 &				&   $\%$ without BDs	 & $\%$ with BD$<2.86$\\
    										&       SDSS--DR7 				&				&					 & 		   		&										\\
  \hline
    $0.001<z<0.1$ 							
    & 	140791					
    &	-			
    &	-	 			
    &	-			
    &   $<0.1$			
    &	2	\\
    
    $0.1<z<0.2$ 								
    & 	70534 					
    &	-			
    &	- 				
    &	-			
    &   $0.3$		
    &	$0.2$	\\
  \hline
\end{tabular}
\label{table:stats}
\end{table*}
\begin{figure}
\includegraphics[scale=0.4]{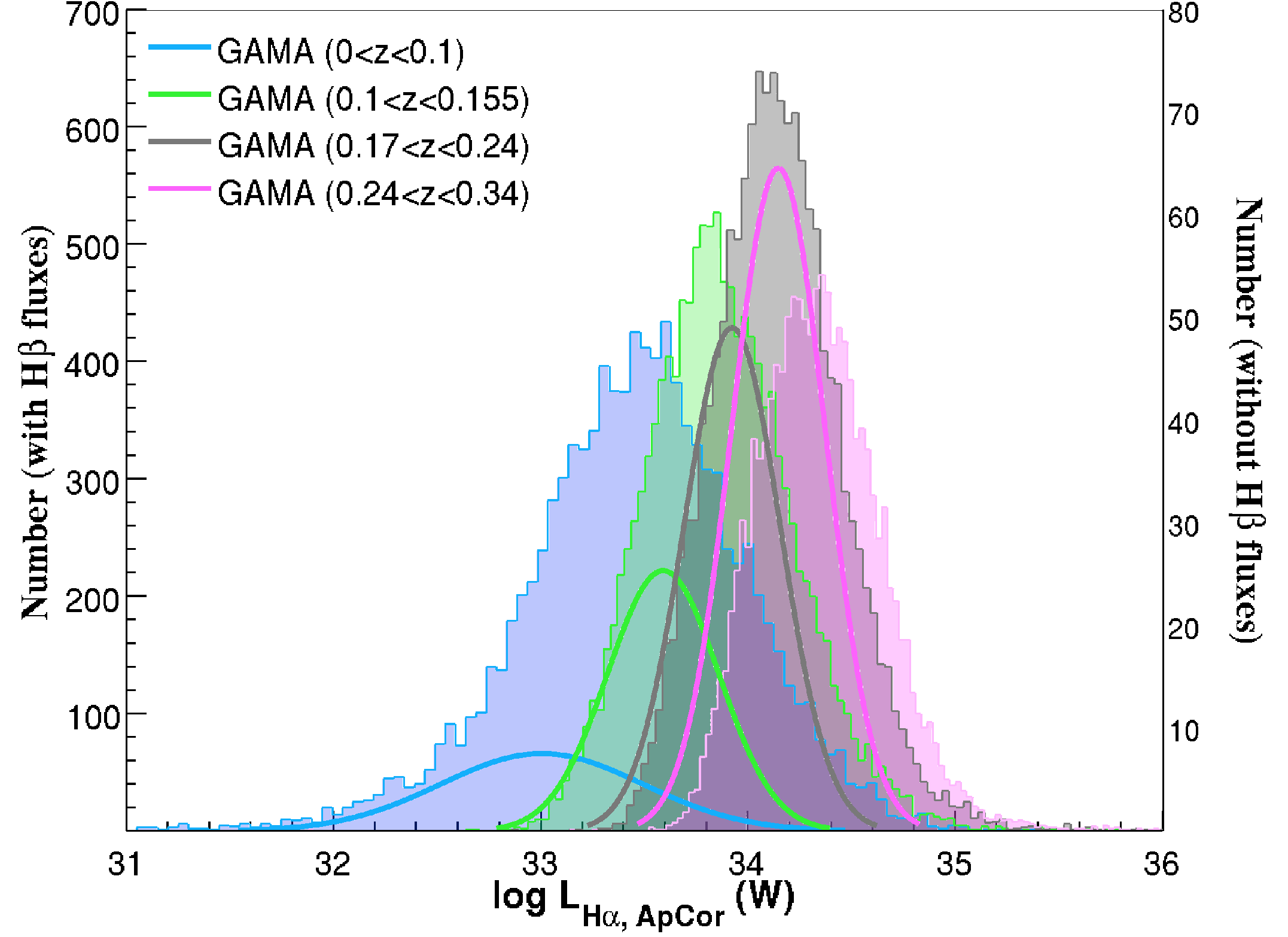}
\caption{\small The distributions of aperture corrected luminosities of galaxies with (histograms, with respect to the left y--axis scaling) and without (solid lines, with respect to the right y--axis scaling) measured Balmer decrements. For the galaxies without measured Balmer decrements, Balmer decrements are estimated using Eq.\,\ref{eq:BDvL_gama}.}
\label{fig:GAMALF_stats}
\end{figure}
The distributions of GAMA star forming galaxies with and without  measured Balmer decrements in several redshift bins are shown in Figure\,\ref{fig:GAMALF_stats}, and detailed in Table\,\ref{table:stats}. As expected the distributions of galaxies without Balmer decrements are skewed towards low--luminosity (weak line) galaxies in all redshift ranges. 

\begin{figure}
\includegraphics[scale=0.55]{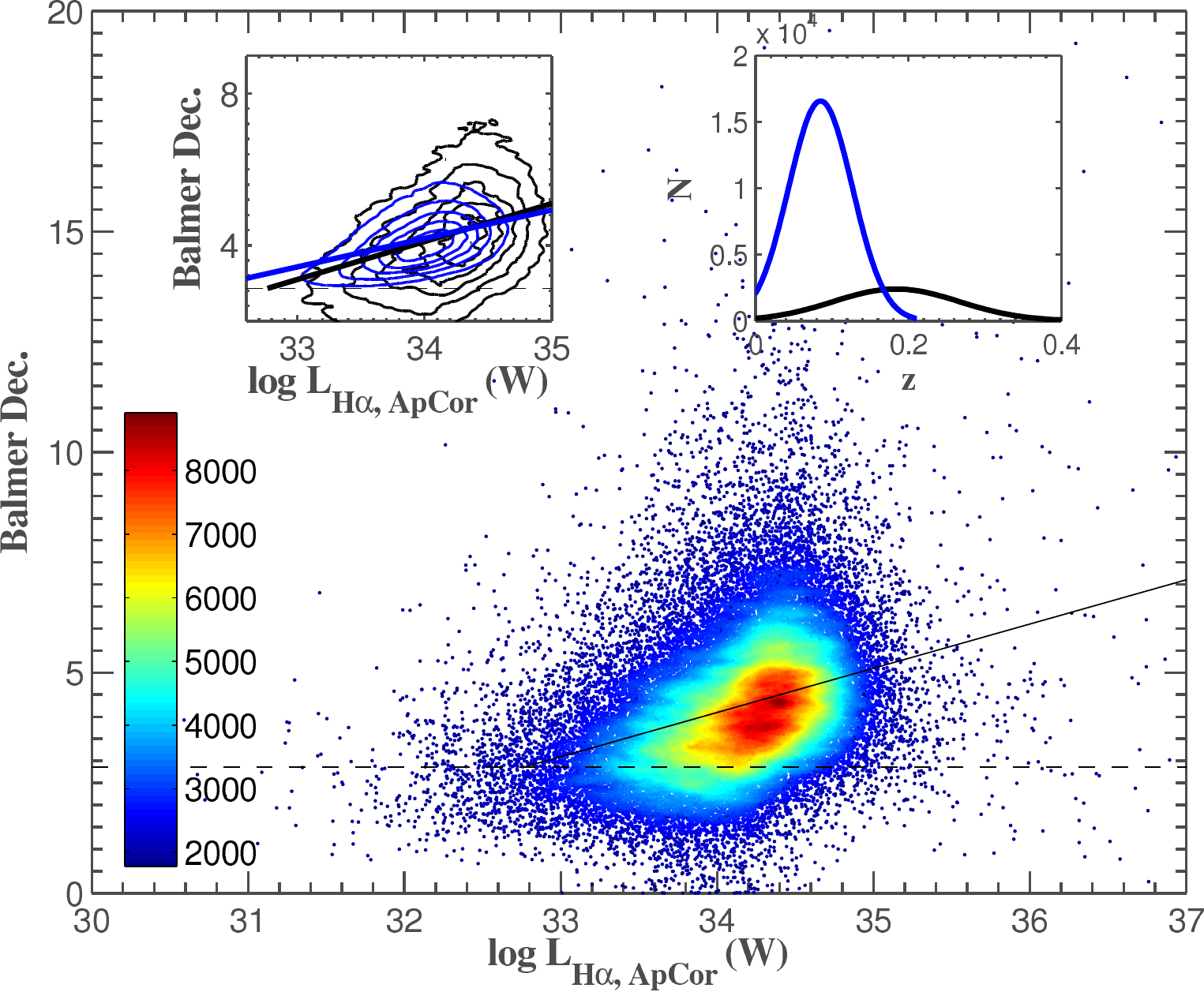}
\caption{Balmer decrement versus aperture corrected luminosity for the GAMA galaxies with measured Balmer decrements. The dashed line indicates the case--B recombination value of 2.86, and the solid line shows the best--fit linear relation to the data (Eq.\,\ref{eq:BDvL_gama}). The colorbar indicates the data density in units of per $\log\,L_{H\alpha, ApCor}$ per Balmer decrement. The two insets compare GAMA (black) and SDSS--DR7 (blue) samples. The left inset shows the Balmer decrement versus aperture corrected luminosity for GAMA and SDSS--DR7 star forming samples, and their respective best--fit linear relations. The difference between the two best fit relations is an indirect consequence of the different redshift distributions of the surveys, as shown in the right inset, leading to a sampling in GAMA of both higher SFR (more obscured) systems at higher redshift, as well as fainter (more obscured) systems at lower redshift.}
\label{fig:BDvL}
\end{figure}

\begin{figure*}
\begin{center}
\includegraphics[scale=0.49]{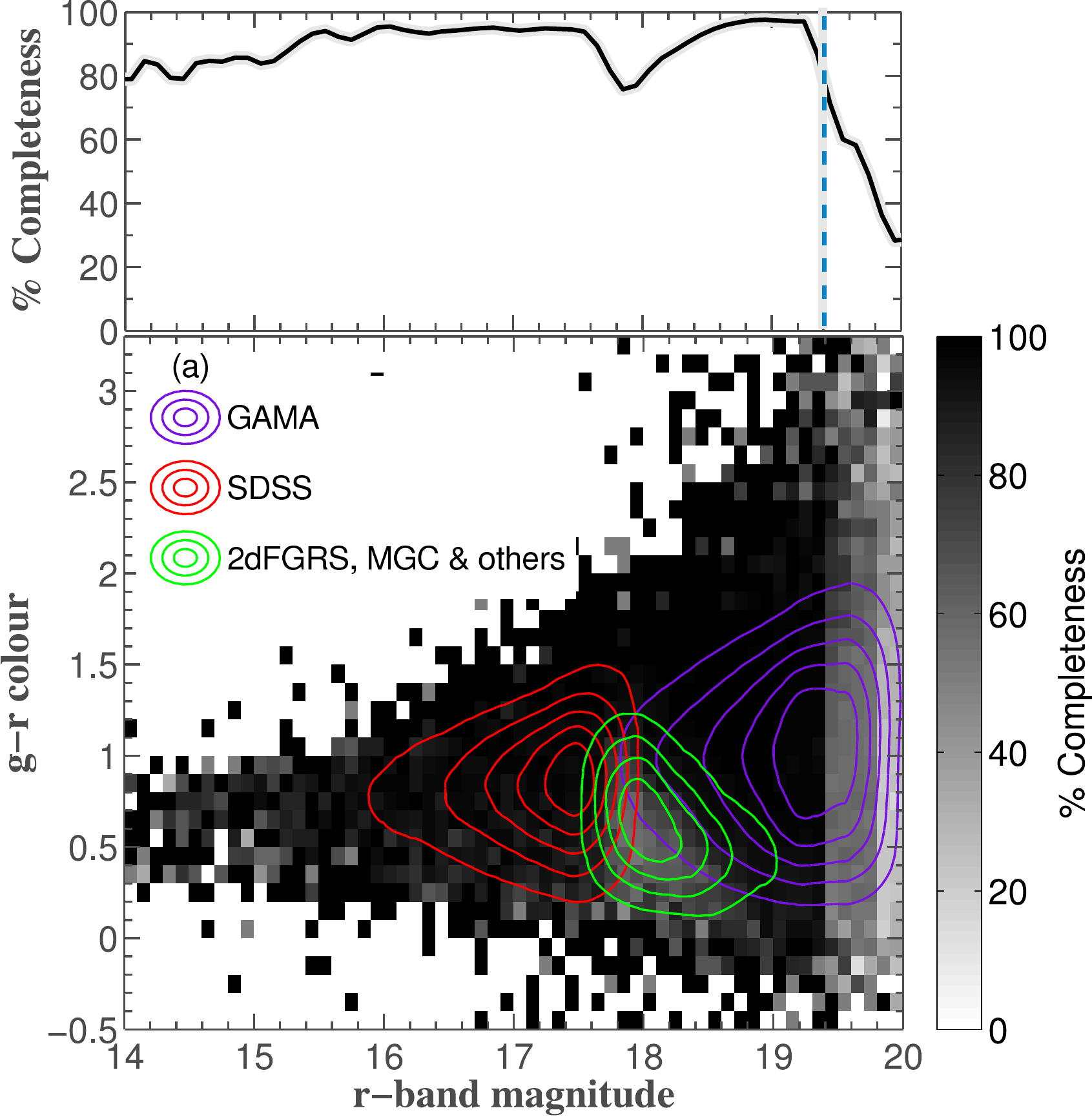}
\includegraphics[scale=0.47]{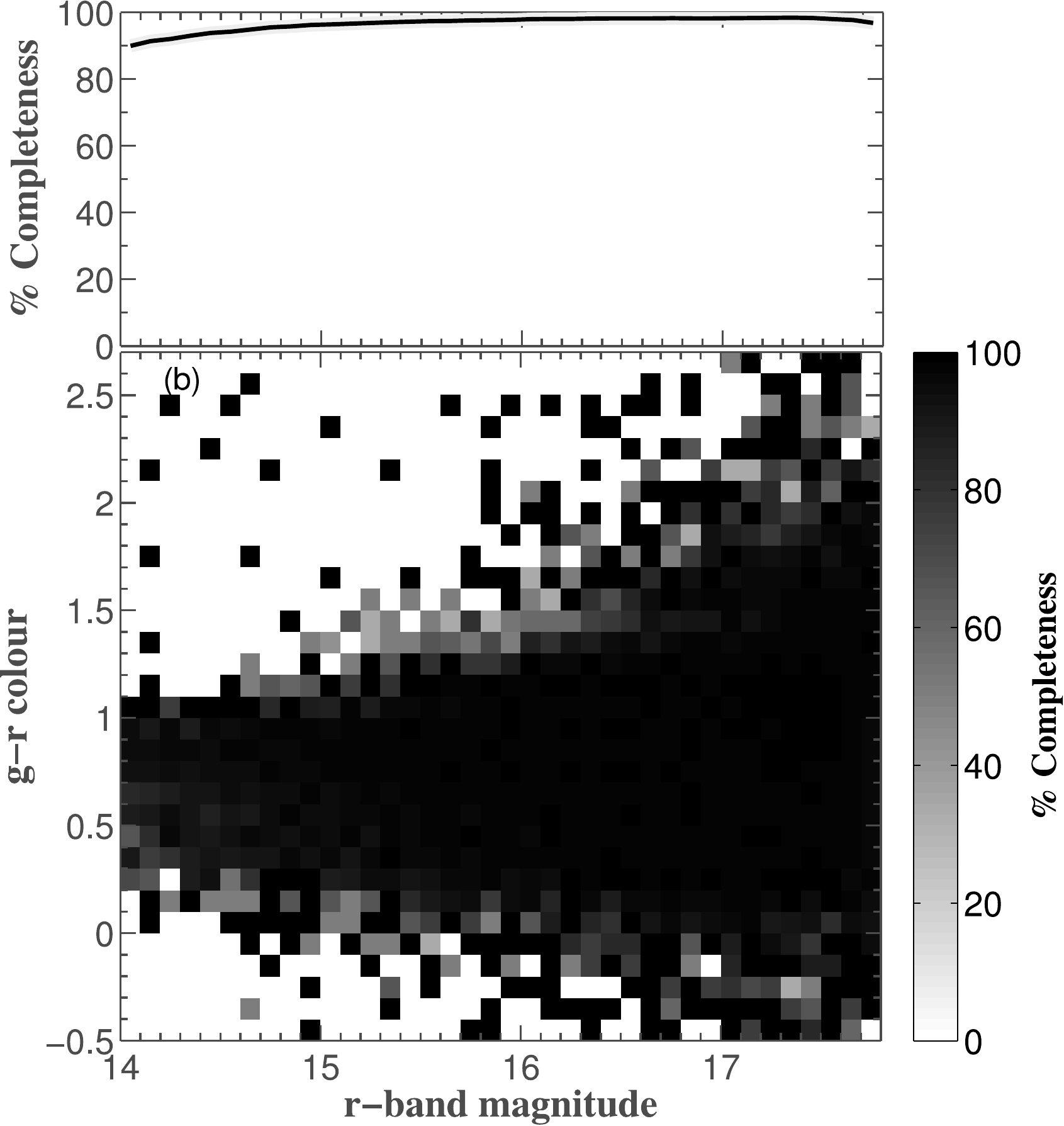}
\caption{A correction ($C_{spec, z}$) based on the distribution of galaxies in $r$--band petrosian magnitude and $g-r$ colour is applied to the LFs to account for the spectroscopic incompleteness and redshift success rate. This correction also takes into account the sample incompleteness due to the lack of 2dFGRS, MGC and 6--degree Field Galaxy Redshift Survey (6dFGRS) data.  The panels (a) and (b)  indicate the percentage completeness (grey scale) at a given $r$--band magnitude and $g-r$ colour bins for the complete GAMA and SDSS--DR7 samples respectively. Bin widths in $r$--band magnitude and $g-r$ colour are $0.1$. The purple and red contours shown in (a) indicate the distributions of objects observed by GAMA and SDSS surveys, and the green contours correspond to the spectra we were not able to measure flux calibrated lumnosities, and are not included in our sample. $C_{spec, z}$ as function of $r$--band magnitude is shown in the top panel.}
\label{fig:completeness}
\end{center}
\end{figure*}
\begin{figure*}
\begin{center}
\includegraphics[scale=0.46]{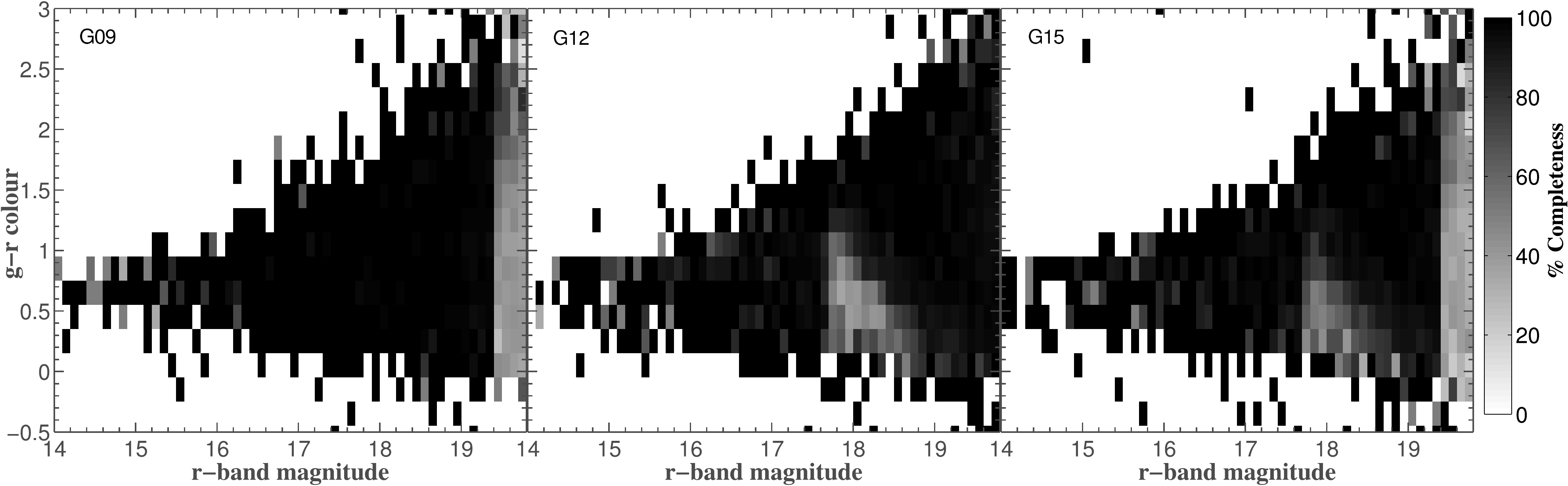}
\caption{$C_{spec, z}$ corrections for each of the GAMA fields separately. The 2dFGRS and MGC survey areas do not overlap with the GAMA 09hr region. The incompleteness around (r, g-r)$\sim$(18, 0.5) in GAMA 12hr and 15hr regions is due to the lack of H$\alpha$ measurements from the 2dfGRS and MGC surveys.}
\label{fig:completeness_G}
\end{center}
\end{figure*}

For these galaxies, we estimate Balmer decrements using the observed relationship between Balmer decrement and aperture corrected luminosity (Figure\,\ref{fig:BDvL}). {The solid line in Figure\,\ref{fig:BDvL} indicates the least absolute deviation fit to the data:
\begin{equation}
{\rm BD_{gama}}= 
\begin{cases} 1.003\times\log L - 30.0 & \text{$\log L \geqslant 32.77$,}
\\
2.86 &\text{$\log L < 32.77$.}
\end{cases}
\label{eq:BDvL_gama}
\end{equation}
A similar relationship is derived for the SDSS--DR7 sample using their Balmer decrements and aperture corrected luminosities. 
\begin{equation}
{\rm BD_{sdss}}= 
\begin{cases} 0.761\times\log L - 21.7 & \text{$\log L \geqslant 32.27$,}
\\
2.86 &\text{$\log L < 32.27$.}
\end{cases}
\label{eq:BDvL_sdss}
\end{equation}
This relationship is then used to estimate Balmer decrements for the SDSS galaxies without Balmer decrements.}

As the GAMA and SDSS surveys probe different star--forming populations, we do not attempt to determine a single fit to the data by combining GAMA and SDSS--DR7 data sets. In contrast to SDSS, the GAMA sample consists of more dust obscured optically faint galaxies at higher redshift, a single fit would, therefore, under-estimate the empirical Balmer decrement correction needed for GAMA galaxies as the fit would be heavily weighted by the relatively numerous SDSS galaxies.

{\color{black} Furthermore, we empirically estimate Balmer decrements for all the sources with measured Balmer decrements$>10$ to avoid the sample being contaminated by sources with overestimated Balmer decrements, a result of weak H$\beta$ measurements.}

Balmer decrement, as an indicator of dust obscuration, scales with both the SFR and redshift. High SFR galaxies typically have greater obscuration than low--SFR systems \citep{Afonso03, Hopkins01}, and are generally found at higher redshifts. The insets in Figure\,\ref{fig:BDvL} illustrate this point. GAMA galaxies with higher $\langle z \rangle \approx 0.2$ (and therefore higher average SFRs) than SDSS ($\langle z \rangle \approx 0.08$) have relatively high Balmer decrement values. 

While only a small percentage of objects are without BDs in GAMA/SDSS samples, this small percentage consists mostly of low--$z$, low luminosity systems. We demonstrate in \S\,\ref{empBDs} below, that the impact of empirically assigning Balmer decrements on our derived LFs is minimal.
 
\section{The luminosity function}\label{LF}

For the LF estimates we use the $V_{\rm max}$ \citep{Schmidt68} method. In this section, we describe the derivation of $V_{\rm max}$ for galaxies in our sample subject to our selection criteria. We then detail the estimation of the luminosity functions.

\subsection{Derivation of volume corrections} \label{vmax_derivation}

The H$\alpha$ luminosity function, $\Phi(L)$, is defined as the number of star forming galaxies per unit volume per unit luminosity \citep{Schmidt68}, and has the general form,
\begin{equation}
\small
\Phi[\log L(H\alpha)] \times \Delta L = \\ \frac{4\pi}{\Omega} {\sum_i} \frac{1}{V_{i, max}}.
\label{eq:vmax_def}
\end{equation}
In this equation, $V_{i,max}$ represents the maximum volume out to which the $i^{th}$ object would be visible to and still be part of the survey, $ \Delta L$ and $\Omega$ define the assumed luminosity bin width and surveyed solid angle respectively.   

The H$\alpha$ star forming samples used in this study are subject to several selection constraints. For the GAMA sample, these are the two different $r$--band magnitude limits ($r<19.4$ for G09, 15 and $r<19.8$ for G12) of the survey \citep{Driver11}, and the emission--line selection. Similarly for the SDSS--DR7, the emission--line selection and $r<17.77$ magnitude limit. Given these constraints, the definition of $V_{i,\rm max}$ is 
\begin{equation}
\small
V_{i,max} = \min[(V_{i, max, H\alpha}), (V_{i, max, r}), (V_{i, z_{lim}})] \times c_i,
\label{eq:vmax_def2}
\end{equation}
where $V_{i,\rm max}$ is the minimum of the maximum volumes that the $i^{th}$ galaxy would have given the flux limit ($V_{i, {\rm max}, H\alpha}$), and magnitude limit ($V_{i, {\rm max}, r}$) of the surveys, and $c_i$ denotes the completeness correction. 

The completeness corrections are made to each galaxy by weighting object numbers by the known missing fraction brighter than the survey magnitude.  As noted in \cite{Jones01}, this type of a correction accounts for the survey incompleteness relatively accurately provided the observed fraction of galaxies is large. This is certainly the case with GAMA, which has a spectroscopic completeness $>98\%$ \citep{Driver11}. 

The three main sources of incompleteness, as identified by \cite{Loveday11} for the GAMA sample, are imaging incompleteness, spectroscopic incompleteness and redshift success. A correction for the imaging incompleteness ($C_{im}$) is estimated from Figure\,1 of \cite{Loveday11}, while an empirical correction for both spectroscopic incompleteness and redshift success ($C_{spec, z}$) is applied based on the detection probability of a galaxy in the $r$--band petrosian magnitude and $g-r$ colour in a given GAMA field. This correction is estimated relative to the GAMA tiling catalogue \citep{Loveday11, Robotham10} and accounts for the missing sets of data (i.e.\,2dFGRS, MGC), see Figures\,\ref{fig:completeness}(a) and \ref{fig:completeness_G}. The final weighting is given as
\begin{equation}
W = \frac{1}{C_{im}\,C_{spec, z}}.
\end{equation}

A similar completeness correction for the SDSS--DR7 is also implemented. $C_{spec, z}$ correction is based on the SDSS--DR7 main galaxy spectroscopic sample chosen from the photometric catalogues. Similarly to GAMA, $C_{spec, z}$ correction for SDSS--DR7 takes into account 2dFGRS, PSCz, and RC3 sources that are not part of our sample, see Figure\,\ref{fig:completeness}(b). The imaging incompleteness correction ($C_{im}$) for SDSS--DR7 is derived from \cite{Blanton05a}.

\subsubsection{Broadband volume corrections}

The determination of $V_{i, max, r}$ for the SDSS sample is relatively straightforward given the single magnitude limit of the survey. For GAMA galaxies however, we estimate $z_{\rm max}$, at which that galaxy would still satisfy the $r < 19.4$ (for G09 and G15 fields) or $r <19.8$ (for G12) selection criteria. The $z_{\rm max}$ values have been derived using the stellar template spectrum that best fits {\em u,g,r,i,z\/} photometry \citep[StellarMassesv08,][]{Taylor11}.  Note that the values of $z_{\rm max}$ are flow corrected \citep{Baldry12}. $V_{i,{\rm max}, r}$ for GAMA becomes
\begin{equation}
\small
V_{i,max,r}= 
\frac{2}{3} (V_{i,max,r=19.4}) + \frac{1}{3} (V_{i,max,r=19.8}).
\end{equation}

A similar functional form to this is used in the derivation of $V_{i, max, H\alpha}$. 

\subsubsection{Emission line volume corrections} \label{HaFluxlimit_main}

\begin{figure}
\begin{center}
\includegraphics[scale=0.6]{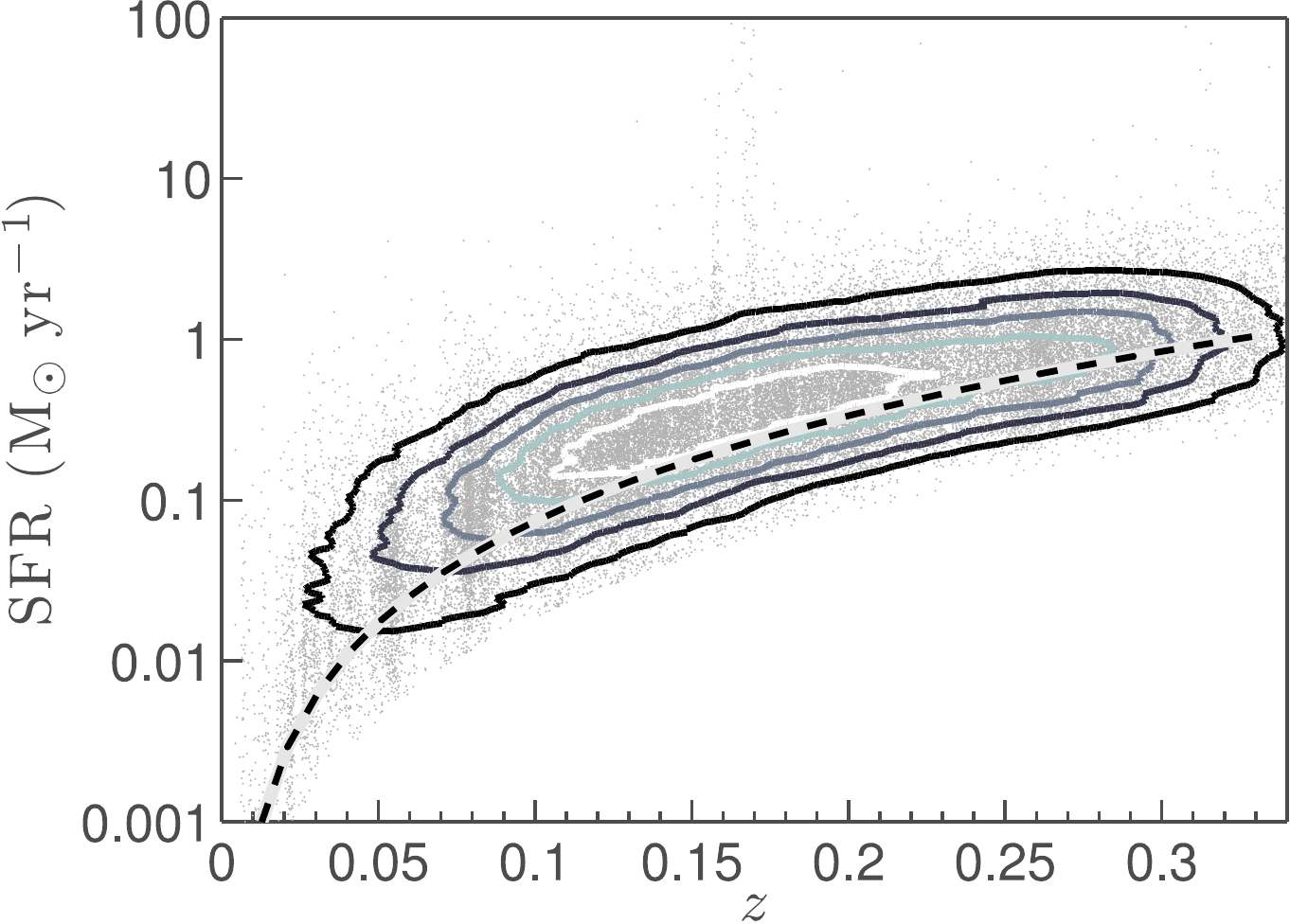}
\caption{The distribution of the observed (i.e.\,not corrected for dust obscuration) H$\alpha$ SFRs (data points and the contours) relative to the SFR corresponding to the assumed flux limit of \mbox{$1\times10^{-18}$W\,m$^{-2}$} (dashed line). Close to half of the sources with detected H$\alpha$ emission lie below the selected H$\alpha$ completeness limit. Note that all the SFR points shown in Figure\,\ref{fig:SFRs} are also shown here. See Figure\,\ref{fig:SFRs} caption for more information.}
\label{fig:obs_sfrs}
\end{center}
\end{figure}

Due to the magnitude--limited nature of the GAMA/SDSS surveys, an approximate H$\alpha$ flux limit of $F(H\alpha) =$ \mbox{$1\times10^{-18}$W\,m$^{-2}$} uncorrected for dust obscuration is assumed for the calculation of $V_{i, max, H\alpha}$ \citep[see][]{Brough11}. This value roughly corresponds to the turn--over in the observed H$\alpha$ flux histogram, and we assume that our sample is incomplete below this limit. Figure\,\ref{fig:obs_sfrs} illustrates the distribution of SFRs in redshift relative to the SFR corresponding to the assumed flux limit.
The impact of our assumptions about the H$\alpha$ flux limit is minimal. This is detailed in \S\,\ref{HaFluxlimit}.

\subsection{H$\alpha$ luminosity functions} \label{HaLFs}

\begin{figure*}
\begin{center}
\includegraphics[scale=0.35]{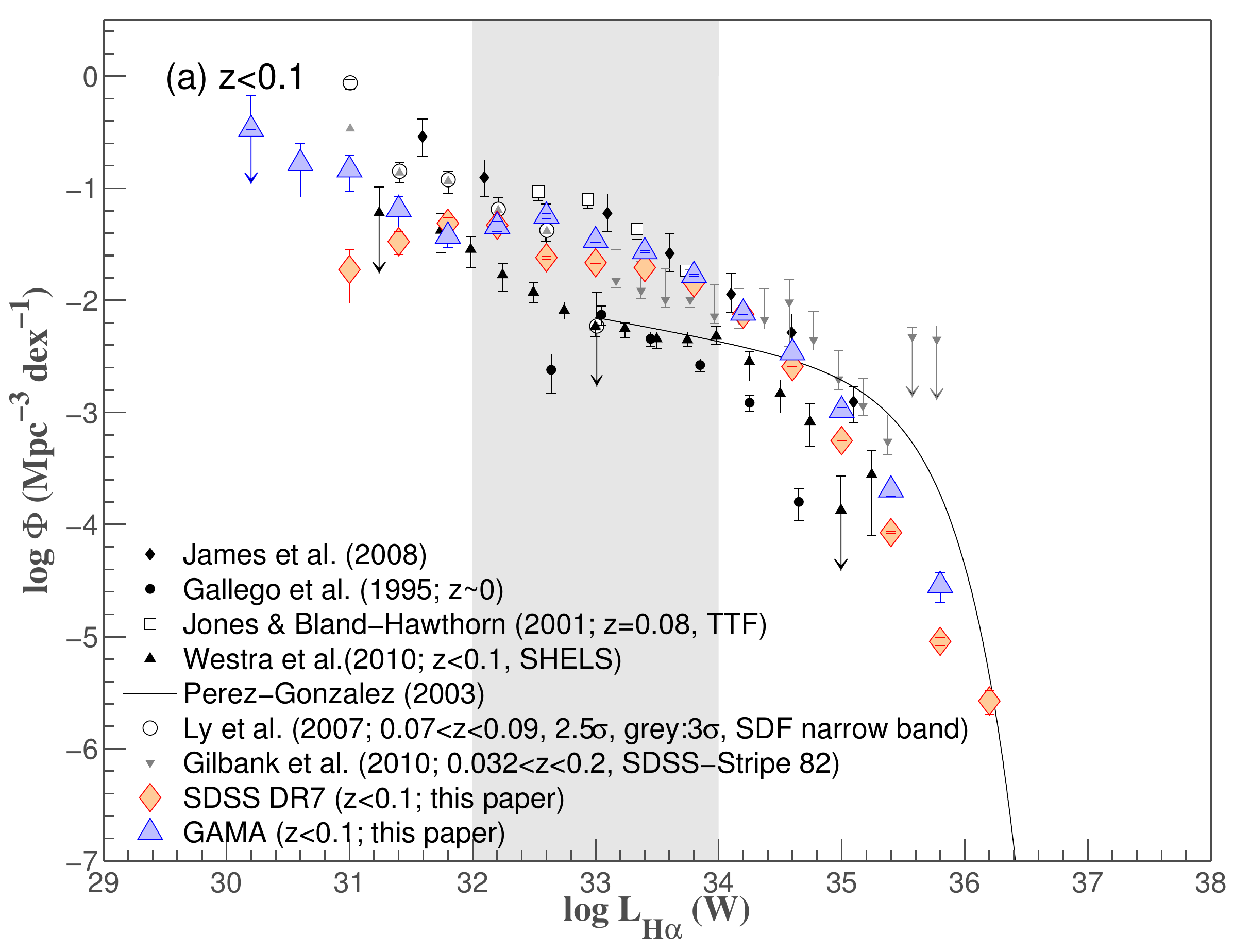}
\includegraphics[scale=0.35]{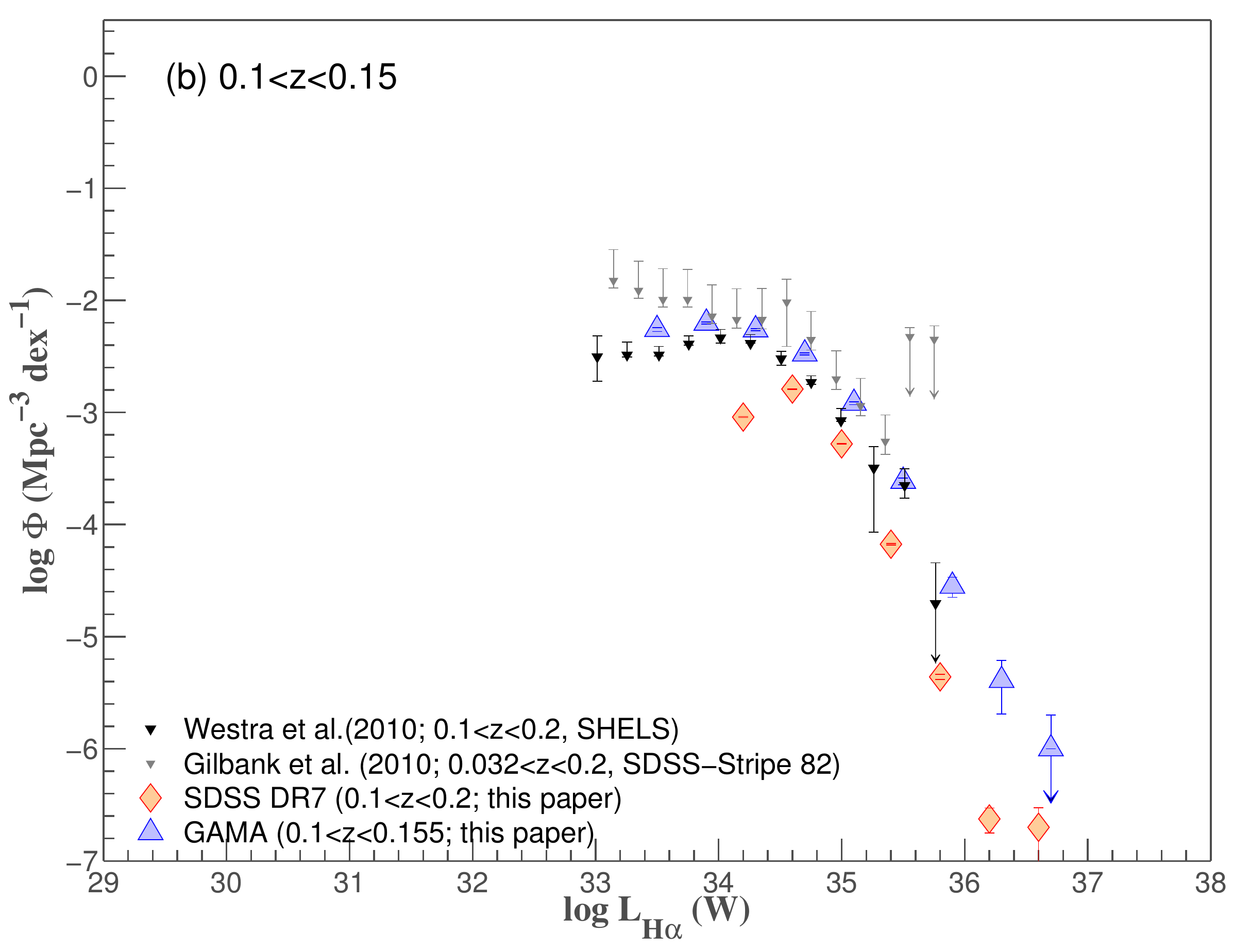}
\includegraphics[scale=0.35]{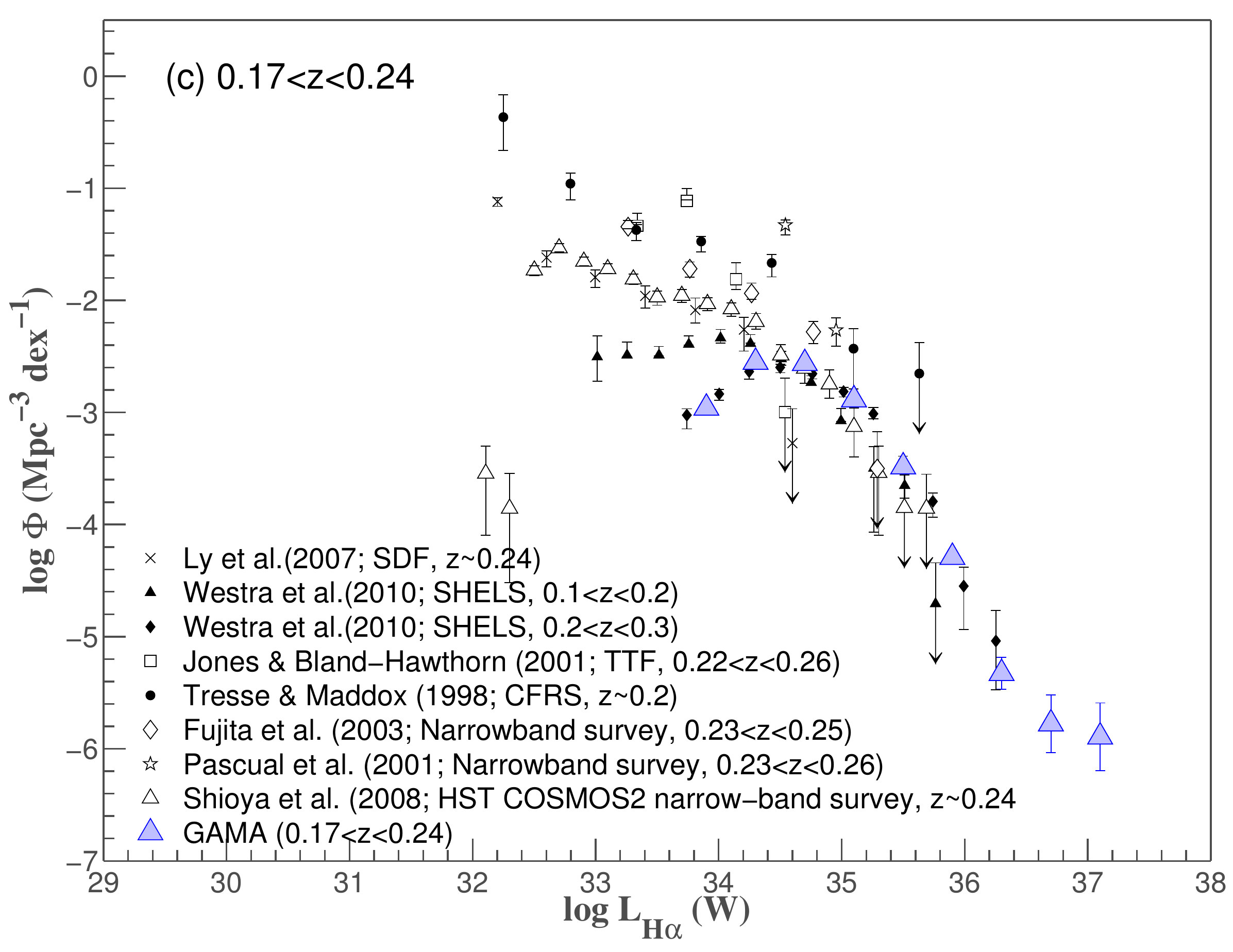}
\includegraphics[scale=0.35]{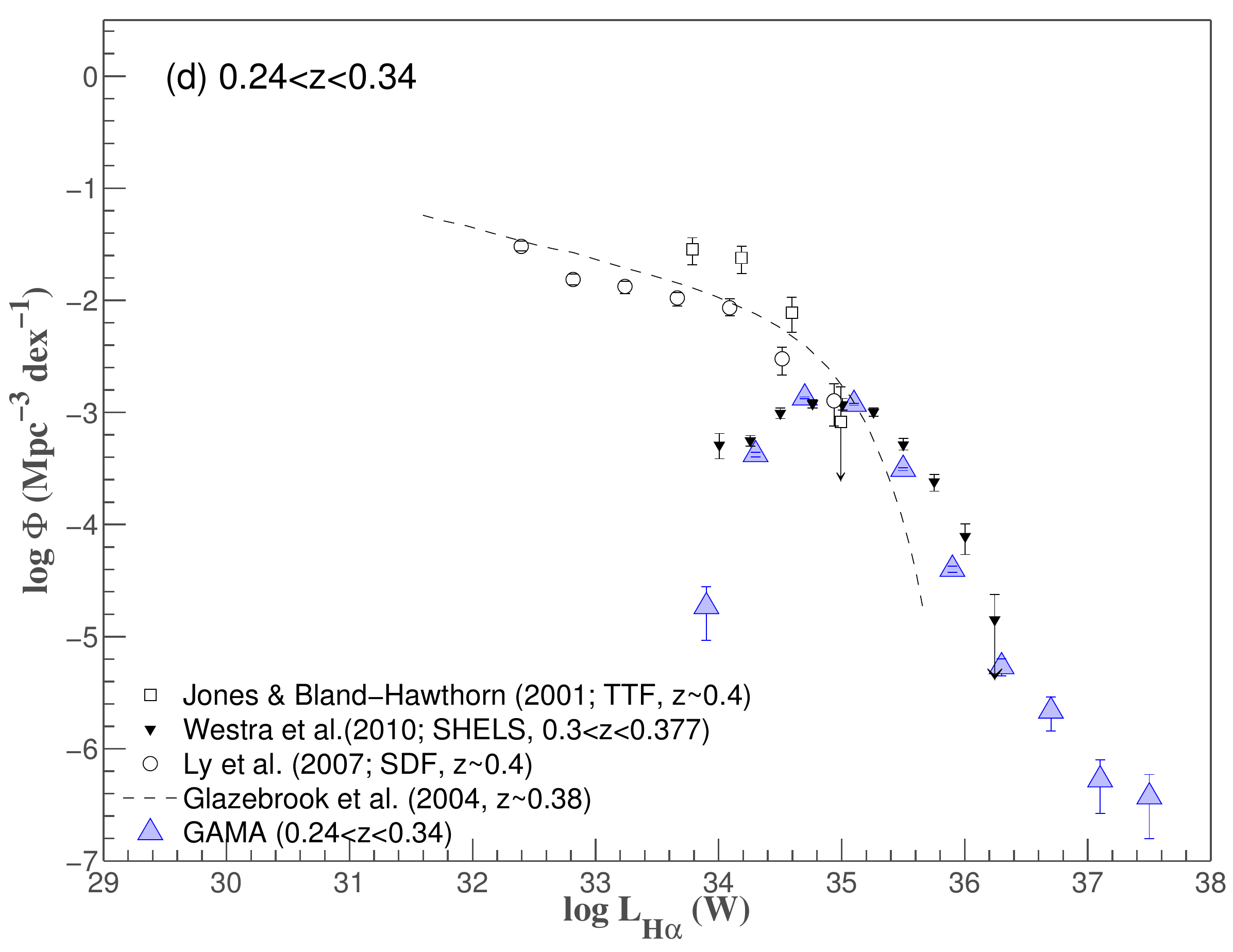}
\caption{ GAMA (blue) and SDSS--DR7 (red) H$\alpha$ luminosity functions in four broad redshift bins (see key in each panel for detailed ranges). The redshift ranges increase from left--to--right, top--to--bottom, covering a total redshift range of $0<z<0.34$ for GAMA and $0<z<0.2$ for SDSS--DR7. The axis ranges in each panel are kept the same to highlight the broad luminosity range sampled by the GAMA LFs. The figure also presents a comparison of our results with published LFs spanning similar redshift ranges. LFs from other authors have been converted to our assumed cosmology. An obscuration correction based on the assumption of a one magnitude extinction in H$\alpha$ \citep{HB06} is applied to correct the observed LFs of \citet{Jones01} and \citet{Gilbank10}. The grey band in (a) highlights the luminosity range over which the discrepancy between \citet{Westra10} and GAMA low--$z$ LFs is largest.}
\label{fig:GAMALFs}
\end{center}
\end{figure*}

H$\alpha$ LFs in several redshift bins are generated using the $V_{\rm max}$ technique described above and are shown in Figure\,\ref{fig:GAMALFs}. The uncertainties in each luminosity bin are Poisson errors. The four panels in Figure\,\ref{fig:GAMALFs} show GAMA LFs in four redshift bins (blue points: $0<z<0.1$, $0.1<z<0.15$, $0.17<z<0.24$, and $0.24<z<0.35$), and SDSS LFs in two redshift bins (orange points: $0<z<0.1$, and $0.1<z<0.2$). The break in redshift between second and third GAMA redshift bins corresponds to the $z\sim0.16$ region where H$\alpha$ measurements are likely to be affected by the atmospheric O$_2$ absorption, see Figure\,\ref{fig:SFRs}. 

All GAMA LFs extend approximately an order of magnitude brighter in luminosity than other published LFs shown in Figure\,\ref{fig:GAMALFs}.
The GAMA low--$z$ LF (Figure\,\ref{fig:GAMALFs} a) extends approximately an order of magnitude in luminosity both fainter and brighter than other published results to date. Furthermore, our result agrees well with other studies in the luminosity range probed by existing data, with the exception of the \cite{Westra10} LF. The disagreement between the GAMA and \cite{Westra10} LFs is largest over the shaded region. This could be due to the relatively small survey area ($\sim4\,$deg$^2$) of \cite{Westra10} sampling an under--dense region.  We demonstrate in \S\ref{Biselection}, however, that there may be a significant impact from the joint $r$--band and emission--line selection, and the assumptions related to H$\alpha$ flux limits for magnitude--limited surveys can contribute to this disagreement. 
These are likely to be the dominant effects. 

The SDSS--DR7 LF explores a similar range in bright luminosities as GAMA and agrees well with both GAMA and published LFs. The turn--over below $L_{H\alpha} \approx10^{31.5}$ W in the SDSS LF is due to the incompleteness arising from the H$\alpha$ line flux limit.

 The GAMA LF over $0.1<z<0.15$ (Figure\,\ref{fig:GAMALFs}b) is in good agreement with the SDSS--Stripe 82 $0.032<z<0.2$ LF of \cite{Gilbank10} within $33\leqslant \log L_{H\alpha} \leqslant 35.5$. The disagreement between the GAMA and SDSS LFs in the second redshift bin is likely due to the brighter SDSS magnitude cut ($r=17.77$) preventing optically faint high--SFR galaxies from entering the SDSS sample (see discussion in \S\,\ref{Biselection}). This assertion is supported by the lack of evolution between $0<z<0.1$ and $0.1<z<0.2$ SDSS LFs. The scatter in published LFs is significant over $0.1<z<0.3$, particularly at relatively low--luminosities, where cosmic (sample) variance, selection and incompleteness issues impact the most. The GAMA LF in the $0.17<z<0.25$ redshift bin certainly provides a better estimate for the bright end of the LF, where other LFs suffer from small number statistics. The final GAMA LF agrees well with \cite{Westra10} at this redshift. This agreement, however, is likely to be a consequence of the bivariate selection of the GAMA sample, as discussed in the next section. The agreement is therefore likely an outcome of both surveys preferentially selecting brighter galaxies at higher redshifts.

\subsection{Bivariate selection} \label{Biselection}
\begin{figure*}
\begin{center}
\includegraphics[scale=0.43]{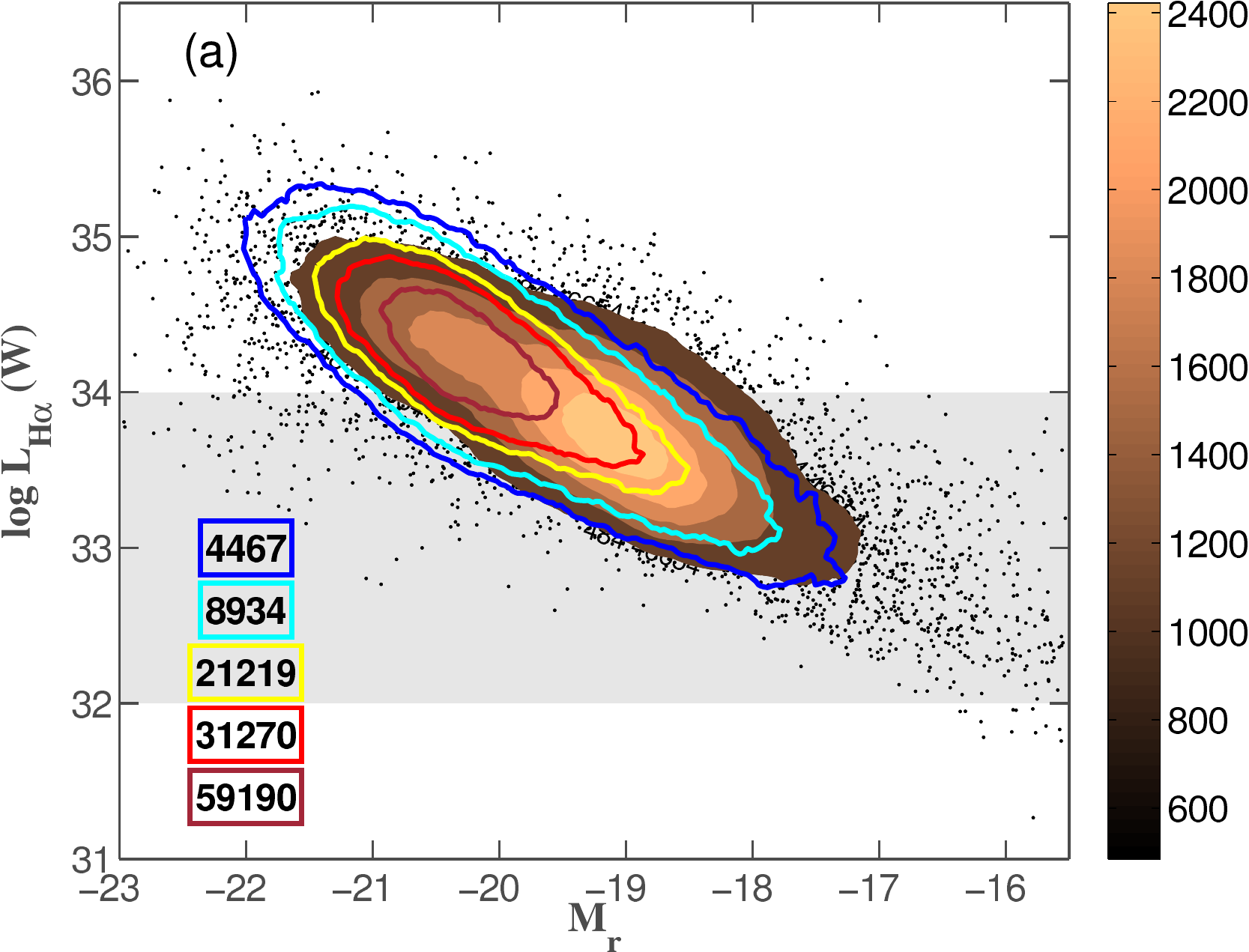}
\includegraphics[scale=0.43]{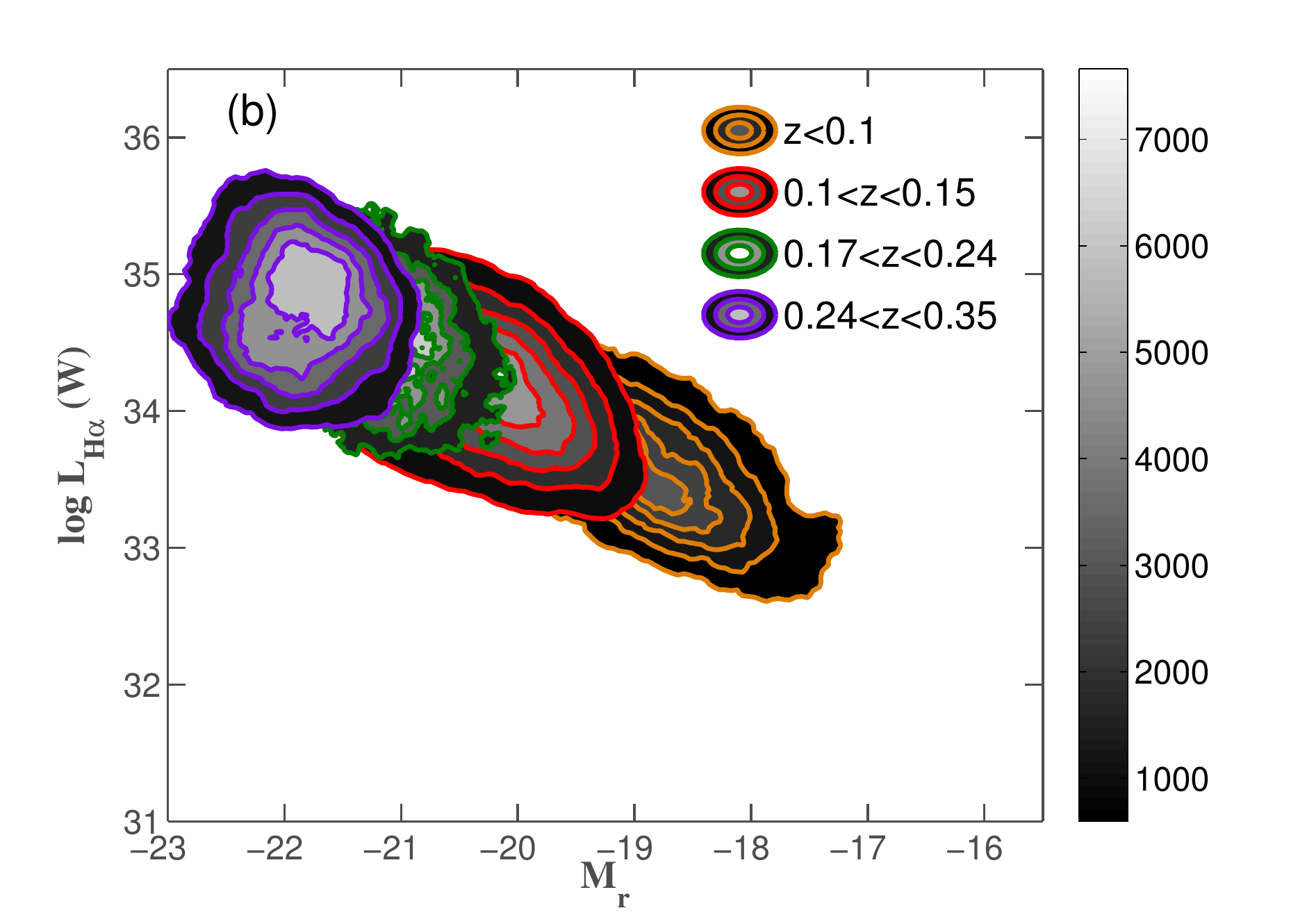}
\caption{GAMA absolute magnitude (M$_r$) versus aperture and obscuration corrected H$\alpha$ luminosity ($L_{H\alpha}$) bivariate distributions. (a) The distribution of GAMA galaxies within $0<z<0.1$ with $r<19.4$ (or $19.8$), compared with SDSS galaxies in the same redshift range with $14.5<r<17.77$. The grey band highlights the same luminosity range as in Figure\,\ref{fig:GAMALFs}a. The GAMA distribution is represented by solid contours and data points, while the SDSS distribution is represented by transparent coloured contours. The brown colour bar indicates the data density colour coding for the GAMA contours, while the data densities corresponding to the SDSS contours are shown in the key. The unit of data density is per M$_r$ per $\log\,L_{H\alpha}$. (b) The H$\alpha$ luminosity versus M$_r$ distributions for the GAMA sample in the four redshift ranges.}
\label{fig:MrDist_GAMA}
\end{center}
\end{figure*}

\begin{figure}
\begin{center}
\includegraphics[scale=0.43]{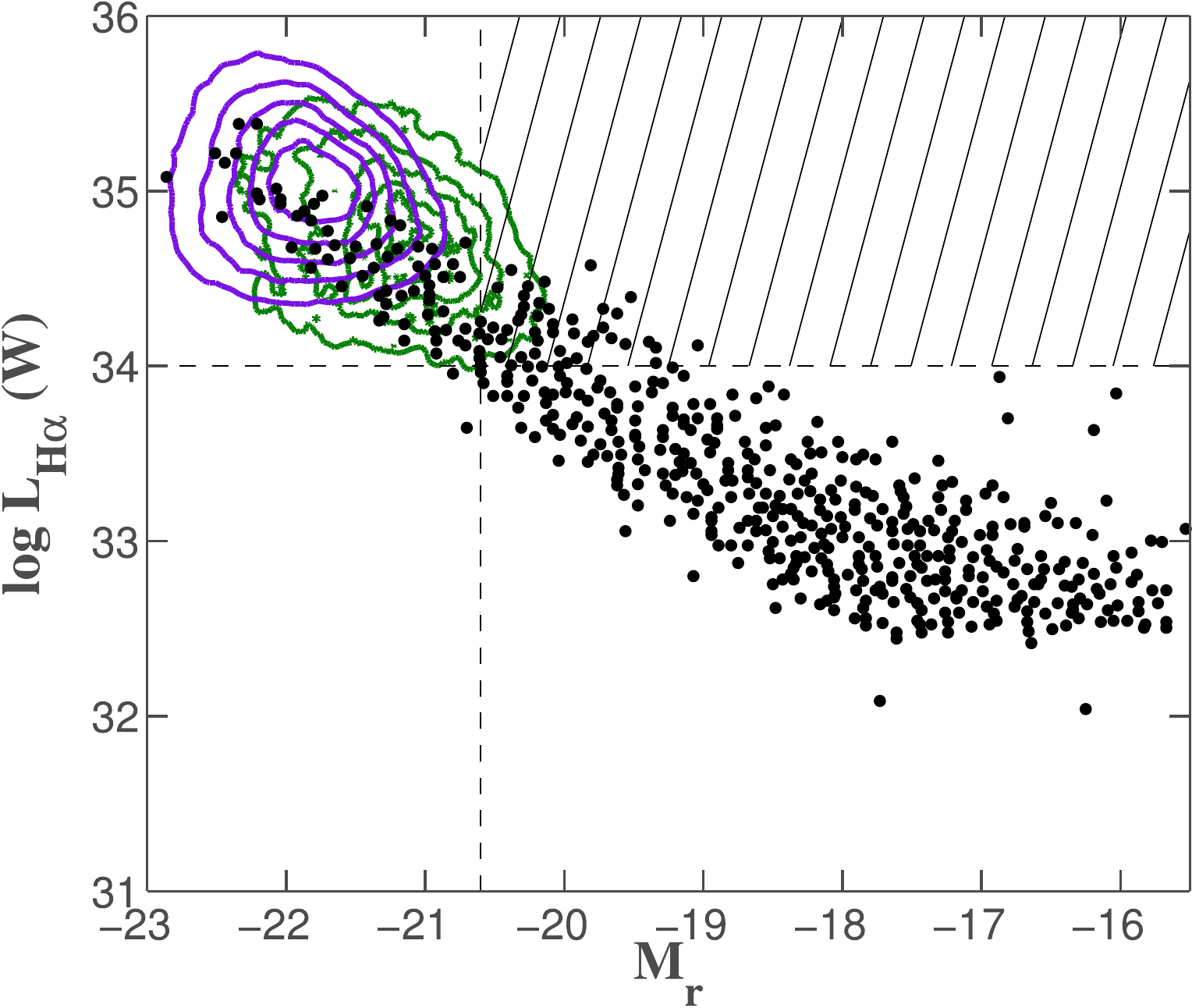}
\caption{GAMA absolute magnitude (M$_r$) versus aperture and obscuration corrected H$\alpha$ luminosity ($L_{H\alpha}$) bivariate distributions for galaxies in $0.17<z<0.24$ (green) and $0.24<z<0.34$ (purple) redshift bins compared to the distribution presented in Figure 9 of \citet{Shioya08}. They used HST COSMOS2 narrowband survey data to construct their H$\alpha$ LF. The redshift coverage of their data is $\sim0.24$. The vertical dashed line indicates the approximate absolute $r$--band magnitude corresponding to $z\sim0.24$ given GAMA's limiting magnitude of $19.8$. The horizontal line marks the approximate luminosity around the `knee' (i.e.\,close to $L^*$) of higher--$z$ LFs. This figure demonstrates that there is a population of optically faint star forming galaxies with $z\sim0.24$ (close to $50\%$) that do not enter either $0.17<z<0.24$ or $0.24<z<0.34$ GAMA samples.}
\label{fig:gama_vs_shioya}
\end{center}
\end{figure}

Both GAMA and SDSS are magnitude--limited surveys and any emission--line sample drawn from such a survey is subject to dual selection criteria. In order to contribute to the LF, a galaxy must satisfy both the broad--band magnitude limit and the emission line flux limit. The completeness corrections applied to the H$\alpha$ LFs account for the incompleteness as a function of broad--band  magnitude and colour, but nonetheless a bias remains. There is a population of bright H$\alpha$ galaxies that do not enter the sample initially as their broad--band magnitudes are too faint, and it is not possible to correct for this effect. We explore the impact of this bias here.

We assume a fiducial H$\alpha$ flux limit of $1\times10^{-18}$W\,m$^{-2}$ for the analysis presented in this paper \citep{Brough11}, which also approximately corresponds to the turn--over in the observed flux histogram, see Figure\,\ref{fig:fluxHist}. As discussed above in \S\,\ref{HaFluxlimit}, the incompleteness increases towards the flux limit, and can be as large as $50\%$ at the limit. The effect of H$\alpha$ incompleteness becomes progressively larger with redshift. As such, the GAMA and SDSS--DR7 low redshift samples are likely to be the most complete, with the higher redshift GAMA and SDSS samples becoming more and more incomplete with increasing redshift.  

In Figure\,\ref{fig:MrDist_GAMA} we show the bivariate H$\alpha$/M$_r$ distributions for our GAMA and SDSS samples. These are not bivariate LFs, which we present in a companion paper (Gunawardhana et al., in prep), but serve to show the distribution of luminosities spanned by the galaxies detected in each sample. 

Figure\,\ref{fig:MrDist_GAMA} (a) shows the bivariate H$\alpha$ luminosity/M$_r$ distribution for both the GAMA and SDSS $z\leqslant0.1$ samples. The overlapping region of the bivariate distributions indicate that the GAMA sample consists of optically faint galaxies with similar SFRs to optically bright SDSS galaxies. This $r$--band faint population is only detected in GAMA, demonstrating the H$\alpha$ incompleteness of SDSS. The grey band in Figure\,\ref{fig:MrDist_GAMA} (a) highlights the same luminosity range emphasised by the shaded region in Figure\,\ref{fig:GAMALFs} (a), where the discrepancy between the \cite{Westra10} and GAMA/SDSS--DR7 LFs is greatest. 

The effects of joint selection on the higher redshift LFs are evident in Figures\,\ref{fig:MrDist_GAMA} (b) and \ref{fig:gama_vs_shioya}. Only the distribution of the low--$z$ sample covers a wide range in both H$\alpha$ luminosity and M$_r$, while the higher redshift distributions become progressively more and more limited in the range of both H$\alpha$ luminosity and M$_r$ probed; each sample is missing a fraction of highly star forming, but optically faint galaxies, and  this missing fraction becomes more significant with increasing redshift. The impact, then, is that our higher redshift LFs remain incomplete, and can potentially be missing as much as $50\%$ of the bright H$\alpha$ population.
This is explored in more detail in  Gunawardhana et al.\,(in prep.), which investigates the evolution of the bivariate H$\alpha$/M$_r$ LF.

\subsection{Lower and upper limits of H$\alpha$ luminosity functions} \label{cases}

In addition to the incompleteness introduced by the bivariate selection, where optically faint star forming galaxies do not enter our sample due to the broadband selection of the survey, further uncertainties arise from the adopted $V_{\rm max}$ definition. Here we investigate a series of $V_{\rm max}$ corrections to the LFs that bracket our best estimated LFs presented in \S\ref{HaLFs}. The aim of this analysis is to identify the (extreme) lower and upper limits to SFR densities.  In subsequent sections we show that the uncertainties related to measurements and systematics fall within theses limits.   

\subsubsection{Best estimate} \label{HaFluxlimit}
\begin{figure*}
\begin{center}
\includegraphics[scale=0.48]{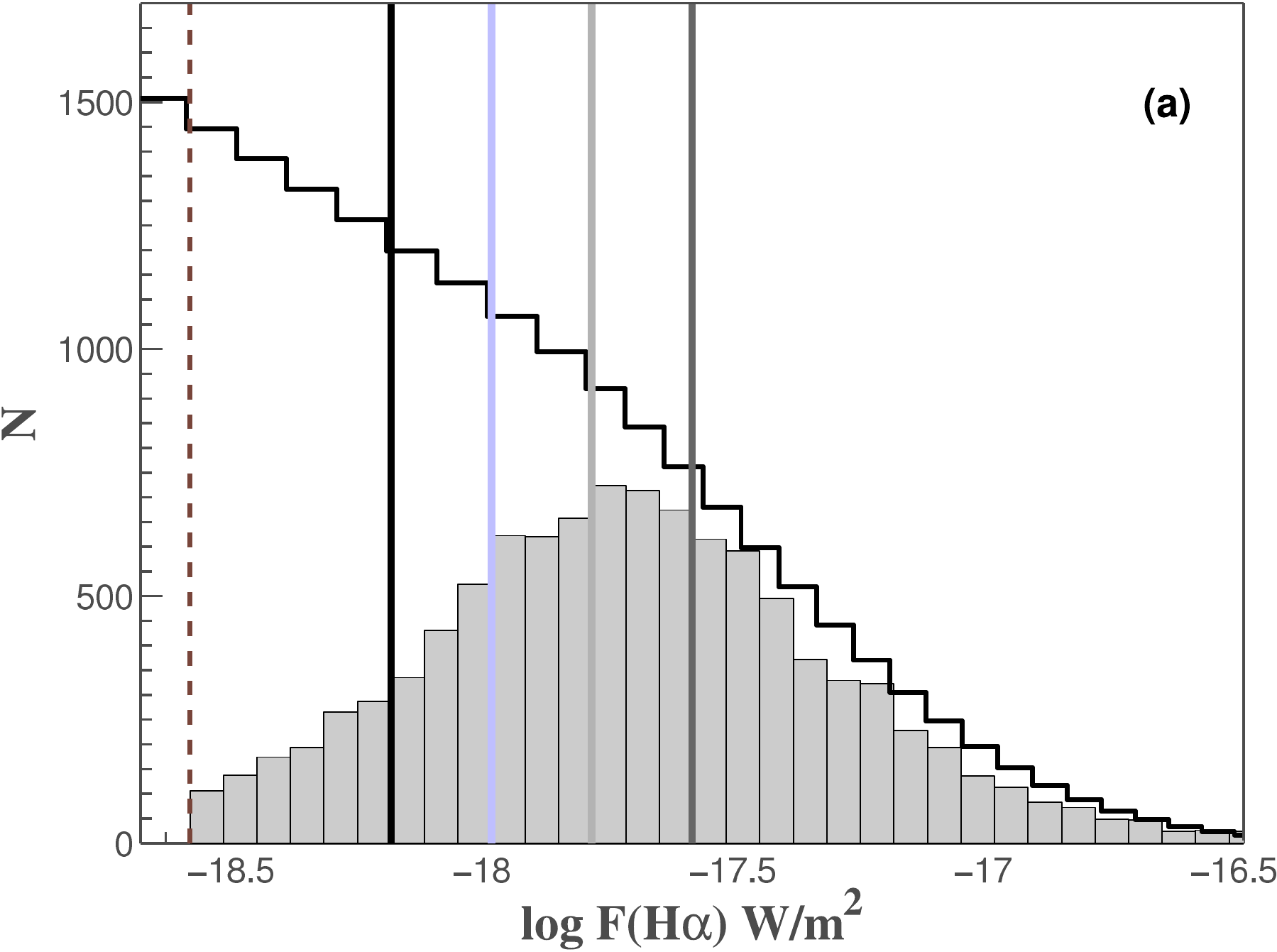}
\includegraphics[scale=0.35]{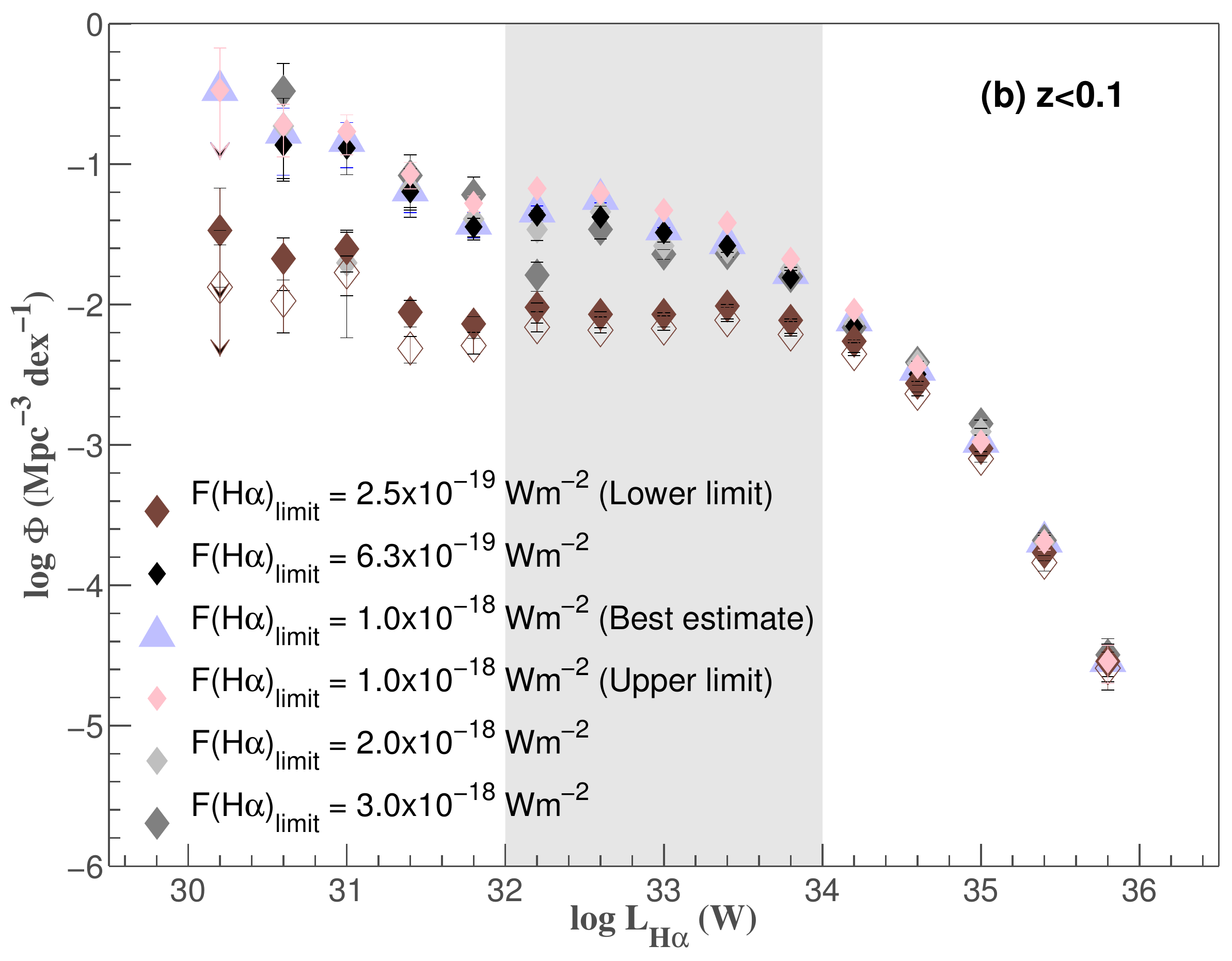}
\caption{The observed versus predicted flux H$\alpha$ flux distributions for low redshift galaxies. The solid histogram shows the observed H$\alpha$ flux histogram for galaxies with $z<0.1$. A simple prediction of the distribution of H$\alpha$ fluxes for low--$z$ galaxies based on our low--$z$ H$\alpha$ LF is shown by the open histogram.  The blue vertical solid line indicates the H$\alpha$ flux limit used in this study, and on the right the resultant low--$z$ LF. The rest of the solid lines indicate different flux limits tested, and on the right the resultant LFs. The low--$z$ LFs corresponding to the two limiting cases are shown here with (filled symbols) and without (open symbols) completeness corrections. The grey band highlights the same luminosity range as in Figure\,\ref{fig:GAMALFs}a}
\label{fig:fluxHist}
\end{center}
\end{figure*}
Firstly, we present a discussion of the true LFs presented in \S\,\ref{HaLFs} that assume a flux limit of \mbox{$1\times10^{-18}$W\,m$^{-2}$}, and explain the discrepancies between GAMA and \cite{Westra10} $0<z<0.1$ LFs shown in Figure\,\ref{fig:GAMALFs}. 

Emission line samples drawn from magnitude--limited surveys, such as GAMA and SDSS, involve assumptions about flux limits. The point at which an observed flux histogram turns over can be taken as a suitable limit. We assume an H$\alpha$ flux limit of   \mbox{$1\times10^{-18}$W\,m$^{-2}$} \citep{Brough11} to produce the results presented in \S\,\ref{HaLFs} following the methodology described in \S\,\ref{vmax_derivation}. This limit roughly corresponds to the peak value of the observed low--$z$ H$\alpha$ flux histogram (Figure\,\ref{fig:fluxHist}\,a). 

 The open histogram in Figure\,\ref{fig:fluxHist} depicts the predicted distribution of H$\alpha$ fluxes over the same redshift range. This distribution is a simple prediction based on the GAMA low--$z$ H$\alpha$ LF presented in \S\,\ref{HaLFs}. This prediction of the flux distribution is derived from the LF that we calculate from the observed flux distribution, and is thus being used merely as a self--consistency test. In the absence of the true underlying flux distribution this is sufficient, though, to explore our expected completeness as a function of H$\alpha$ flux, and we can see that even at the peak of the observed flux distribution, we are only about 75\% complete. At our assumed flux limit of  \mbox{$1\times10^{-18}$W\,m$^{-2}$} we are about 50\% complete. To investigate how our assumptions about the H$\alpha$ flux limit influence the shape of the LF, we reproduce the low--$z$ GAMA H$\alpha$ LF assuming several different flux limits indicated in Figure\,\ref{fig:fluxHist} (a). The resultant LFs are shown in Figure\,\ref{fig:fluxHist} (b), and it can be seen that the changes are primarily at the fainter end of the LF ($\log\, L \lesssim 34$). 
 
We assert that the differences in the assumed H$\alpha$ flux limit and in the formulation of $V_{\rm max}$ between this analysis and \cite{Westra10} contribute to some of the discrepancies between LFs shown in Figure\,\ref{fig:GAMALFs}(a). \cite{Westra10} have used a lower H$\alpha$ flux limit to construct their LFs. A low flux limit yield a larger volume over which an object could be detected, resulting in a lower LF normalisation. We demonstrate this in Figure\,\ref{fig:fluxHist}(b) by varying the flux limit. Note that in this study, a flux limit lower than \mbox{$1\times10^{-18}$W\,m$^{-2}$} yield a $V_{\rm max}$ limited by $r$--band magnitude (see Eq.\,\ref{eq:vmax_def2}). These differences in methods along with the uncertainties arising from the cosmic (sample) variance could explain the discrepancy between GAMA low--$z$ and \cite{Westra10} LFs (Figure\,\ref{fig:GAMALFs}a).

\subsubsection{Identifying a lower limit}

In order to identify a lower limit to SFR density, we set the H$\alpha$ flux limit to be equal to our H$\alpha$ detection limit of \mbox{$2.5\times10^{-19}$W\,m$^{-2}$}. This is an unrealistically low limit as the observed H$\alpha$ flux histogram in comparison to that predicted indicates close to 90\% incompleteness in H$\alpha$ detections (Figure\,\ref{fig:fluxHist}a). Additionally, we apply no $r$--band $V_{max}$ constraint or completeness corrections to the LFs. 

Intentionally neglecting the $r$-band volume limits and completeness correction ensure that the resulting LF will underestimate the true values, and should be a strong lower limit (Figure\,\ref{fig:fluxHist}b). The integral of this resulting LF in turn will give a strong lower limit to the SFR density. The H$\alpha$ LFs constructed this way are shown in Figures\,\ref{fig:fluxHist} (b) and \ref{fig:LFfits}. The lower limit number densities indicated by data points in Figure\,\ref{fig:LFfits} are generally lower than that predicted by the best estimate LFs at a given luminosity. This is, however, not always true as can be seen in Figure\ref{fig:LFfits} (d) where the red points at $\log L\approx34$ (W) and $\log L>37$ (W) indicate higher number densities than the best estimate LF points. As a result of the low flux limit, a large number of low H$\alpha$ flux detections enter the sample. Most of these objects are low luminosity galaxies such that the number of galaxies contributing to the lower limit LF point at $\log L\approx34$ (W) is relatively larger than the number contributing to the best estimate LF data point. The total number of galaxies contributing to the best estimate LF at $\log L>37$ (W) are both lower and their $V_{\rm max}$ are closer to $V_{i, zlim}$ (see Eq.\,\ref{eq:vmax_def2}), whereas the number contributing to the lower limit LF at $\log L>37$ (W) is slightly larger with $V_{\rm max}$ approximately equal to $V_{i, zlim}$.
 
\subsubsection{Identifying an upper limit}

An upper limit to SFR density is determined by including all H$\alpha$ detections down to our detection limit. The $V_{max, H\alpha}$ for objects with fluxes between the assumed detection limit and flux limit are set to equal to their comoving volumes. The resultant LFs are shown in Figures\,\ref{fig:fluxHist}(b) and \ref{fig:LFfits}. Note that this is not a substantial increase over our best estimate LF. The addition of all reliable H$\alpha$ detected sources, those below the nominal flux limit, does not significantly increase the LF or the corresponding SFR density.

\section{Functional fitting}\label{functions}

Galaxy LFs are usually fit with a \cite{Schechter76} function \citep[e.g.][]{Loveday92, Blanton03},
\begin{equation}
\Phi(L) dL = \Phi^* \left(\frac{L}{L^*}\right)^\alpha \exp \left(-\frac{L}{L^*}\right) d \left(\frac{L}{L^*}\right), 
\end{equation}
where $L$ is the galaxy luminosity, and $\Phi(L)\,dL$ is the number of galaxies in luminosity range $L+dL$ per cubic Mpc. The parameters $\alpha$, $L^*$ and $\Phi^*$, determined empirically, describe the shape of the fit, the slope of the LF at faint luminosities, the characteristic Schechter luminosity, and the normalisation factor at $L^*$ respectively.

The same functional form is generally used to fit star forming LF data \citep[e.g.][]{Gallego95, Jones01}. However, in contrast to broad--band optical LFs, our measured H$\alpha$ LFs are inconsistent with an exponential drop in number density for $L>L^*$. Most of the published H$\alpha$ LFs for star forming galaxies only probe a limited range in luminosity centred around $L^*$. Within this narrow range probed, the Schechter function provides a good fit. The much larger volumes probed by the GAMA and SDSS--DR7 LFs allow us to sample a wide range in H$\alpha$ luminosities. This enables us for the first time to study both faint and bright ends of the H$\alpha$ LF. For these LFs, the Schechter function is clearly not the best representation. This can be best seen in Figure\,\ref{fig:GAMALFs} (a) by comparing GAMA and SDSS LFs with the \cite{Perez03} LF, the exponential drop of the Schechter function is too steep to match the LFs presented in this paper. 

We find that the functional form presented in \cite{Saunders90}  provides a more suitable fit to the GAMA LFs. 
\begin{equation}
\Phi(L) dL = C \left(\frac{L}{L^*}\right)^{\alpha} \exp \left[-\frac{1}{2\sigma^2}\log^2\left(1+\frac{L}{L^*}\right)\right] d \left(\frac{L}{L^*}\right).
\end{equation}

Motivated by the power--law shape of the far--infrared $60\,\mu m$ LFs for $L>L^*$, \cite{Saunders90} introduced the above function, which behaves as a power law for $L<L^*$ and as a Gaussian in $\log L$ for  $L>L^*$  with a Gaussian width given by $\sigma$, and a normalisation factor at L$^*$ given by $C$. The SFR density is estimated from numerically integrating the \cite{Saunders90} function. This functional form is widely used to describe the LFs of far--infrared, and radio star forming populations \citep[e.g.][]{Hopkins98, Rowan93}. 

In particular, it is encouraging that this functional form can be used consistently to reproduce the LFs  for SFR tracers at each of the radio, far--infrared, and now H$\alpha$ wavelengths. We highlight here that while Schechter functions have been used in the past to fit the shape of the H$\alpha$ LF, the surveys in question have all probed relatively small volumes compared to GAMA and SDSS. Only with a sufficiently large volume is the bright end of the star forming population able to be sufficiently well--sampled to reliably measure the rare extreme star--forming population. Even with a Saunders parametrisation, it is still difficult to describe both the lowest SFR galaxies contributing to the faint--end rise in $\Phi$ and the highest SFR galaxies that diverge from a Gaussian decline in $\Phi$. Therefore, in order to constrain the functional fits to the LFs, the outlying GAMA LF points, shown as open symbols in  Figure\,\ref{fig:LFfits}, are excluded from the fitting. The variation in H$\alpha$ luminosity density with H$\alpha$ luminosity as traced by the LFs and the Saunders functional fits to data is shown in Figure\,\ref{fig:LD}. See \S\,A3 for comparisons with the Schechter functions. The fact that a Saunders functional form seems to be the most
appropriate form for each of radio, far-infrared and H$\alpha$ LFs suggests that the same should be true for ultraviolet LFs probing star formation in galaxies. \cite{Salim12} also find that the SFR distribution cannot be adequately described by a Schechter form.
This could have a potentially signficant impact on the very highest redshift estimates of SFR density, where UV LFs are often fit by Schechter functions \citep[e.g.][]{Bouwens11, Bouwens10} to data measured over a comparatively narrow range of observed luminosity. 

\subsection{Fits to the data}
\begin{figure*}
\includegraphics[scale=0.33]{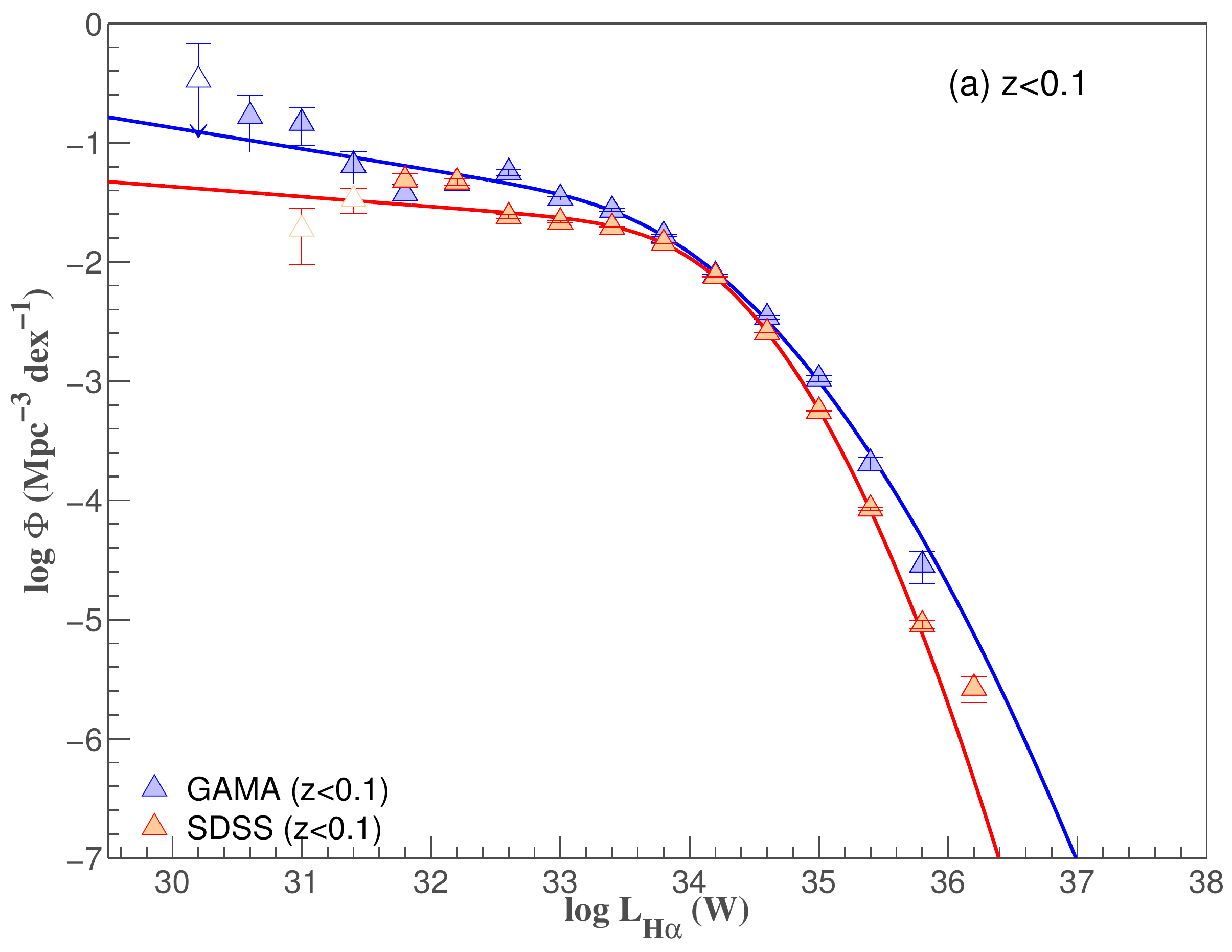}
\includegraphics[scale=0.33]{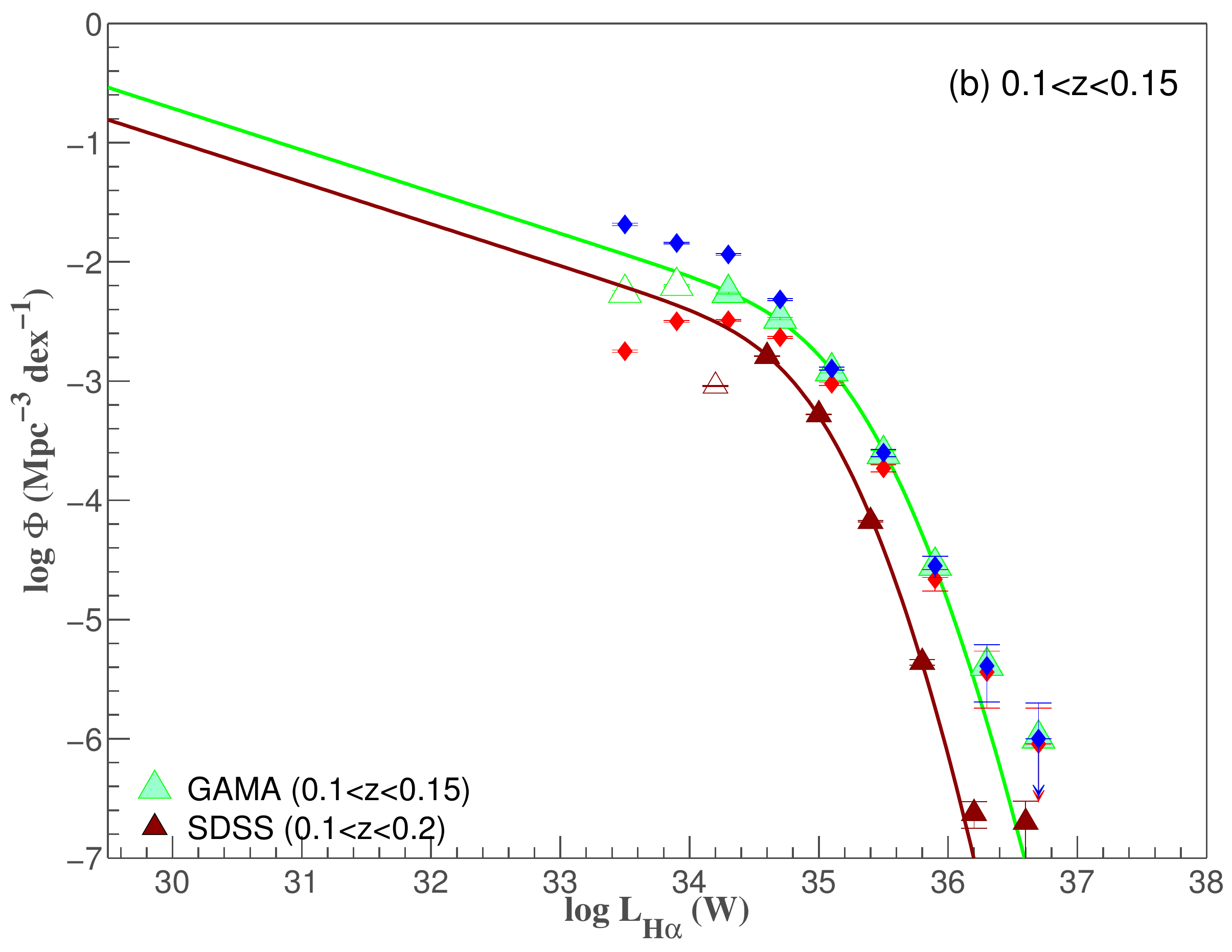}
\includegraphics[scale=0.33]{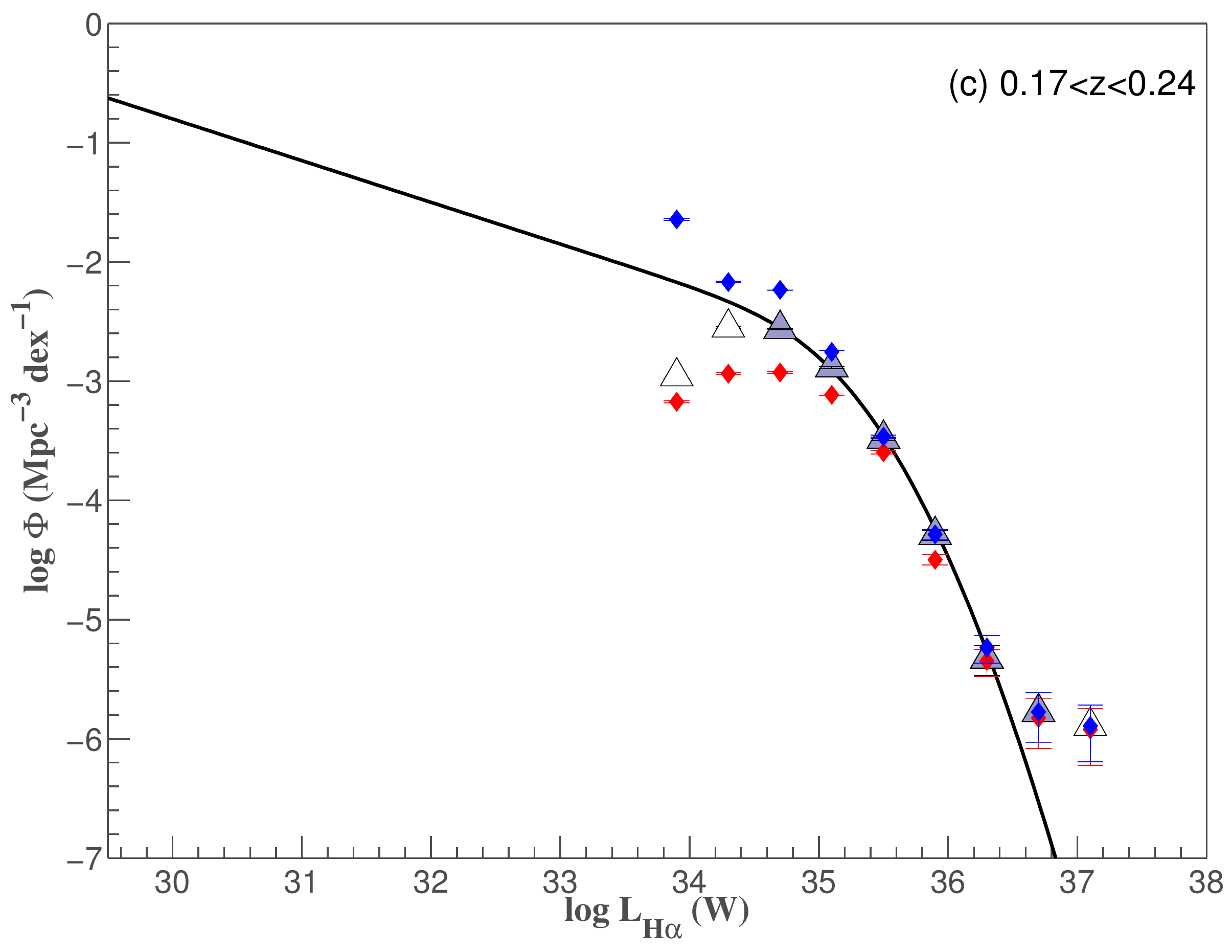}
\includegraphics[scale=0.33]{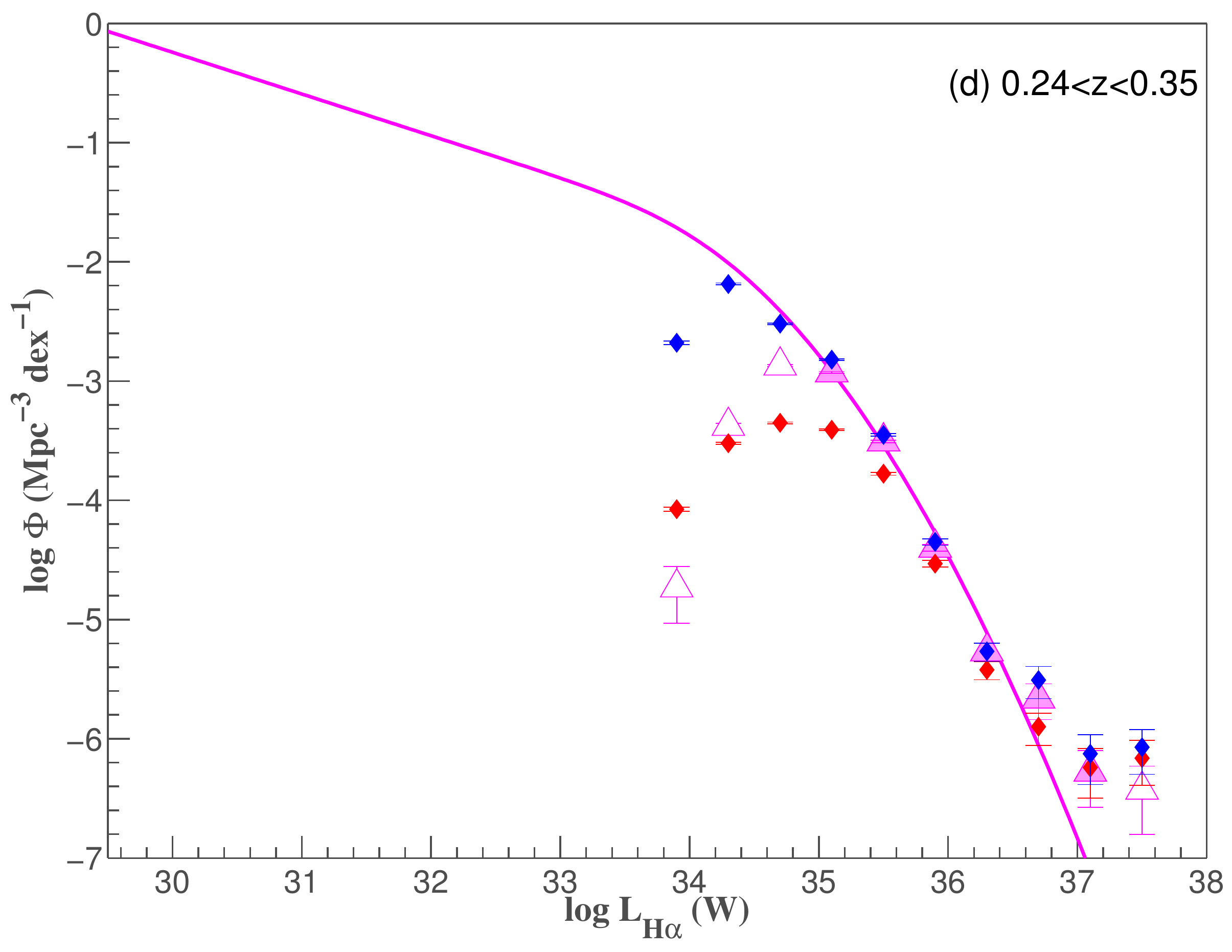}
\caption{The best--fit Saunders functions for all the luminosity functions.  The redshift increases from left--to--right, top--to--bottom. The triangles denote $\Phi$ values in each luminosity bin, with filled triangles showing the points used for the fit. The best--fit $\alpha$ value from the \citet{Shioya08} LF is used to constrain the faint--end slope of GAMA LFs beyond $z>0.1$. The blue and red diamonds indicate the LFs corresponding to lower and upper limiting $V_{max}$ cases discussed in \S\,\ref{cases}, and the resulting SFR densities are shown in Figure\,\ref{fig:CSFH}. The best--fit Saunders parameters are given in Table\,\ref{table:LFfits}. }
\label{fig:LFfits}
\end{figure*}

\begin{table*}
\begin{minipage}{28cm}
\caption{ The best fit Saunders parameters for the LFs, corresponding SFR densities and their Poisson uncertainties.}
\begin{tabular}{|l|c|c|c|c|||cl}
\hline
         $z$ 	 & 	$\log$ L$^*$ 	&   $\log$ C 	&  $\alpha$     & $\sigma$        &  log SFR density \\
                 & 	(W)			&   			& 		       & 			&  (M$_{\odot}$yr$^{-1}$Mpc$^{-3}$)\\
                               
\hline
\hline
	$z<0.1$ (GAMA, Figure\,\ref{fig:LFfits}a)						
&	$33.00\pm0.41$ 		
&  $-1.77\pm0.16$		
& $-1.16\pm0.07$ 		
& $0.84\pm0.09$ 
& $-2.068\pm0.010$\footnote{An uncertainty of $\sim15\%$ needs to be incorporated to this value to account for GAMA's known under--density \citep{Driver11}}		 \\

  $z<0.1$ (SDSS--DR7,  Figure\,\ref{fig:LFfits}a)					
&	$33.43\pm0.53$  		
&  $-2.02\pm0.23$		
& $-1.08\pm0.22$		
& $0.61\pm0.08$ 
& $-2.244\pm0.010$			\\

 $0.1<z<0.15$ (GAMA,  Figure\,\ref{fig:LFfits}b)		
&	$34.55\pm0.85$ 		
&  $-2.67\pm0.83$		
& $-1.35$\footnote{$\alpha$ is fixed to be $-1.35$ \citep{Shioya08}} 		
& $0.47\pm0.18$ 
&  $-2.096\pm0.004$			\\

 $0.1<z<0.2$ (SDSS--DR7, Figure\,\ref{fig:LFfits}b)\footnote{The fit cannot be constrained as the LF turns--over at around L$^*$. This leads to very large uncertainties in the fitted parameters} 					
&	$34.42\pm1.46$ 				
&  $-2.89\pm1.99$				
& $-1.35^b$ 				
& $0.42\pm0.25$ 
& $-2.522\pm0.002$			\\

 $0.17<z<0.24$ (GAMA, Figure\,\ref{fig:LFfits}c)		
&	$34.53\pm0.74$ 		
&  $-2.75\pm0.73$		
& $-1.35^b$ 		
& $0.55\pm0.17$ 
& $-2.118\pm0.004$			\\

 $0.24<z<0.35$ (GAMA ,  Figure\,\ref{fig:LFfits}d)$^{c}$	   	
&  $33.47\pm3.56$ 		
&  $-1.82\pm4.14$		
& $-1.35^b$ 		
& $0.81\pm0.52$ 
& $-1.896\pm0.004$  			\\
\hline
\end{tabular}
\label{table:LFfits}
\end{minipage}
\end{table*}

We use a Levenberg--Marquardt method for finding the minimum $\chi^2$ fit to the binned LF data points. The resultant Saunders functional fits and the best--fit parameters to the GAMA/SDSS LFs are presented in Figure\,\ref{fig:LFfits} and Table\,\ref{table:LFfits}.

All GAMA LF data points and most of the SDSS LF data points belonging to the lowest--$z$ bins are used in the fitting, see Figure\,\ref{fig:LFfits}.
None of the rest of our LFs cover the same wide range in luminosity that the GAMA low--$z$ LF covers, largely due to the $r$--band flux limit in the survey selection (Figure\,\ref{fig:gama_vs_shioya}).  As a consequence, we can only constrain the LF over a narrow range in luminosity at the higher redshifts. For instance, the range in luminosity sampled by the second GAMA LF is less than half the range sampled by the GAMA low--$z$ LF (see Figure\,\ref{fig:GAMALFs}). The lack of $L\ll L^*$ LF data is a significant drawback in determining $\alpha$ accurately.  To overcome this difficulty, we investigated two alternative approaches to fitting the Saunders functional forms to LF data.
\begin{figure*}
\includegraphics[scale=0.23]{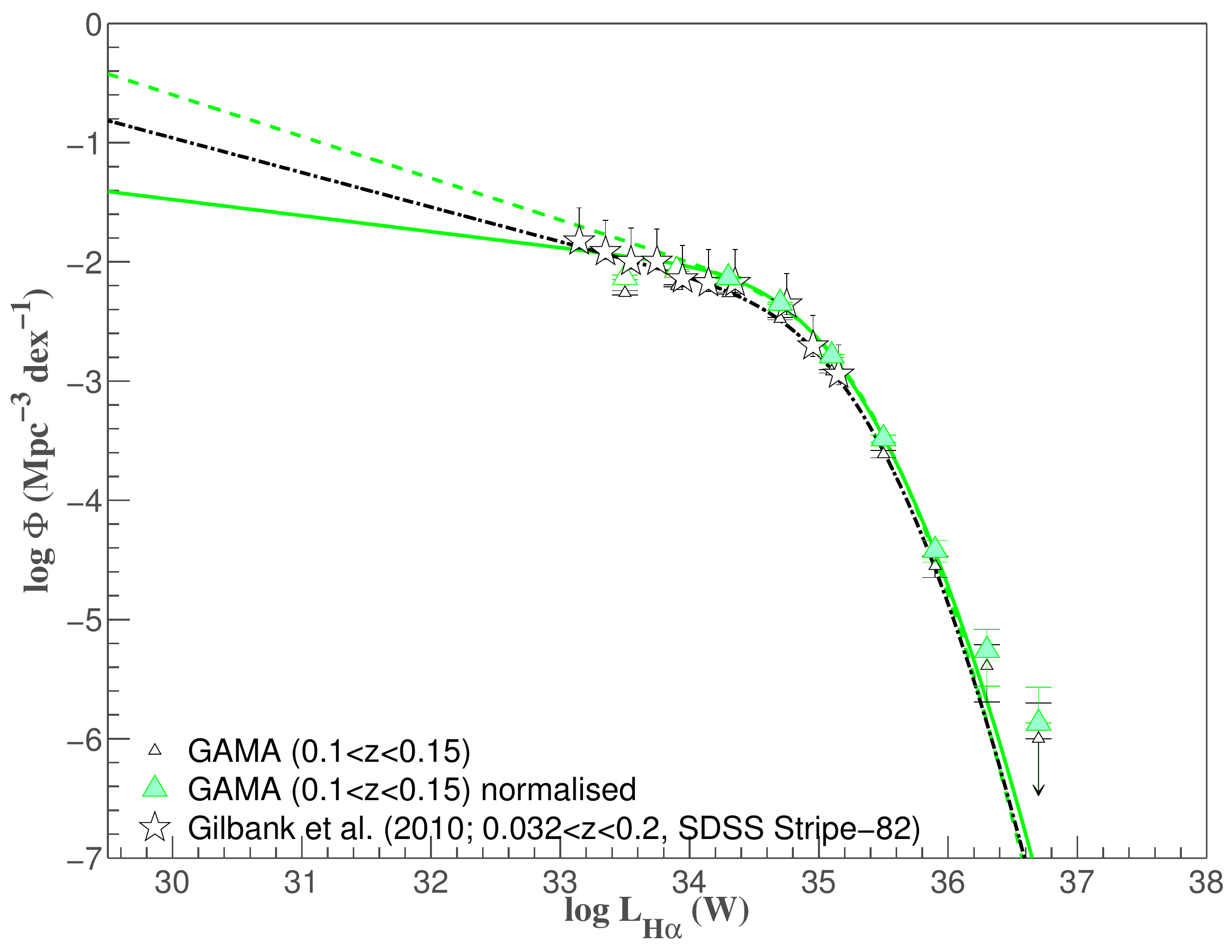}
\includegraphics[scale=0.23]{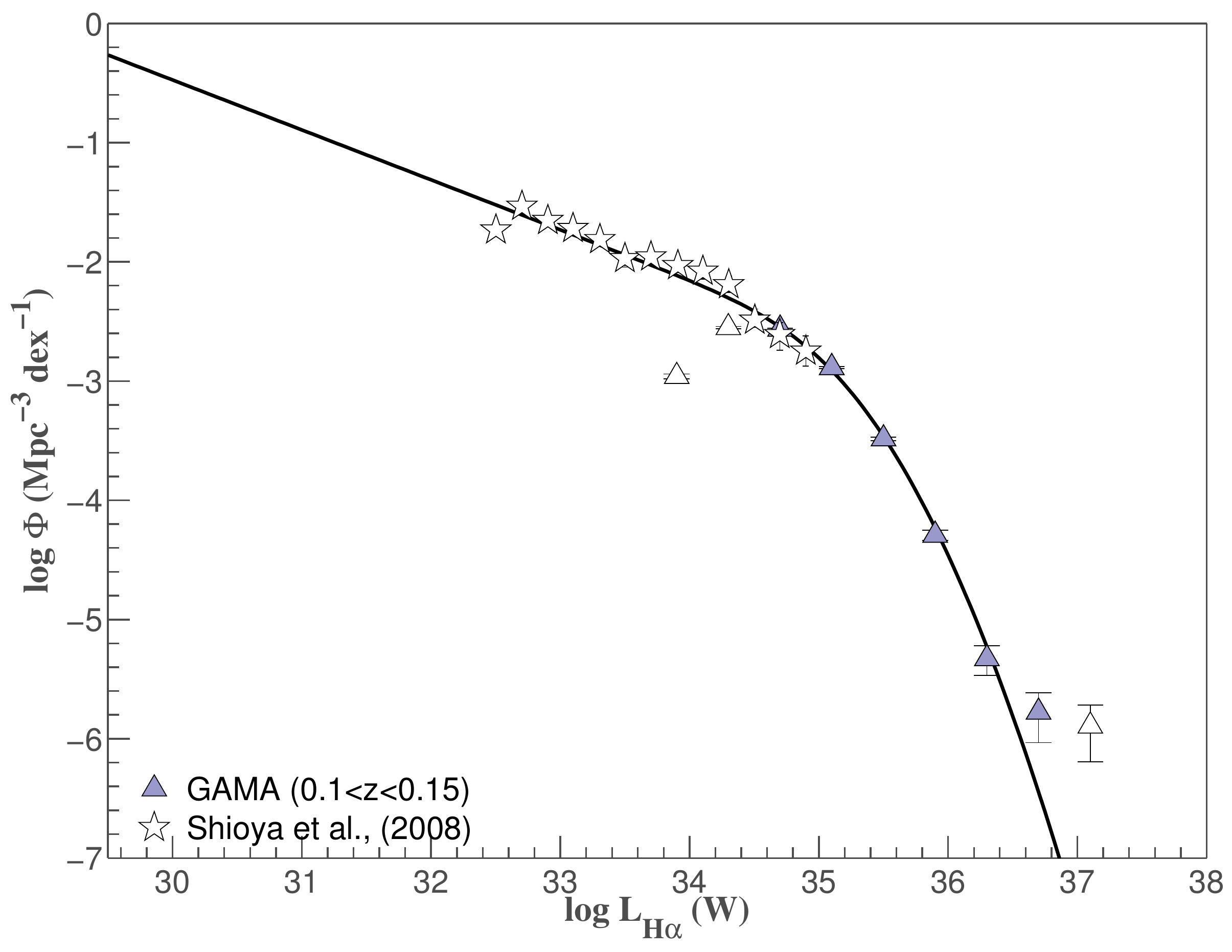}
\includegraphics[scale=0.23]{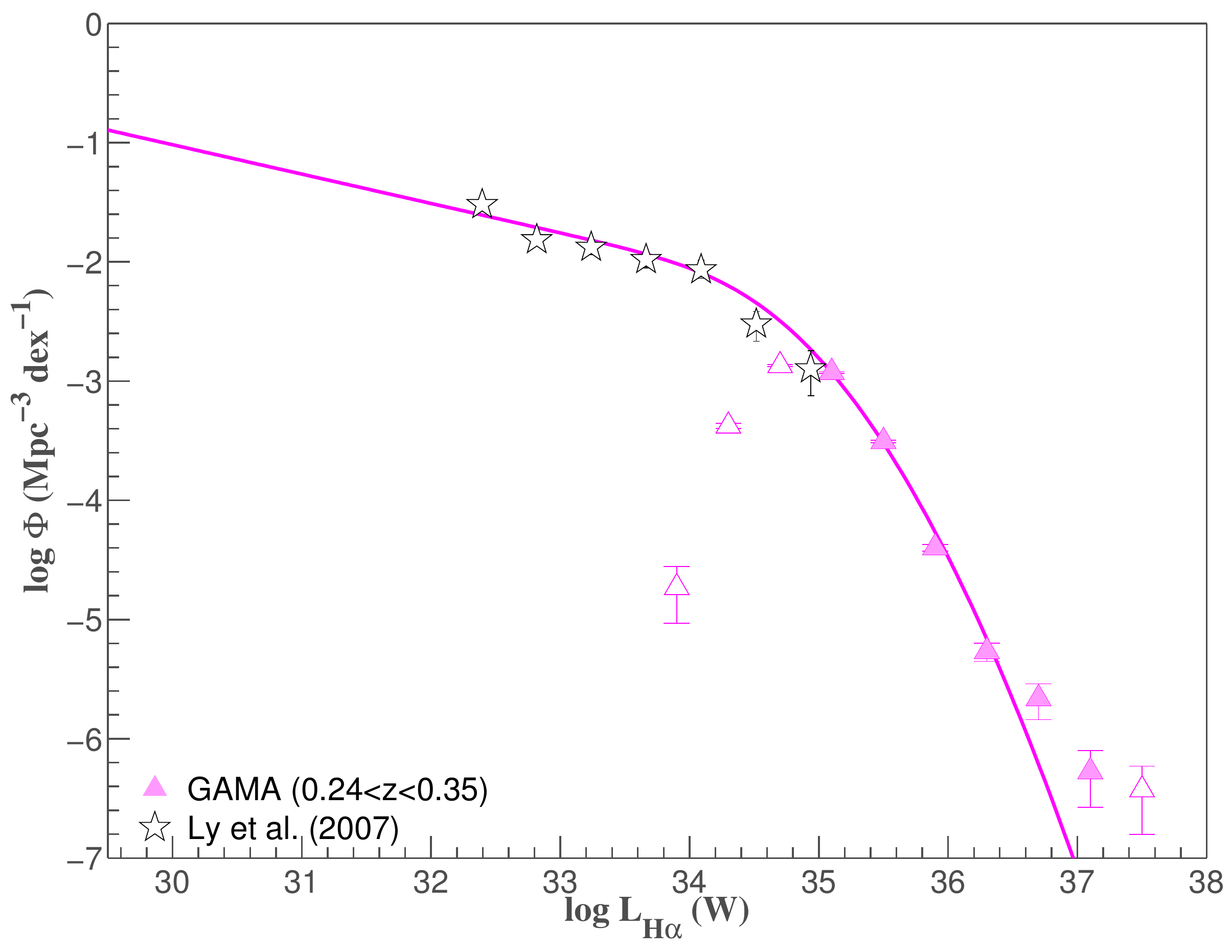}
\caption{The best--fit Saunders functions for the higher--$z$ luminosity functions.  The triangles denote GAMA $\Phi$ values in each luminosity bin, with filled triangles showing the points used for the fit. We use the published luminosity functions of \citet[][narrow--band survey, $0.233<z<0.249$]{Shioya08}, and \citet[][narrow--band survey, $0.382<z<0.418$]{Ly07} to get a better estimate of the faint--end evolution of GAMA $0.17<z<0.24$ and $0.24<z<0.34$ LFs. For the GAMA $0.1<z<0.15$ LF, we have used the \citet[][based on magnitude limited SDSS data, $0.032<z<0.2$]{Gilbank10} LF data as there are no wide-area narrowband measurements are available. These points are shown as open stars in all panels. The normalisation factor determined for $0.1<z<0.15$ range using \citet{Gilbank10} is the largest ($0.12\,$dex). This could be due to the differences in the redshift ranges probed the two LFs. The normalisation determined for the $0.17<z<0.24$ range using \citet{Shioya08} measurements is negligible, and a normalisation cannot be determined for the final redshift bin as \citet{Ly07} and GAMA $0.24<z<0.34$ LF data do not overlap. The colours correspond to those in Figure\,\ref{fig:LFfits}. The best--fit Saunders parameters are given in Table\,\ref{table:LFfits2}. For the GAMA $0.17<z<0.24$ combined LF we provide the functional fits determined using \textit{normalised} GAMA LF data combined with \citet{Gilbank10} data and fixing the faint--end slope ($\alpha$) at $-1.35$ (dashed line), fitting to the faint--end (solid line), and the original GAMA $0.1<z<0.15$ LF data combined with \citet{Gilbank10} data and fitting to the faint--end slope (dot--dashed line).  }
\label{fig:LFfits2}
\end{figure*}
\begin{table*}
\begin{minipage}{28cm}
\caption{The best fit Saunders parameters and SFR densities for the combined LFs presented in Figure\,12.}
\begin{tabular}{|l|c|c|c|c|||cl}
\hline
         $z$ 	 & 	$\log$ L$^*$ 	&   $\log$ C 	&  $\alpha$     & $\sigma$        &  log SFR density \\
                 & 	(W)			&   			& 		       & 			&  (M$_{\odot}$yr$^{-1}$Mpc$^{-3}$)\\
                               
\hline
\hline
 $0.1<z<0.15$	 (dashed green line in Figure\,12)
&  $34.62\pm0.37$ 	
&  $-2.58\pm0.26$	
& $-1.35$\footnote{$\alpha$ is fixed to be $-1.35$ \citep{Shioya08}} 		
& $0.45\pm0.09$	
&  $-1.973\pm0.004$ 		\\

 $0.1<z<0.15$	(solid green line in Figure\,12)	
&  $34.31\pm0.61$	
&  $-2.42\pm0.35$	
& $-1.13\pm0.38$		
& $0.51\pm0.09$	
&  $-1.996\pm0.004$			\\

 $0.1<z<0.15$ (dot--dashed black line in Figure\,12)
&  $34.48\pm0.63$	
&  $-2.62\pm0.48$	
& $-1.29\pm0.36$		
& $0.48\pm0.10$	
&  $-2.105\pm0.004$			\\

 $0.17<z<0.24$	
&  $34.54\pm0.32$ 		
&  $-2.74\pm0.24$		
& $-1.42\pm0.08$ 		
& $0.57\pm0.08$ 
& $-2.077\pm0.004$			\\

 $0.24<z<0.35$
&  $33.94\pm0.88$ 		
&  $-2.35\pm0.53$		
& $-1.25\pm0.34$ 		
& $0.68\pm0.12$ 
& $-2.056\pm0.004$  			\\
\hline
\end{tabular}
\label{table:LFfits2}
\end{minipage}
\end{table*}

\textit{Fixed the faint--end slope of the LF.} This is the approach shown in Figure\,\ref{fig:LFfits2}. We use the best--fit $\alpha$ parameter from the \cite{Shioya08} narrowband LF ($z\sim0.24$) to fix the faint end of the higher--$z$ GAMA/SDSS LFs. This value is chosen instead of that estimated using GAMA low--$z$  LF or other narrowband LFs estimates as the redshift range probed by the \cite{Shioya08} LF roughly corresponds to the redshift range of the GAMA higher--$z$ LF, and it probes a relatively larger luminosity range at that redshift. Narrowband surveys, although only covering a comparatively small sky area, have the advantage of being complete down to a given H$\alpha$ luminosity, and at modest redshifts, they successfully extend substantially below $L\ll L^*$ compared to those from surveys initially selected with a broad--band magnitude limit. 

The LF data points that indicate a turn--over in number density as a result of higher--$z$ sample incompleteness are excluded from the functional fits. These excluded points are denoted by open symbols in Figure\,\ref{fig:LFfits}. Even though the faint--end slope of the LF is fixed, the bright end of the LF is affected by the bivariate selection effects, which progressively become significant with redshift (Figure\,\ref{fig:MrDist_GAMA}b). This results in lower integrated SFR densities for the higher--$z$ LFs as the bivariate selection prevents optically faint high H$\alpha$ luminosity objects from entering the higher--$z$ samples, and affecting the overall normalisation of the LF. The fitting for the highest--$z$ LF cannot be constrained as only the points above the knee of the LF can be used.   

\textit{Normalised to match narrowband LF data.}

 Another way of determining the faint--end slope of the H$\alpha$ LFs is to normalise our data to match narrowband LFs. Narrowband surveys are complete down to a given H$\alpha$ luminosity and consist of relatively large number of faint H$\alpha$ emitters (Figure\,\ref{fig:gama_vs_shioya}). As a result their LFs are more complete below $L<L^*$ than those based on a broadband selected galaxy sample. In contrast, magnitude--limited surveys covering a large sky area consist of relatively large numbers of bright H$\alpha$ sources, and the respective LFs are likely more complete above $L>L^*$ than those based on narrowband data sets.
We therefore use published narrowband LF data to estimate the evolution of the faint end of higher--$z$ ($0.17<z<0.24$ and $0.24<z<0.34$) GAMA LFs. For the GAMA $0.1<z<0.15$ LF, where no wide-area narrowband measurements are available, we combined GAMA LF data with \cite{Gilbank10} data to determine the faint--end slope of the LF. 

\begin{figure}
\begin{center}
\includegraphics[scale=0.3]{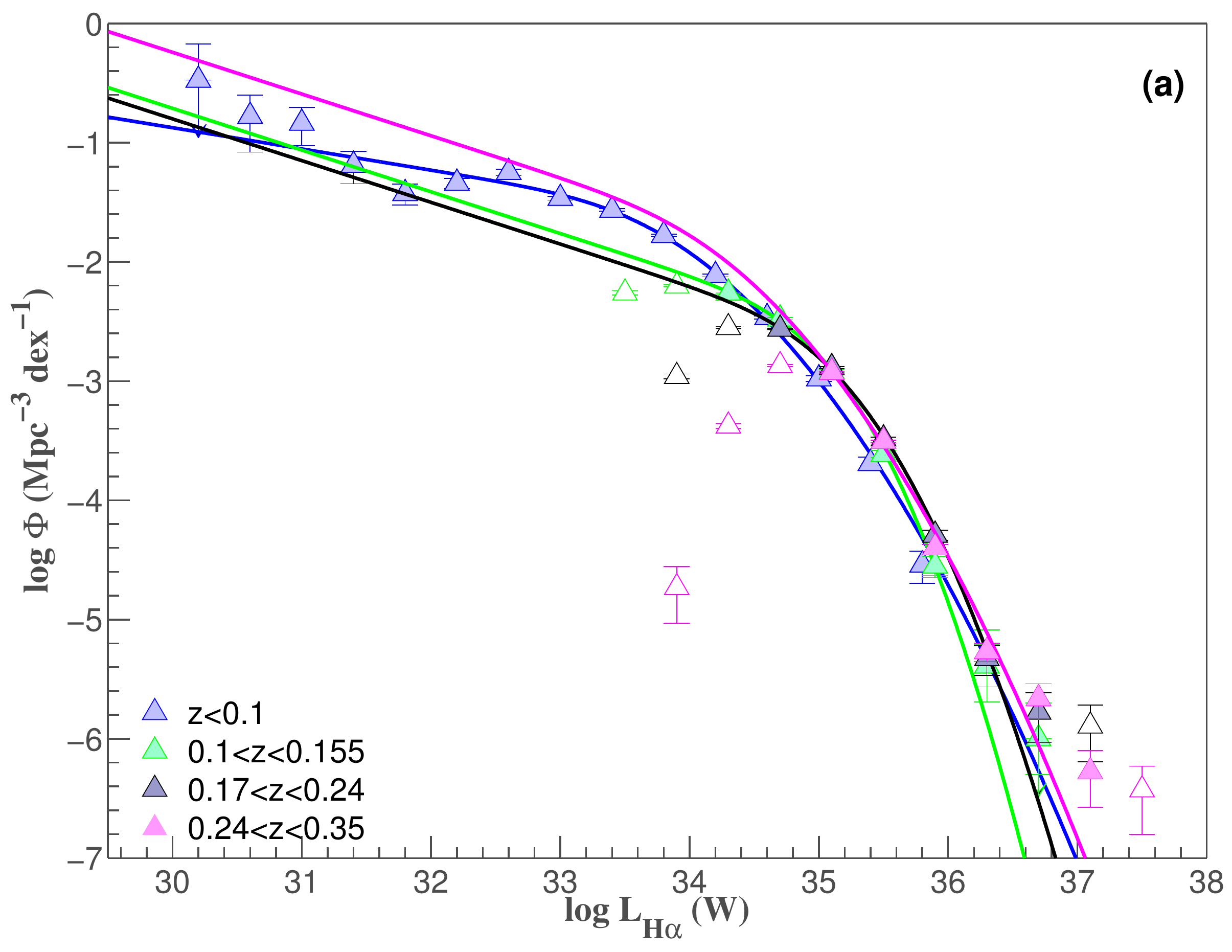}
\includegraphics[scale=0.3]{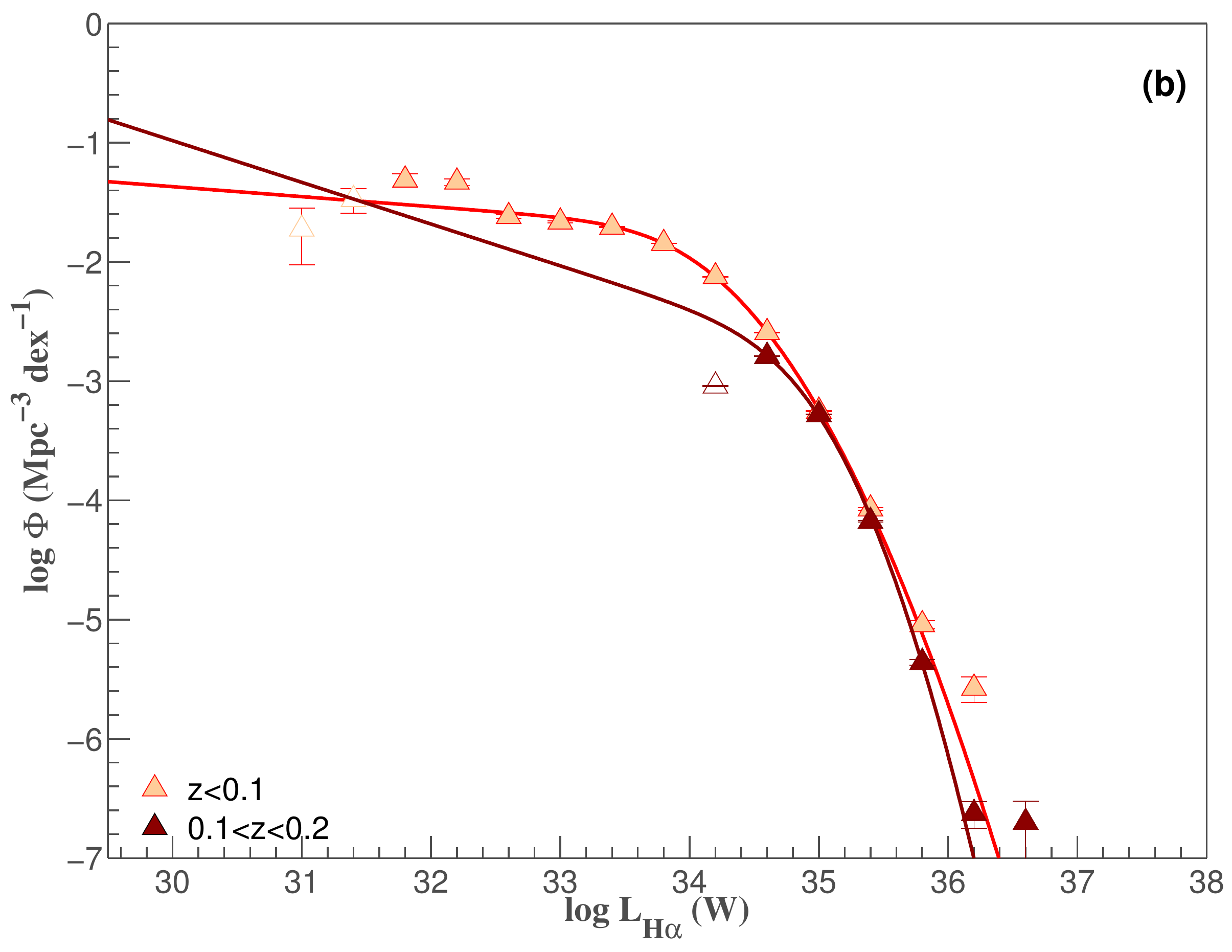}
\caption{ A comparison of GAMA (a) and SDSS (b) H$\alpha$ LFs, and their functional fits, demonstrating the modest evolution over the observed redshift range. Again, the colour scheme corresponds to that from Figure\,\ref{fig:LFfits}.}
\label{fig:allLFfits}
\end{center}
\end{figure}

The overlapping LF data from \cite{Gilbank10} and \cite{Shioya08} are used to normalise the higher--$z$ ($0.1<z<0.15$ and $0.17<z<0.24$) GAMA LFs. However, such a normalisation could not be achieved for the highest--$z$ ($0.24<z<0.34$) GAMA LF due to the lack of overlap between GAMA and \cite{Ly07} LF. The normalisation factors estimated using the approximately overlapping \cite{Gilbank10} and GAMA $0.1<z<0.15$ LF data in $34<\log L(W)<35.5$ range (see Figure\,\ref{fig:GAMALFs} b) is $\sim0.12$\,dex, and the factor using overlapping \cite{Shioya08} and GAMA $0.17<z<0.24$ LFs is negligibly small as the two LFs agree very well (Figure\,\ref{fig:GAMALFs} c). We note that the larger normalisation required to match GAMA $0.1<z<0.15$ LF data with \cite{Gilbank10} LF could be a result of the different redshift ranges probed by the LFs. It is likely that \cite{Gilbank10} LF indicates some evolution as it covers a larger redshift range than the respective GAMA LF. The functional fits to the combined LFs are shown in Figure\,\ref{fig:LFfits2}, and the best fit functional parameters are given in Table\,\ref{table:LFfits2}. 

The modest level of evolution demonstrated by these LFs is highlighted in Figure\,\ref{fig:allLFfits}.  The largest change is seen between the first and second redshift bins, with minimal measurable change thereafter. The lack of evolution here is most likely due to the high incompleteness of higher redshift samples, a result of the joint selection in both broad--band magnitude and emission line flux. Even though there is some evolution in the LF over this redshift range, it is difficult to quantify the extent accurately without accounting for the impact of the sample selection. This has been outlined above in \S\ref{LF}, and is explored in more detail in an analysis of the bivariate H$\alpha$/M$_r$ LF in Gunawardhana et al.\,(in prep).

\begin{figure}
\begin{center}
\includegraphics[scale=0.35]{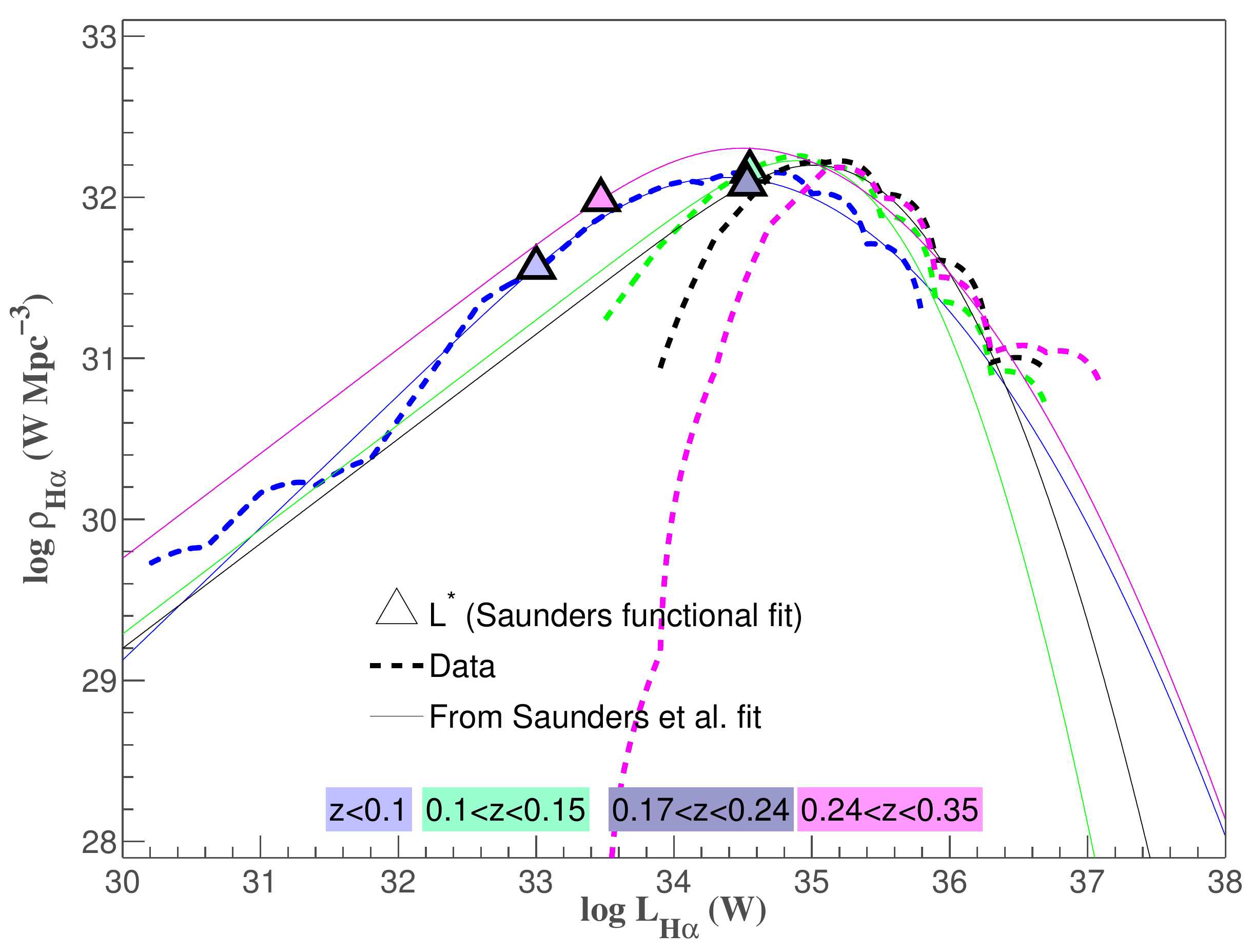}
\caption{The H$\alpha$ luminosity density as a function of H$\alpha$ luminosity, and its evolution. This illustrates that the bulk of the luminosity density comes from galaxies with luminosities close to L$^*$. The solid  lines indicate the luminosity densities derived from the Saunders fits shown in Figures\,\ref{fig:LFfits} and \ref{fig:allLFfits}, and dashed lines indicate luminosity densities derived directly from the data. The solid symbols indicate L$^*$ values obtained from Saunders (triangles; best fit $L*$, $\log C$, $\alpha$, and $\sigma$ are estimated for the lowest--$z$ LF, $\alpha$ is fixed to be $-1.35$ and the best fit  $L*$, $\log C$, and $\sigma$ are estimated for the rest). Note that the best--fit L$^*$ values corresponding to a Saunders functional fit are smaller than that obtained from a Schechter functional fit. }
\label{fig:LD}
\end{center}
\end{figure}

Finally, the H$\alpha$ luminosity density at a given H$\alpha$ luminosity is given by, 
\begin{equation}
\rho_{H\alpha}(L) = L_{H\alpha} \times \Phi(L).
\label{math:LD}
\end{equation}
The  luminosity density versus luminosity distributions  are shown in Figure\,\ref{fig:LD}. The peak luminosity density occurs approximately at $L^*$, demonstrating both that it is typically galaxies close to $L^*$ that dominate the luminosity density of the universe at low redshift, and also the modest evolution in H$\alpha$ luminosity density with redshift.  Although not shown in Figure\,\ref{fig:LD}, the best--fit L$^*$ values from the Schechter functional fits to the LFs are always larger than those corresponding to the Saunders fits. GAMA and SDSS LFs indicate a Gaussian--like decrease in number density with increasing luminosity owing to the large range in luminosity sampled. Therefore, fitting a Schechter function to our data is clearly not appropriate, and results in an overestimation of L$^*$, see Figure\,\ref{fig:LFfits2}.  This is discussed further in Appendix\,\ref{biases}, and illustrated in Figure\,\ref{fig:AGNLFs}.

\section{The cosmic history of star formation}\label{CSFH}

The cosmic star formation history (SFH) is a fundamental component in understanding galaxy formation and evolution. The observed SFH encompasses the imprint of all the underlying physical processes such as mergers, feedback processes, accretion, etc.\,that shape a galaxy, and is a crucial constituent in constraining galaxy formation/evolution models \citep[e.g.][]{HB06}.

\begin{figure*}
\begin{center}
\includegraphics[scale=0.4]{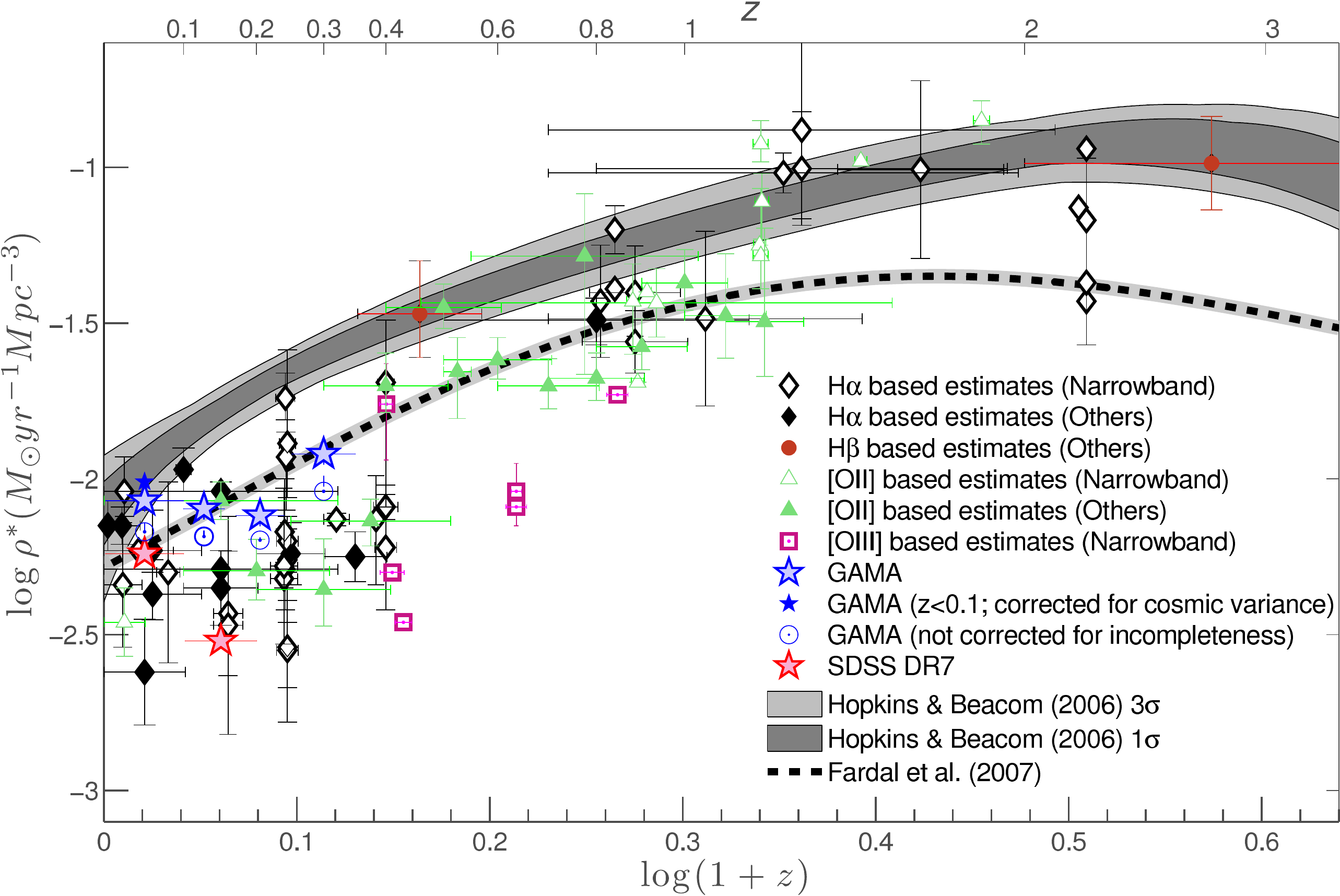}
\includegraphics[scale=0.37]{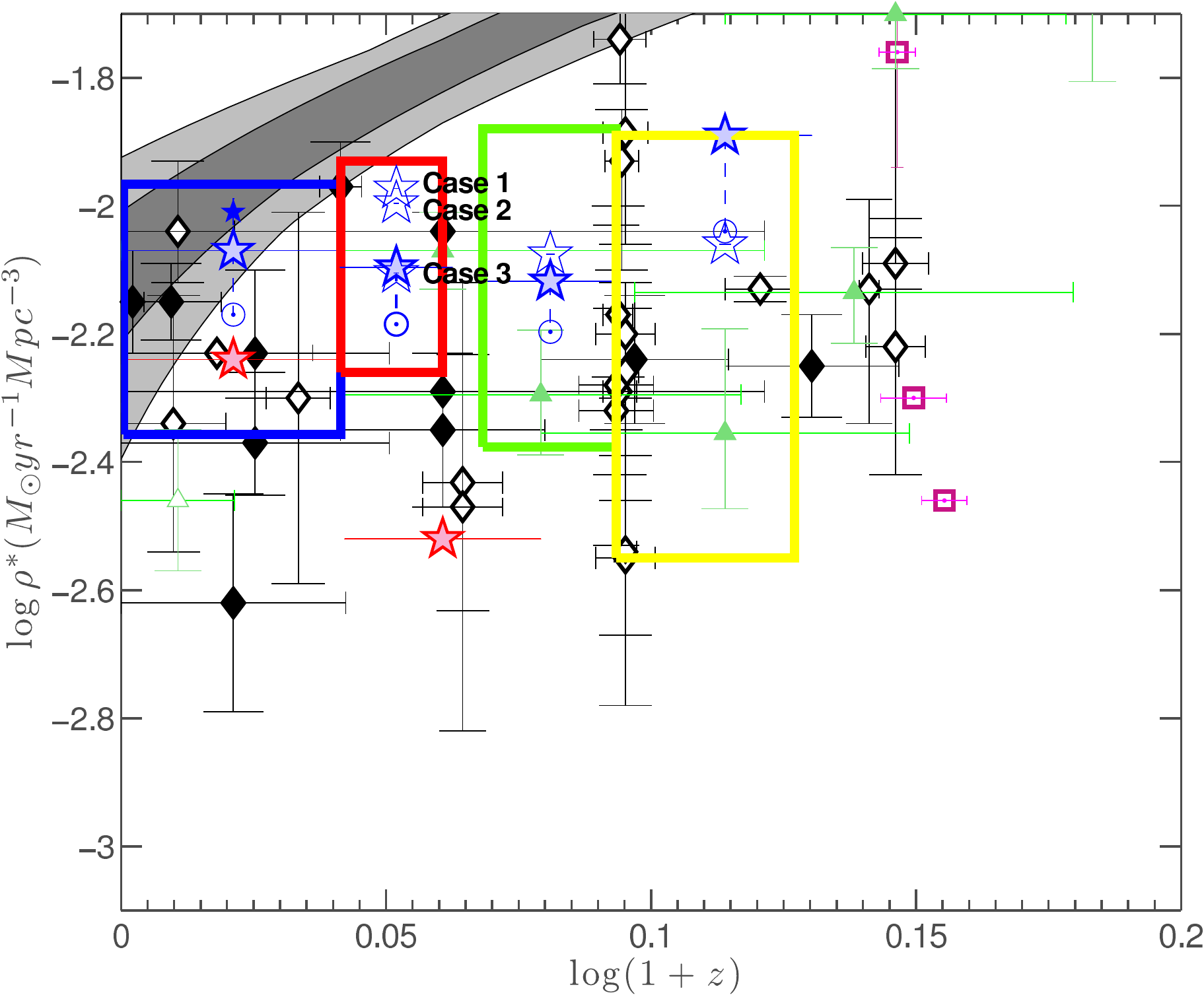}
\caption{The cosmic history of star formation. The SFR densities estimated using GAMA and SDSS--DR7 H$\alpha$ LFs are shown as blue and red stars.  The filled light blue stars denote the completeness corrected GAMA SFR densities corresponding to the LFs shown in Figure\,\ref{fig:LFfits}, with the blue open circles indicating the drop in SFR density if no correction is applied (see Figure\,\ref{fig:MonteCarlo_BDs} for the LFs). The dark blue filled star indicates the $z<0.1$ SFR density corrected for GAMA's known under--density \citep{Driver11}. The GAMA and SDSS points are compared with other narrowband/slitless spectroscopy (open symbols) and  magnitude--limited and other types (filled symbols) surveys using H$\alpha$ (black symbols), H$\beta$ (brown), [\ion{O}{ii}]~$\lambda$3727 (green), and [\ion{O}{iii}]~$\lambda$5007 (magenta). The shaded regions and the dashed line denote the best--fit cosmic SFRs derived by \citet{HB06}, and  \citet{Fardal07} respectively using available observational data. All comparisons presented here assume concordance cosmology, and the SFR calibration given in Eq.\,\ref{eqn:SFR_calib}. The right panel shows zoomed--in comparison. The boxes that bracket the GAMA SFR densities indicate the uncertainties associated with the measurements. The lower/upper edges of the boxes indicate the lower/upper SFR density limits described in \S\,\ref{cases} and shown in Figure\ref{fig:LFfits}. Note that these limits, particularly the lower limits, are limits obtained by intentionally calculating unrealistically extreme bounds for the LFs. The overlap between the highest--$z$ SFR density point and the upper limit (yellow box) is due to the functional fit to the LF data being overestimated. This is a result of the small number of LF data that can be used for the functional fitting as the highest--$z$ LF becomes incomplete (i.e.\,turns over) at luminosities $>L^*$. The open blue stars in the right panel indicate the SFR density measurements calculated by combining higher--$z$ GAMA LFs with other LFs in the literature. The combined LFs and their functional fits are shown in Figure\,\ref{fig:LFfits2} and the best fit parameters are given in Table\,\ref{table:LFfits2}. The three entries for $0.1<z<0.15$ given in Table\,\ref{table:LFfits2} are shown as case 1 (dashed green line in Figure\,\ref{fig:LFfits2}), 2 (solid green line in Figure\,\ref{fig:LFfits2}) and 3 (dot--dashed black line in Figure\,\ref{fig:LFfits2}).  }
\label{fig:CSFH}
\end{center}
\end{figure*}

Figure\,\ref{fig:CSFH} shows the derived SFR densities from H$\alpha$ luminosities for the GAMA and SDSS--DR7 samples, compared against a variety of published measurements derived from SFR--sensitive emission lines (H$\alpha$, H$\beta$, [\ion{O}{iii}]~$\lambda$5007, [\ion{O}{ii}]~$\lambda$3727). Where necessary the data are corrected to the cosmology assumed in this paper using the approach of \cite{Hopkins04}. If the published measurements do not already correct for obscuration, we apply a simple correction assuming one magnitude extinction in H$\alpha$. The tables in Appendix A list the published data used in this study. 

The GAMA SFR density estimate for the $0.001<z<0.1$ range is in agreement with the results \cite{Nakamura03}. They have used optically selected and morphology--classified bright galaxies from the SDSS northern stripe to estimate SFR density at $z\sim0.1$. We should expect to see an increase in the SFR density over the redshift range probed by GAMA. The GAMA data, shown as filled light blue stars in Figure\,\ref{fig:CSFH}, however, indicate essentially no evolution in SFR density. These SFR densities correspond to the completeness corrected LFs shown in Figure\,\ref{fig:GAMALFs}, with the blue circles indicating the reduction in SFR density if no completeness correction is applied to the LFs. The lack of evolution in SFR density we see is mainly due to the bivariate selection effects discussed in \S\,\ref{Biselection}. The SFR density measurements corresponding to $z<0.1$ GAMA LF (Figure\,\ref{fig:LFfits}) and $0.1<z<0.15$ GAMA LF combined with \cite{Gilbank10} LF points (Figure\,\ref{fig:LFfits2}) indicate some evolution. This dictates that normalising GAMA LFs to narrowband or other magnitude-limited LF data provides a better estimate of  the faint--end of the LF, thus increasing the reliability of the final measurement. However, none of the GAMA higher--$z$ LFs normalised to narrowband LF data indicate any evolution. This is a direct result of the bivariate selection introducing a significant incompleteness to $L>L*$ LF points (see \S\,\ref{Biselection} and Figure\,\ref{fig:gama_vs_shioya}). We see a similar lack of evolution in higher--$z$ SFR density measurements (i.e.\,the two higher--$z$ data points, see Table B1) from \cite{Westra10}. This is likely due bivariate sample selection as their sample is also drawn from a magnitude--limited survey. The other data points in Figure\,\ref{fig:CSFH} indicate the emission--line estimates of SFR densities at different epochs. Emission line measurements, as direct indicators of on--going star formation in a galaxy, are ideally the best tracers of the evolution of SFR density, and yet this figure shows considerable scatter, almost an order of magnitude, between different surveys. 

Despite the spread in SFR densities due to different indicators, the scatter is still present within the SFR densities estimated from individual indicators, in particular H$\alpha$ emission. Part of this scatter can be explained by  the inconsistencies between and biases within the different samples. For instance, most of the data shown here are mainly from narrowband filter, H$\alpha$ imaging and broad--band magnitude--selected surveys. The spectroscopy of optically selected emission line samples is biased by the bivariate selection discussed in \S\,\ref{Biselection}. As such, a galaxy sample drawn from a magnitude selected survey tends to be incomplete. Narrowband filter surveys, although not subject to this effect, suffer from cosmic (sample) variance issues, uncertainties due to dust corrections, and the blending of H$\alpha$ and [\ion{N}{ii}] in narrowband filters, unless spectroscopic data are available \citep[e.g.][]{LSE08}. Such surveys are currently limited in area to at most few square degrees, and consequently are only able to probe a narrow range in the LF, e.g.\,$\log\,L_{H\alpha}\approx31-33$ (W) over $z\approx0.065-0.095$ \citep{Ly07} and $\log\,L_{H\alpha}\approx32.5-33.5$ (W) over $z\approx0.08$ \citep{Jones01}. At high redshifts narrowband surveys are more complete as they become less sensitive to cosmic (sample) variance, reducing the scatter in SFR density measurements at these redshifts. 

In view of the biases introduced into our sample through differences in survey selection criteria, the local star formation history measured by GAMA (blue stars) is a lower limit. Nonetheless, GAMA provides currently the best galaxy sample to investigate star formation in the local universe, and therefore (currently) the best estimates of the SFR densities at low--$z$. The SFR densities at higher redshift ranges are underestimated as a result of the joint selection imposed on our GAMA star forming sample. As we showed in \S\,\ref{Biselection}, this incompleteness is a result of drawing a star--forming galaxy sample from a magnitude limited survey, introducing a bias to the sample against optically faint star--forming systems. 

The SFHs of \cite{HB06}, the best--fit to FUV and IR observational data, and \cite{Fardal07}, the best--fit to UV, emission line and IR observational data, are also shown in Figure\,\ref{fig:CSFH}. Most of the low--redshift ($z\lesssim1$) FUV SFR density estimates used by \cite{HB06} are based on $u$--band luminosity, a reasonable alternative to FUV luminosity \citep{Hopkins03}. Also, the $u$--band luminosity has two advantages over FUV, the availability of more data for better statistics, and being less affected by extinction \citep{Prescott09}. In this context the $u$--band luminosity has the additional advantage of not being subject to a bivariate selection. For these reasons, the \cite{HB06} SFH is likely somewhat more complete than both emission--line based measurements and the \cite{Fardal07} SFH. This is consistent with the emission line based measurements being on average lower than those from the combination of UV and IR. 

The sensitivity of various star formation indicators to different time scales must also be considered. Emission line indicators are sensitive to shorter time scales of typically $\leqslant10$\,Myr than UV estimators, $\geqslant100$\,Myr--$1$\,Gyr \citep{Moustakas06, Gilbank10, Koribalski09, LSK+12}. $u$--band measures are likely contaminated by the flux from old stellar populations, and consequently caution must be used in order not to overestimate the derived SFR densities \citep{Cram98, Kennicutt98, Hopkins03}, although an $u$--band - SFR relationship seems to be valid for starburst galaxies \citep{LS10}.

\subsection{Impact of the assumptions on cosmic star formation history}

A number of assumptions are made in order to calculate the SFR densities presented in this paper. Here we summarise the impact of some of those assumptions on the cosmic SFR density. 

In \S\,\ref{HaFluxlimit} we discuss how the LF varies if the assumed H$\alpha$ flux limit is varied. That analysis indicates that the change in L$^*$ can be as much as $\sim0.4\,$dex if we increase our assume F${_H\alpha}$ limit to  \mbox{$3\times10^{-18}$W\,m$^{-2}$}. While this may seem like a significant effect, this change introduces only a $\sim10\%$ variation to the integrated SFR density. 

The Balmer line measurements for the GAMA sample are corrected for the underlying stellar absorption by assuming a constant correction (see \S\,\ref{PhysicalProp}). The impact of this assumption on the shape of the LFs is discussed in \S\,\ref{constEWc}, the effect on SFR density is minimal. 

The uncertainties arising from the completeness corrections (see \S\,\ref{LF}) are investigated in \S\,\ref{unccorr} by constructing the LFs without applying any corrections for incompleteness.  Figure\,\ref{fig:CSFH} shows how the SFR densities would be underestimated if no corrections for incompleteness are applied.

\section{cosmic (sample) variance}
\begin{figure*}
\begin{center}
\includegraphics[scale=0.32]{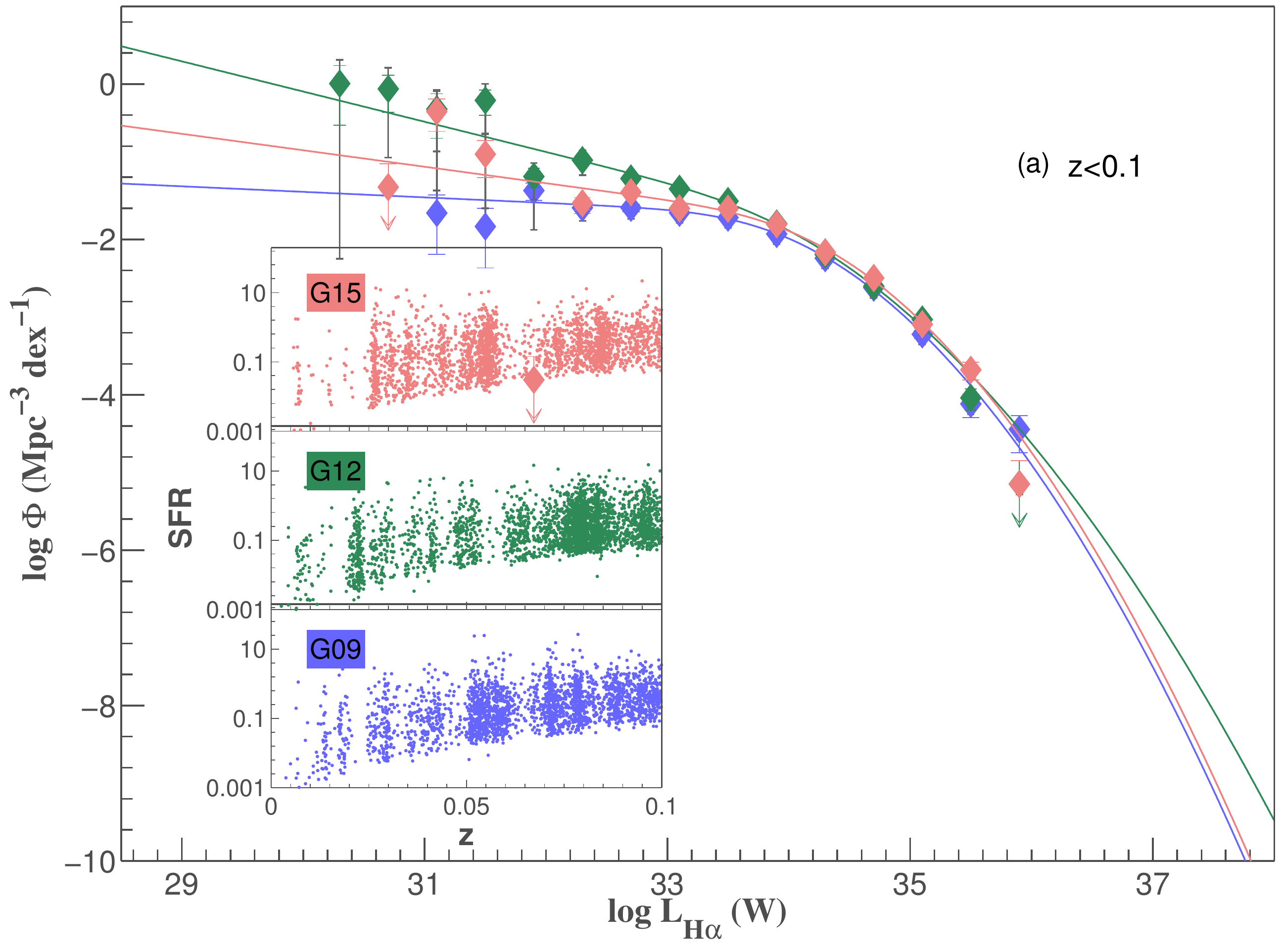}
\includegraphics[scale=0.32]{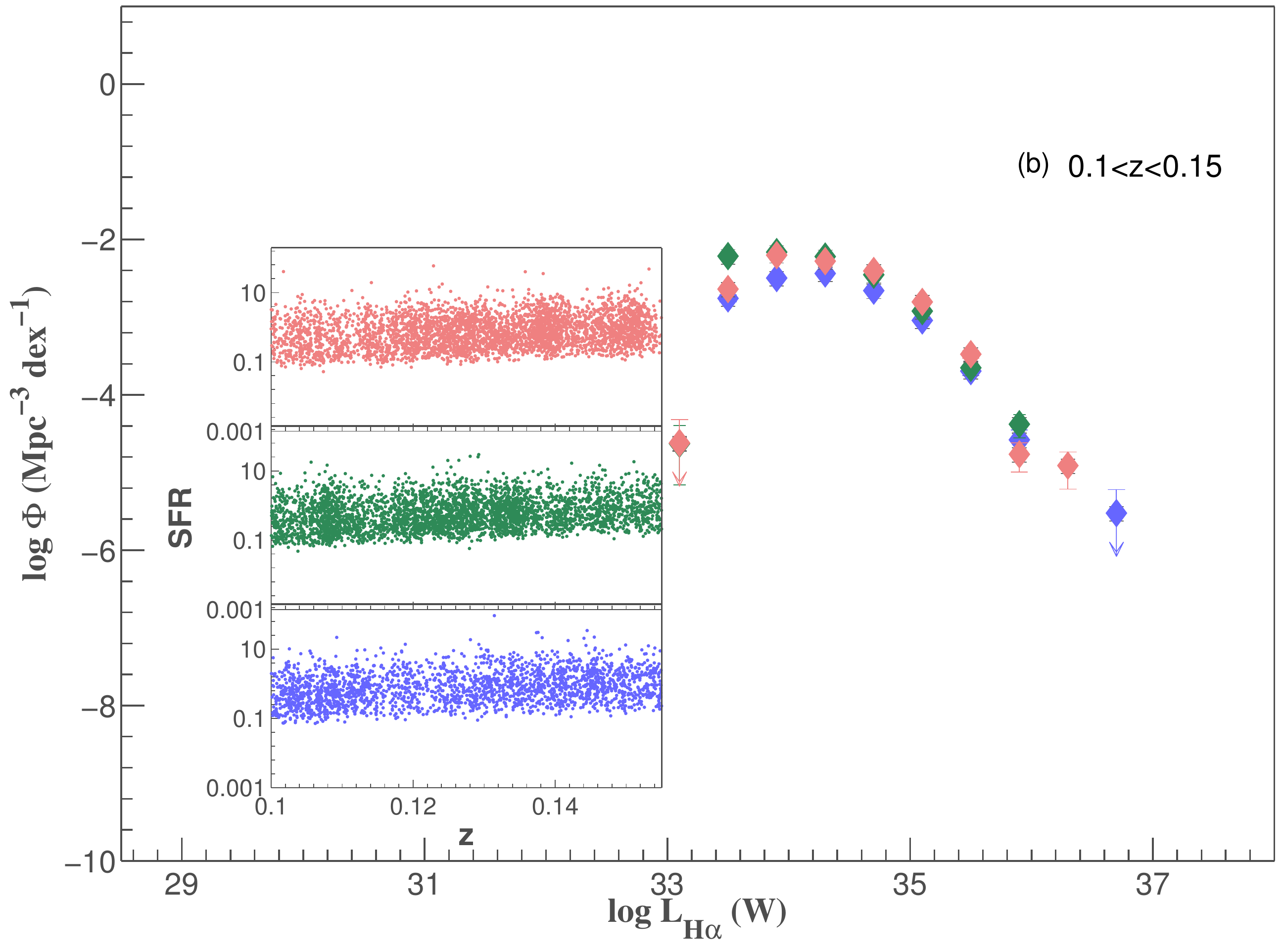}
\includegraphics[scale=0.32]{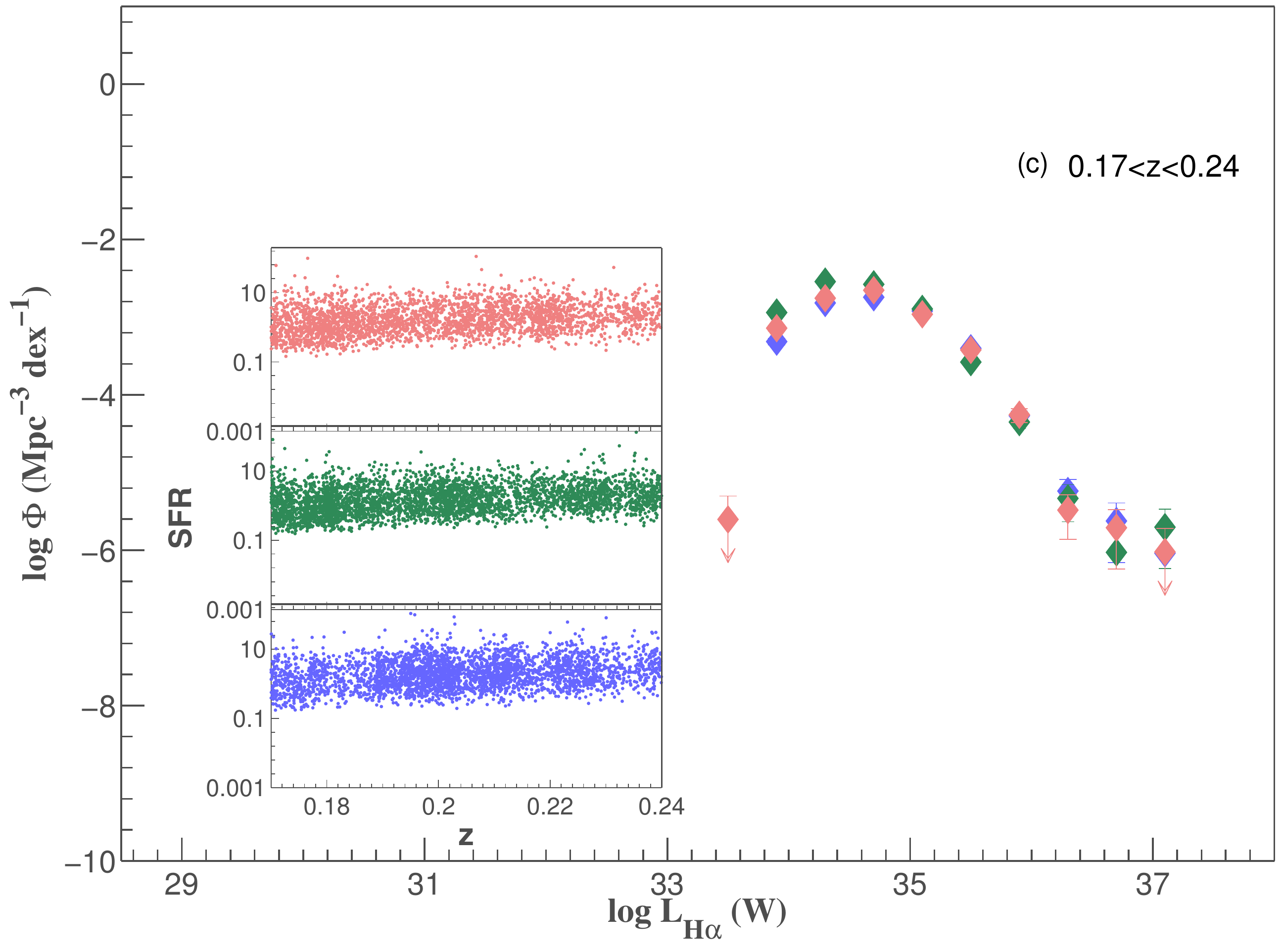}
\includegraphics[scale=0.32]{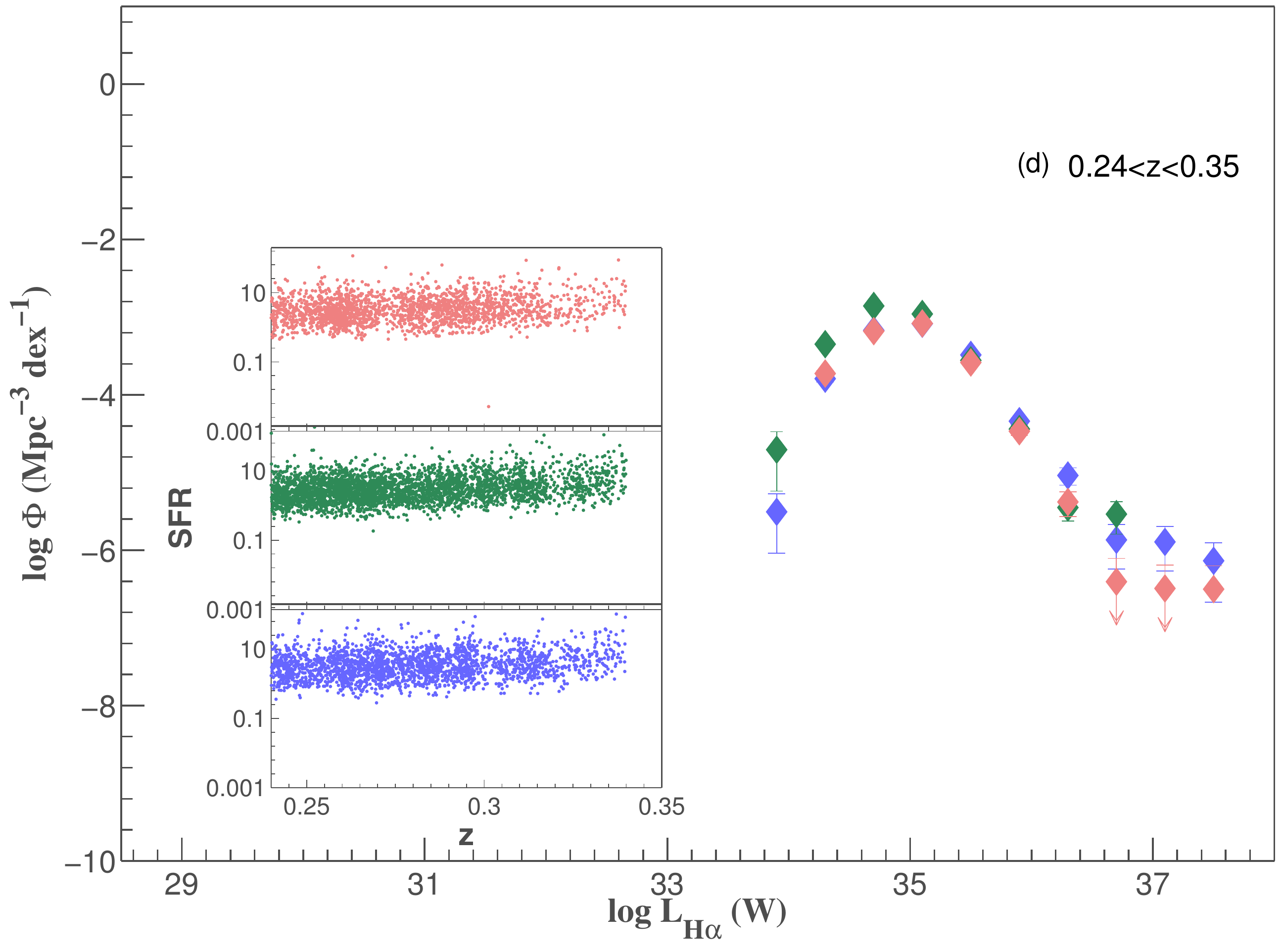}
\caption{The observations of three independent fields, and the availability of a large number of galaxies gives an opportunity to investigate the effects of cosmic (sample) variance on the star forming LFs. H$\alpha$ luminosity functions of each GAMA field in four redshift bins. The error bars shown in black, which are mostly smaller than the Poisson errors, correspond to the cosmic (sample) variance estimates. The insets in each panel show the distribution of SFR of galaxies in each GAMA field over the redshift range considered. }
\label{fig:LFs}
\end{center}
\end{figure*}
\begin{table*}
\begin{minipage}{28cm}
\caption{The best fit Saunders parameters for the $z<0.1$ LFs in the three independent GAMA fields.}
\begin{tabular}{|l|c|c|c|c|||}
\hline
         GAMA field 	 & 	$\log$ L$^*$ 	&   $\log$ C 	&  $\alpha$     & $\sigma$       \\
                 		& 	(W)			&   			& 		       & 			\\
                               
\hline
\hline
 GAMA-09h	
&  $33.14\pm0.49$ 	
&  $-1.97\pm0.17$	
& $-1.07\pm0.13$ 		
& $0.76\pm0.10$	\\

GAMA-12h	
&  $33.12\pm1.07$ 	
&  $-1.66\pm0.61$	
& $-1.38\pm0.15$ 		
& $0.91\pm0.28$	\\

GAMA-15h	
&  $33.39\pm1.06$ 	
&  $-1.93\pm0.42$	
& $-1.21\pm0.16$ 		
& $0.75\pm0.30$	\\
\hline
\end{tabular}
\label{table:GAMA_fields}
\end{minipage}
\end{table*}

Cosmic (sample) variance has been widely cited as a prominent contributor to the scatter present between published LF/SFR density measurements \citep{Westra08, Ly07}, see also Figures\,\ref{fig:GAMALFs} and \ref{fig:CSFH}. Several authors \citep{Moster11, Driver10b, Somerville04} have provided prescriptions on addressing cosmic (sample) variance issues.

\cite{Driver11} and \cite{Driver10b} provide a quantitative description of sample variance issues related to the GAMA survey. In short, the three GAMA fields overall are $15\%$ under--dense compared to a $5000$\,deg$^2$ region of SDSS--DR7 for $z$ out to $0.1$. Beyond $z>0.1$, an internal comparison between the three fields indicates that the cosmic (sample) variance is significant between the fields with the GAMA 09h field being particularly under-dense. Table 2 of \cite{Driver10b} provides the cosmic (sample) variance values for GAMA over several redshift  intervals. Using their method, we estimate cosmic (sample) variance values for the redshift ranges corresponding to the LFs presented in this paper (Table\,\ref{tab:cv}).  Although the cosmic (sample) variance is significant per GAMA field, it is largely mitigated overall as the sampling variance is inversely related to the number of distinct fields observed. 
\begin{table}
\caption{Sampling variance estimates for GAMA in redshift ranges considered in this study. These estimates are based on the prescription of \citet{Driver10b}.}
\begin{center}
\begin{tabular}{|c|c|c|}
\hline
$z$ range        		 &         sampling variance ($\%$)  & sampling variance \\
                                     &                                                        &   per field ($\%$)  \\
\hline 
\hline
$0<z<0.1$        		&         15                                            &  26 \\
$0.1<z<0.155$        &         12                                             &  21\\
$0.17<z<0.24$        &         8                                                &  14 \\
$0.24<z<0.35$        &         6                                               &  10 \\
\hline
\end{tabular}
\end{center}
\label{tab:cv}
\end{table}%

Given the large sample size and the GAMA observations of three independent fields, we are well--placed to investigate the effects of cosmic (sample) variance on star forming galaxy LFs. The H$\alpha$ LFs in each redshift bin are generated for each GAMA field (Figure\,\ref{fig:LFs}). The three insets in each panel show the distribution of SFR for galaxies contributing to the three LFs. The under-density of sources in the GAMA 09h field is clearly evident in the SFR distributions, and from the LFs shown in the first panel.  This local structure is also identified and explored by \cite{Driver11}.

The sampling variance for each field given in Table\,\ref{tab:cv} is an overall estimate of cosmic (sample) variance for the redshift range considered. These estimates translated to uncertainties are small compared to the Poisson errors. This is not to say the effect of large--scale structure is negligible, but the impact of such effects is most significant at low--$z$. An error based on overall sampling variance over a relatively large redshift range does not necessarily represent the large--scale effect influencing the faint--end of the LF. Instead, we estimate a cosmic (sample) variance error for each LF data point, using the \cite{Driver10b} prescription. These uncertainties, shown as black error bars in Figure\,\ref{fig:LFs}, are only indicative and subject to the limitations described in \cite{Driver10b}. These have a measurable effect only at the lowest luminosity end of the lowest--redshift bin.

 Finally, we have explored the dispersion in low redshift ($z<0.1$) SFR density measurements that may arise from cosmic (sample) variance effects by dividing two GAMA regions, GAMA--09h (a known under--dense region) and GAMA--12h (the deepest GAMA field) into 8 separate regions each $12$ square degrees, and calculating LFs and corresponding SFR densities for each of the sub--region. This provides a direct indication of the significance of cosmic (sample) variance as we are comparing LFs and SFR densities estimated using a single data set. In other words, the SFR densities corresponding to the $12$ sub--regions are not influenced by the assumptions about different surveys, measurements and corrections. 

The results indicate that the dispersion between measurements due to cosmic (sample) variance at low redshift can be as large as $0.4$\,dex. The SFR densities estimated from the LFs constructed from the four sub--regions within GAMA--09h field, a known under--dense field, indicate the largest variation. The results of this analysis highlight that a survey covering a large sky area (greater than $12$\,deg$^2$) is needed to reduce the non--negligible influence of cosmic (sample) variance. 

\section{Summary}

We have used large samples of GAMA and SDSS galaxies covering a wide range in SFR to construct the H$\alpha$ LFs in several redshift bins. Owing to the deep spectroscopic observations of GAMA combined with the area of the survey, both the faint and bright ends of the low redshift ($z<0.1$) star forming LF are explored in detail in this study. 

The key results are:
\begin{itemize}
\item{The \cite{Saunders90} functional form, which is used to fit the observed radio and far--infrared LFs for star forming galaxies in the literature, now proves to be a  good representation of the H$\alpha$ LF. This is an important result demonstrating that a consistent functional form reproduces the LF of the star forming galaxies at a variety of different SFR--sensitive wavelengths. \\}

\item { Using GAMA data we extend the observed H$\alpha$ LF by $\sim1$ order of magnitude in luminosity towards both fainter and brighter luminosities than other published results. The low--$z$ GAMA and SDSS LFs indicate an increasing number density of star forming galaxies at faint luminosities. While this result is qualitatively in agreement with the LFs of \cite{Westra10} and \cite{James08}, we observe this effect at fainter luminosities than they reach. The nature of this faint population has been examined further in \cite{Brough11}. \\}

\item{We investigate the effects of bivariate selection and find that it introduces an incompleteness that is difficult to account for, excluding optically faint but H$\alpha$ bright systems. We find that the SFR density estimates from emission line measures are affected strongly by bivariate selection, leading to the large scatter seen in the SFH.\\}

\item {We have investigated the comic (sample) variance effects on GAMA LFs by dividing two GAMA regions (GAMA--09h and GAMA--12h) into $12$ square degree regions, and calculating LFs and SFR densities for each sub--region. We find that the dispersion in SFR densities due to cosmic (sample) variance can be between factors of two to three. \\}

\item{We exhaustively test a number of potential biases, systematics and limitations such as the assumption of a constant stellar absorption and completeness corrections, the empirical estimation of Balmer decrements, cosmic (sample) variance issues etc., on the calculation of the LFs, and find that our results are robust to all of these.\\}

\item{The bivariate M$_r$/H$\alpha$ selection imposed on the GAMA and SDSS emission line galaxies make the star forming samples somewhat incomplete. As a consequence, the SFR densities we derive can only be lower limits. Nonetheless our measurements are the best estimates to date of the low redshift H$\alpha$ LFs, and the corresponding luminosity density arising from H$\alpha$. \\}

\end{itemize}

\section*{Acknowledgments}

We thank Russell Jurek and Eduard Westra for valuable discussions. We also thank the anonymous referee for extensive comments that have helped to refine our analysis and discussion.

GAMA is a joint European-Australasian project based around a spectroscopic campaign using the Anglo-Australian Telescope. The GAMA input catalogue is based on data taken from the Sloan Digital Sky Survey and the UKIRT Infrared Deep Sky Survey. Complementary imaging of the GAMA regions is being obtained by a number of independent survey programs including GALEX MIS, VST KIDS, VISTA VIKING, WISE, Herschel-ATLAS, GMRT and ASKAP providing UV to radio coverage. GAMA is funded by the STFC (UK), the ARC (Australia), the AAO, and the participating institutions. The GAMA website is http://www.gama-survey.org/. 

Funding for the SDSS and SDSS-II has been provided by the Alfred P. Sloan Foundation, the Participating Institutions, the National Science Foundation, the U.S. Department of Energy, the National Aeronautics and Space Administration, the Japanese Monbukagakusho, the Max Planck Society, and the Higher Education Funding Council for England. The SDSS Web Site is http://www.sdss.org/.

The SDSS is managed by the Astrophysical Research Consortium for the Participating Institutions. The Participating Institutions are the American Museum of Natural History, Astrophysical Institute Potsdam, University of Basel, University of Cambridge, Case Western Reserve University, University of Chicago, Drexel University, Fermilab, the Institute for Advanced Study, the Japan Participation Group, Johns Hopkins University, the Joint Institute for Nuclear Astrophysics, the Kavli Institute for Particle Astrophysics and Cosmology, the Korean Scientist Group, the Chinese Academy of Sciences (LAMOST), Los Alamos National Laboratory, the Max-Planck-Institute for Astronomy (MPIA), the Max-Planck-Institute for Astrophysics (MPA), New Mexico State University, Ohio State University, University of Pittsburgh, University of Portsmouth, Princeton University, the United States Naval Observatory, and the University of Washington.

M.L.P.G.\ acknowledges support provided through the Australian Postgraduate Award and Australian Astronomical Observatory PhD scholarship. J.L.\,acknowledges support from the Science and Technology Facilities Council [grant number ST/I000976/1]. 

\appendix
{\section{Biases, systematics and limitations}\label{biases}}

In this section we explore a number of potential biases to identify the level of uncertainty they introduce to the LFs, and SFR densities presented in this paper. 

\subsection{Low--SFR galaxies}

\subsubsection{Constant stellar absorption corrections} \label{constEWc}
\begin{figure}
\begin{center}
\includegraphics[scale=0.35]{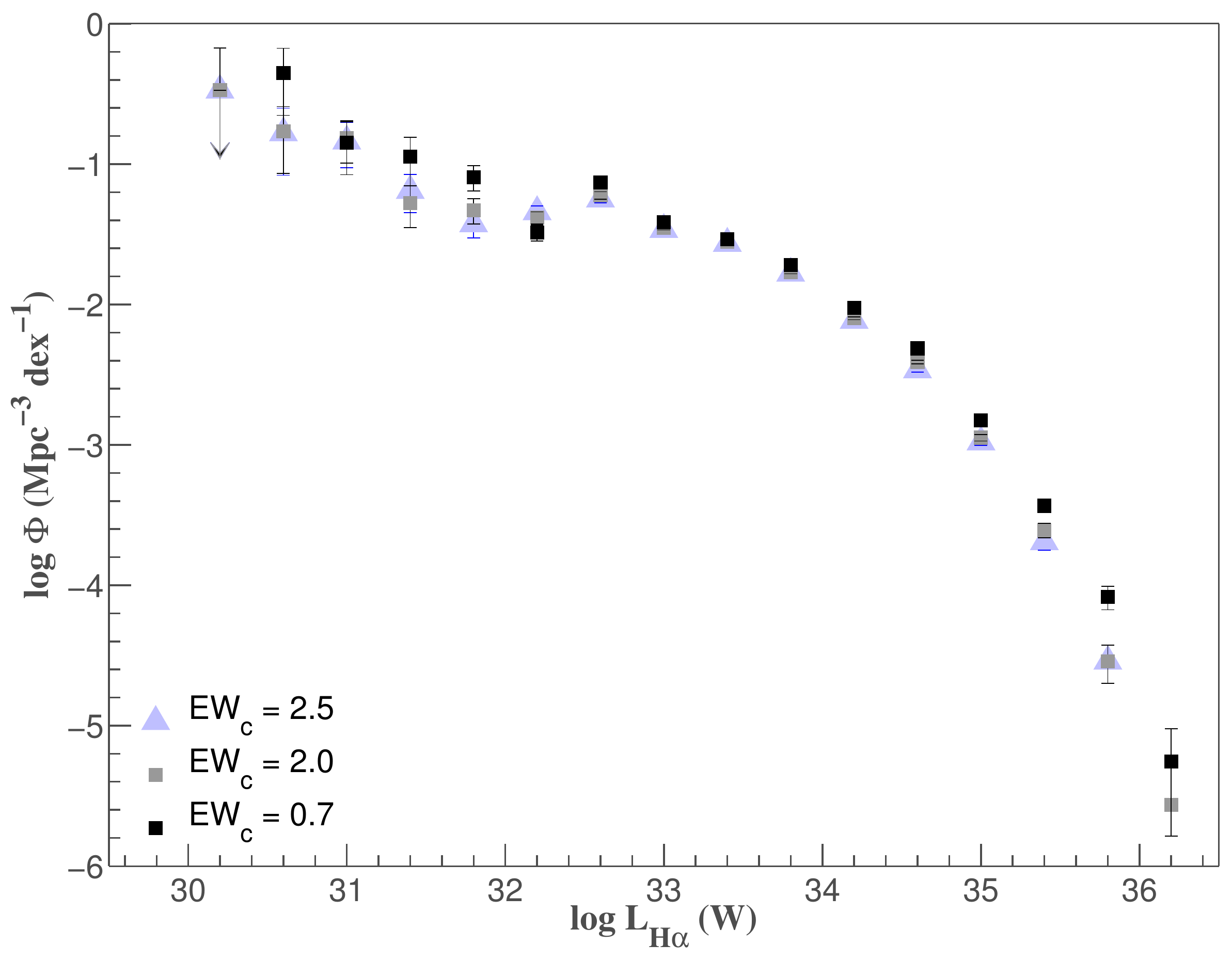}
\caption{ We reproduce the GAMA low--$z$ LF assuming different stellar absorption corrections (EW$_c$ in \AA) to investigate how the assumption of a constant EW$_c$ affects the shape of the LF at faint luminosities.}
\label{fig:GAMAlowLF_stellar}
\end{center}
\end{figure}

\begin{figure*}
\begin{center}
\includegraphics[scale=0.44]{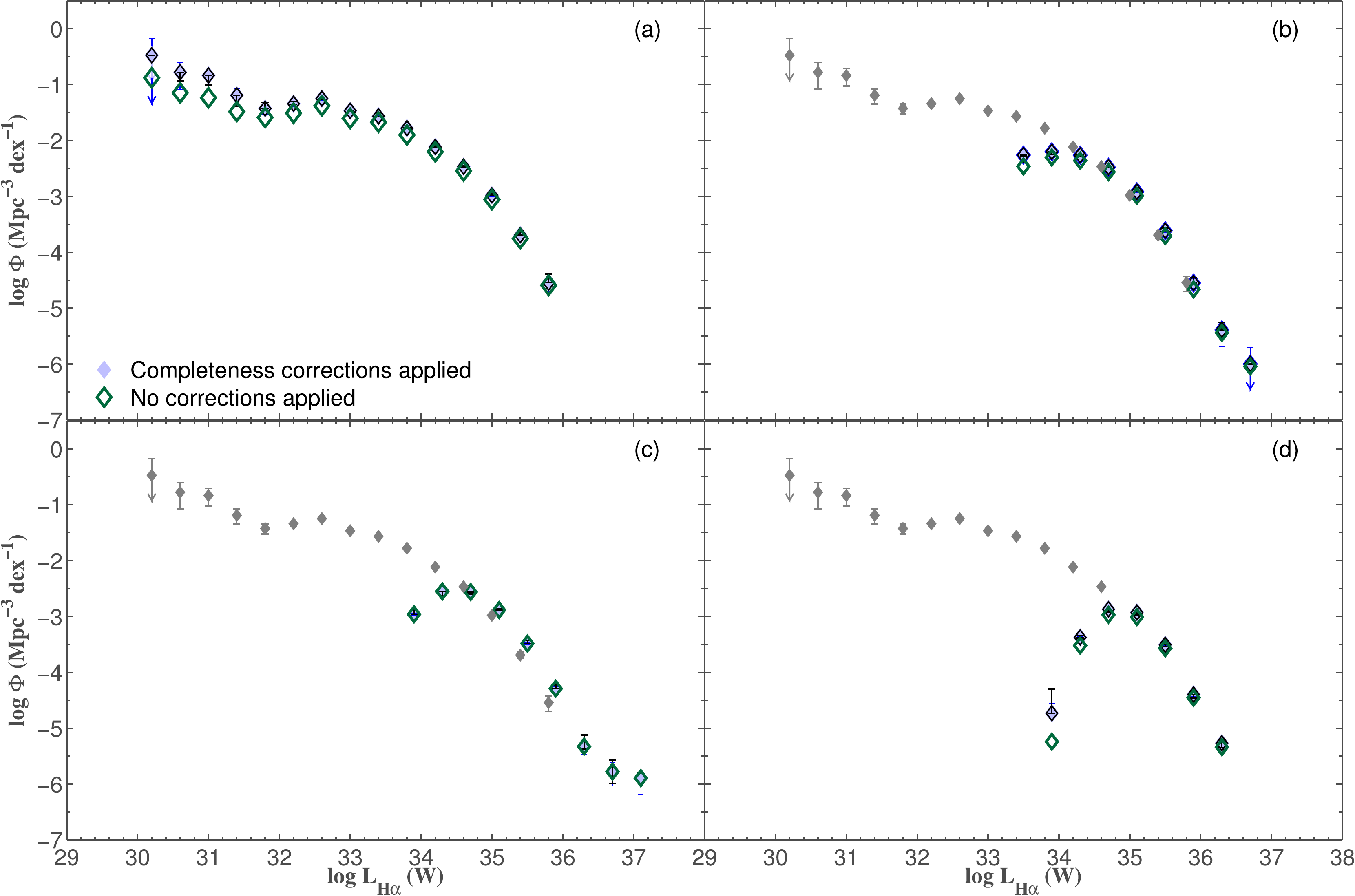}
\caption{ The effects of the completeness correction and the uncertainties arising from the empirical estimation of BDs for galaxies without measured H$\beta$ fluxes are investigated here. The error bars shown in black in each panel are estimated using a Monte Carlo method and show the full range of the values measured, and the Poisson uncertainties are shown in blue. The filled and open symbols in each panel show the LFs with and without completeness corrections respectively. Not surprisingly, the application of completeness corrections has little effect on improving the shape of the GAMA LFs as the survey is $>98\%$ complete. Also shown in (b), (c) and (d) is the low--$z$ LF (faint small symbols) for comparison.}
\label{fig:MonteCarlo_BDs}
\end{center}
\end{figure*}

\cite{Brough11} investigate the properties of low--luminosity galaxies contributing to the rise in $\Phi$ shown in Figure\,\ref{fig:GAMALFs}(a). Here we investigate how the assumption of a constant stellar absorption correction in the derivation of H$\alpha$ luminosities affects the GAMA low--$z$ luminosity function. \cite{Hopkins03} argue a stellar absorption correction of $1.3$\AA\, is sufficient. We find that a stellar absorption correction in the range $0.7-1.3$\AA\, for GAMA galaxies in $0<z<0.35$ causes a negligible change to the majority of the H$\alpha$ luminosities and SFRs, although the lowest luminosity systems are the most affected \citep{Gun11}. For the analysis presented in this paper, we have assumed a fairly conservative stellar absorption correction of $2.5$\AA.

Here we investigate quantitatively how different stellar absorption corrections affect the low--$z$ LF. 

Figure\,\ref{fig:GAMAlowLF_stellar} shows the variation in the GAMA low--$z$ LF if we assume different stellar absorption corrections. We perform this analysis only for the GAMA sample, since the emission--line measurements for SDSS galaxies in GAMA fields are taken from the MPA--JHU database, which are already corrected for stellar absorption. Also, the SDSS galaxies contributing to the GAMA low--$z$ LF are the bright galaxies in our sample. Typically, any uncertainty arising from stellar absorption corrections affects primarily the lowest H$\alpha$ luminosity galaxies, and weak line systems. For the lowest--$z$ LF, the assumption of low EW$_c$ values only affects the faintest end of the H$\alpha$ LF, and even then only in a modest way. In the case of higher--$z$ LFs, the assumption of low EW$_c$  values act to increase the integrated SFR density. 

\subsubsection{Uncertainties from completeness corrections}\label{unccorr}

The determination of completeness corrections for each galaxy is described in \S\,\ref{LF}. Here we investigate the effects of the uncertainties propagated through the application of completeness corrections, and their influence on the shapes of the LFs. A comparison of the LFs before and after the application of completeness corrections (see  Figure\,\ref{fig:MonteCarlo_BDs}) indicates that the completeness corrections have a low impact on GAMA LFs. Omitting the correction mostly affects the low-SFR galaxies in each redshift bin. This result is not surprising as the GAMA survey currently has a {spectroscopic followup} completeness of $\sim98\%$,  and  faint systems are most likely affected by any incompleteness of the survey \citep{Loveday11, Driver11}.

As mentioned before we have not attempted to apply any completeness corrections to the SDSS--DR7 sample. Nonetheless, based on the GAMA results we assume that the shapes of the SDSS LFs presented in this paper are unlikely to change significantly. 

\subsection{Empirical estimation of BDs}\label{empBDs}

In addition to the examination of the effects of the assumption of a constant stellar absorption correction, we also investigate the effects of empirically estimating Balmer decrements for the galaxies without measured H$\beta$ fluxes. 

A Monte Carlo experiment is performed using the distribution of Balmer decrements as a function of $L_{H\alpha, ApCor}$. For each galaxy without a measured Balmer decrement, rather than assigning it from Eq.\,\ref{eq:BDvL_gama}, we randomly assign a Balmer decrement from the observed distribution from galaxies of similar $L_{H\alpha, ApCor}$. This process is repeated $\sim100$ times, and the variation in the resulting LF is indicated by a second set of errors for the LFs. These errors are shown in  Figure\,\ref{fig:MonteCarlo_BDs} in black, the error estimates are simply the highest and the lowest $\Phi$s derived in the MC experiment. The uncertainties due to the empirical estimation of BDs becomes more important for high redshift LFs, as those have the highest fraction of galaxies without H$\beta$ flux measurements. Also, the uncertainties estimated from this analysis for the low--$z$ LF are small both because this $z$ range has the lowest fraction of galaxies without BDs, and the number of galaxies with BDs is particularly low at low $L_{H\alpha, ApCor}$ (Figure\,\ref{fig:BDvL}). Furthermore, the assumption of a flat BD versus  $L_{H\alpha, ApCor}$ relation above the average luminosity of the sample, for example, resulted in errors smaller than Poisson errors of the sample. Therefore, the effects of the empirical estimation of BDs for the $14\%$ of galaxies without measured H$\beta$ fluxes in the sample  are minimal. 

\subsection{AGN contamination}
\begin{figure}
\begin{center}
\includegraphics[scale=0.24]{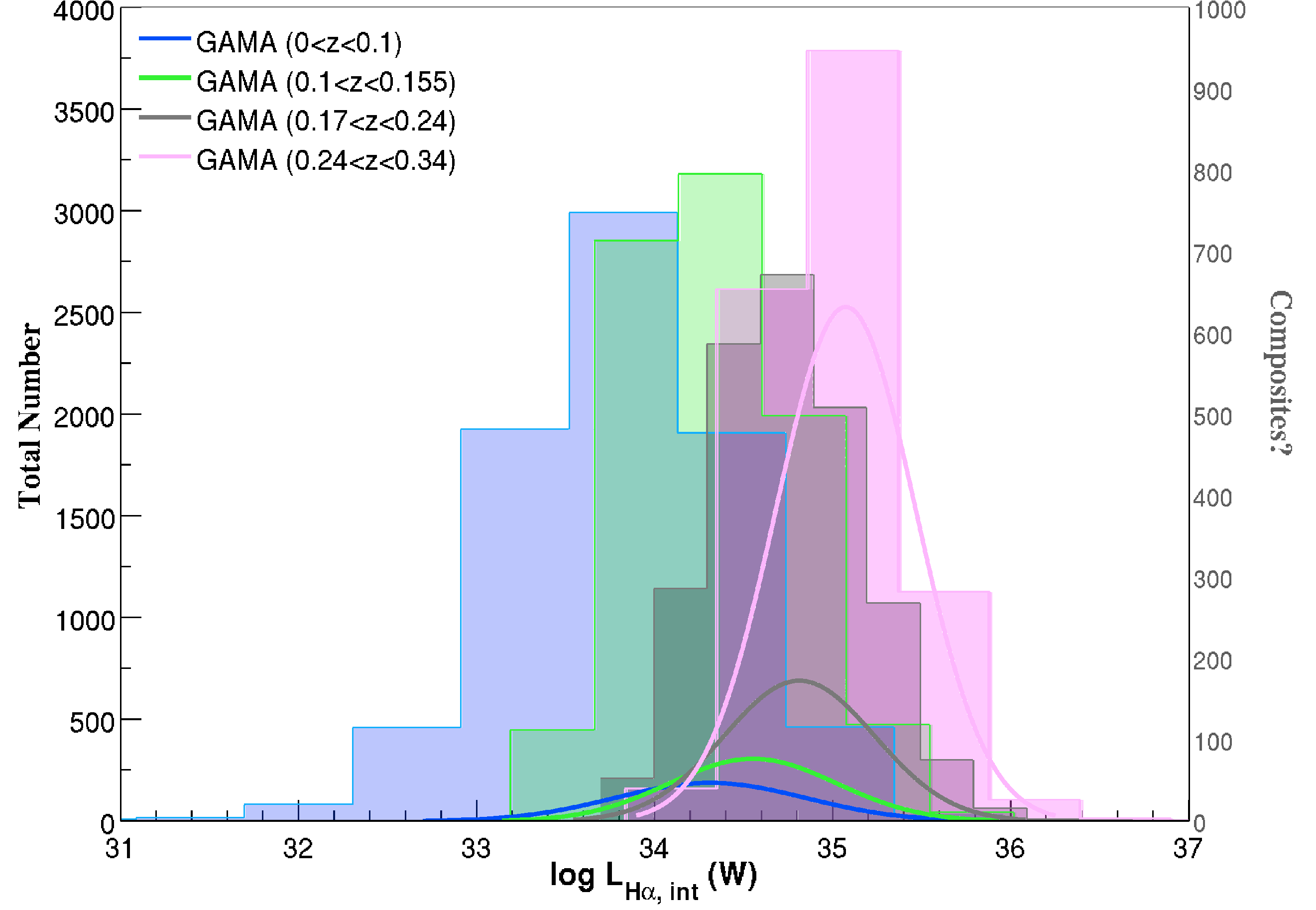}
\includegraphics[scale=0.34]{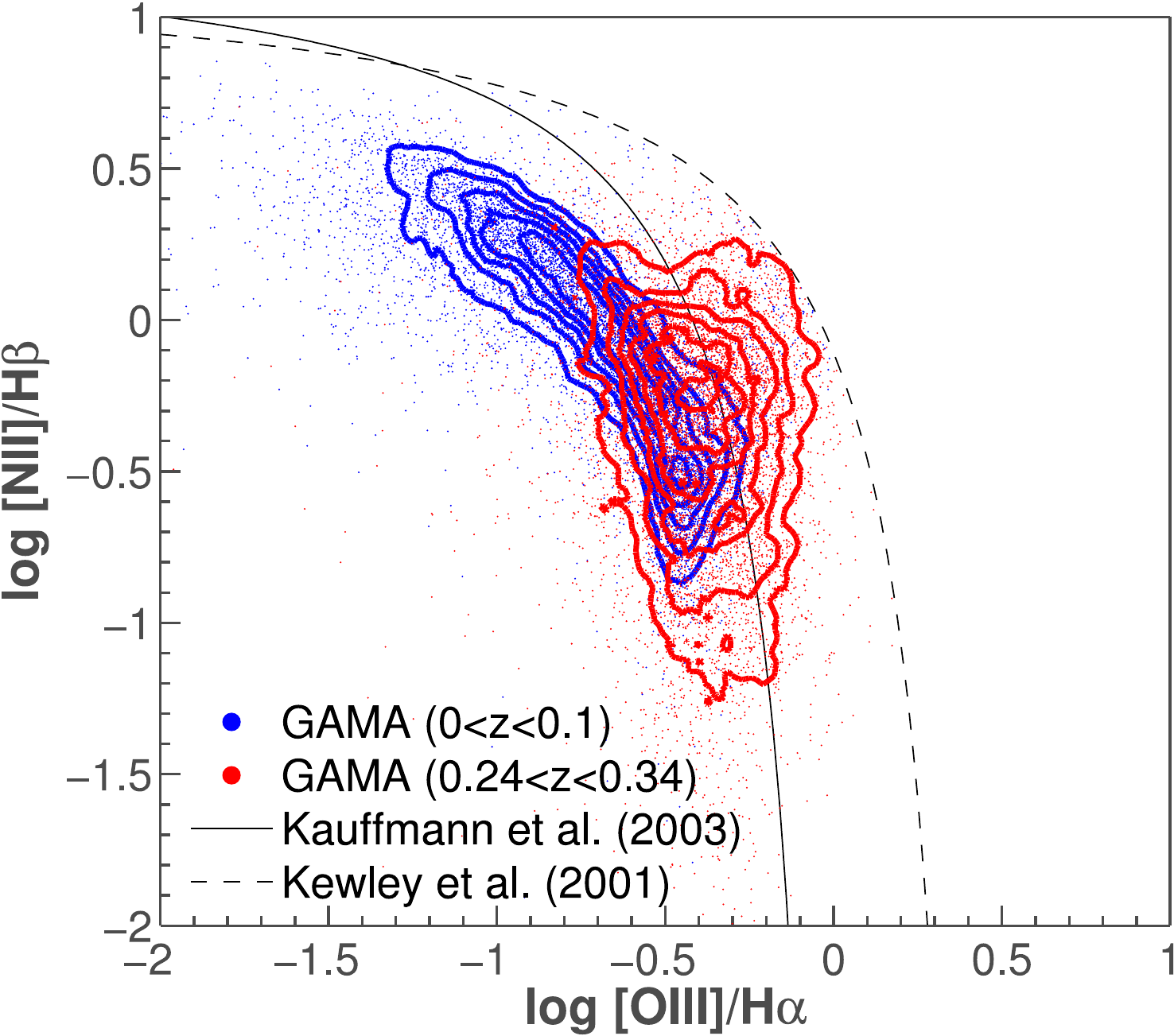}
\caption{Top panel:  The distribution of intrinsic H$\alpha$ luminosities in four redshift bins compared to the distribution of luminosities of objects classified as composites based on \citet{Kauffmann03} diagnostic over the same redshift ranges. Bottom panel: The BPT diagnostics for the lowest ($z<0.1$) and highest ($0.24<z<0.34$) redshift samples. The solid line indicates the \citet{Kauffmann03} relation.}
\label{fig:AGNs_dist}
\end{center}
\end{figure}
The \cite{Kewley01} AGN/star--forming diagnostic is used exclude AGNs from the LFs presented in the main paper. Alternatively, the \cite{Kauffmann03} relation can be used to identify pure star forming galaxies. Figure\,\ref{fig:AGNs_dist} shows the distribution of luminosities in four redshift bins compared to the distribution of luminosities of objects classified as composites based on the \cite{Kauffmann03} relation. The lowest redshift ($z<0.1$) sample consists of relatively small number of composites, and this number increase with redshift. A number of studies \citep[e.g.][]{Xue10, Best05} have found that the AGN fraction increases with the stellar mass. \cite{Hopkins13} show the BPT diagnostics for the GAMA galaxy sample as a function of both redshift and stellar mass, demonstrating the increase in AGN fraction with stellar mass.

\begin{figure}
\begin{center}
\includegraphics[scale=0.35]{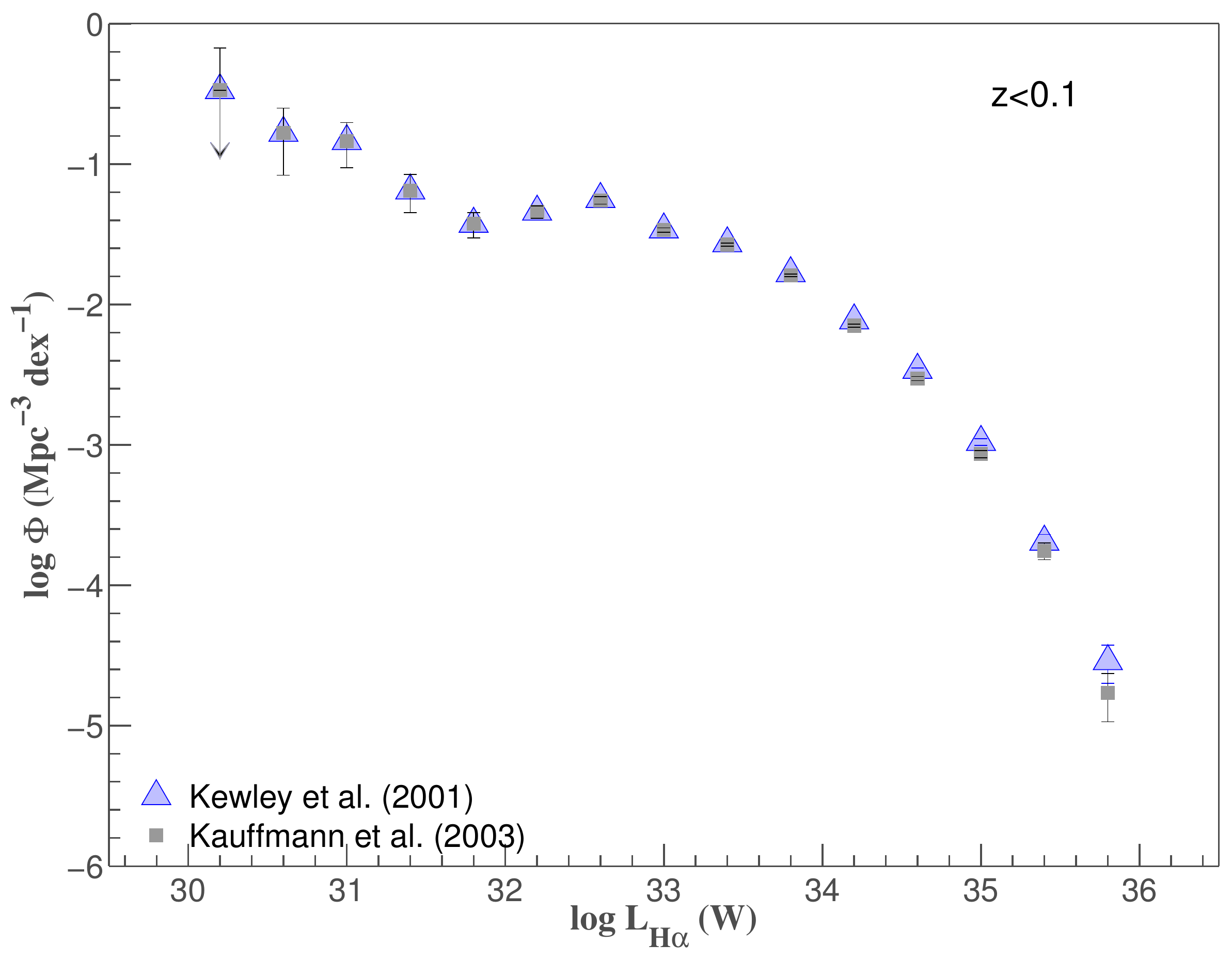}
\includegraphics[scale=0.35]{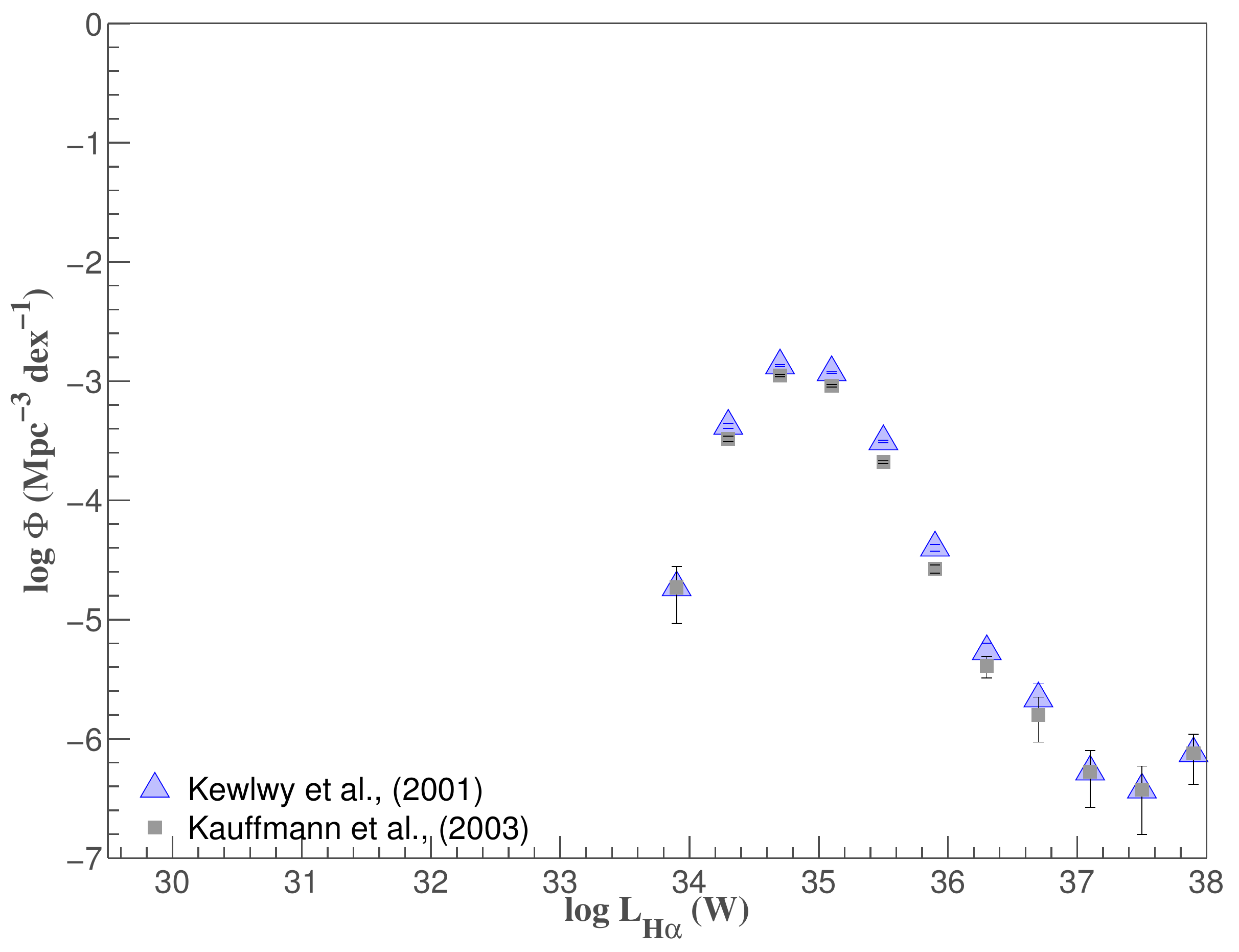}
\caption{ The lowest and highest redshift GAMA LFs shown in Figure\,\ref{fig:GAMALFs} compared to the LFs over the same redshift ranges constructed by excluding objects classified as composites based on \citet{Kauffmann03} diagnostics (grey points).}
\label{fig:AGNLFs} 
\end{center}
\end{figure}

The effects of AGN contamination on the lowest ($z<0.1$) and highest ($0.24<z<0.34$) GAMA LFs presented in \S\,\ref{HaLFs} are shown in Figure\,\ref{fig:AGNLFs}. The effects on $z<0.1$ H$\alpha$ LF and the respective SFR density is negligible. Even though the highest--$z$ LF constructed by removing these composites indicate a small drop in number density, the difference that makes to the integrated SFR density is less than $10\%$. 

\clearpage
\onecolumn
\begin{landscape}
\small
\section{Compilation of SFR densities from the literature}{\label{AppA}}
In \S\,\ref{CSFH} we present measurements of the SFR density as a function of redshift. Here we tabulate the measurements from the literature that are shown in Figure~\ref{fig:CSFH}, detailing the survey type, area, selection methods if appropriate, and various corrections, following the approach of \cite{Hopkins04}. Note that the SFR densities given in this table assume a Sapeter IMF. In order to convert these values to those presented in Figure\,\ref{fig:CSFH} simply add a factor of $-0.43$ to $\log\,\dot{\rho}_{*, int}$ measurements.
\setlength\LTleft{0pt}
\setlength\LTright{0pt}
\begin{longtable}{lllllllllll}
			
\caption[]{Compilation of H$\alpha$, H$\beta$ emission--line SFR density measurements} \label{table1} \\
\hline
Reference  & Redshift  ($z$) & Area & Selection\footnotemark[1] & N\footnotemark[2]   &  C$_1$\footnotemark[3]      &  $\log\,\dot{\rho}_{*, obs}$\footnotemark[4]     &$\log\,\dot{\rho}_{*, int}$\footnotemark[5]   \\ [0.75ex]
 						&     ($z$)                 & (deg$^2$)  &    &    &     &     (M$_{\odot}$ yr$^{-1}$ Mpc$^{-3}$)          & (M$_{\odot}$ yr$^{-1}$ Mpc$^{-3}$)\\  [0.75ex]

\endfirsthead

\hline \hline \hline

\cite{Gallego95}     		& $0<z<0.045$ 		& 471.40 				&	SLS H$\alpha$ selected, UCM 			& 176  		  	& 1.37     				& ...		  			&  -1.91 $\pm$ 0.20  \\[0.75ex]

\cite{Tresse98}       		& $0<z<0.3$   			&  0.12		 		&	$I$--selected, CFRS					& 138(SF=110)    	& 1.17  	& ...		  						& -1.61 $\pm$ 0.03 \\ [0.75ex]
				
{\cite{Yan99}}   & $0.8<z<1.8$   		&  $\sim$0.024			&	SLS H$\alpha$ selected				& 33 				 	  & 0.837	& -0.96$^{+ 0.09}_{-0.11}$			& -0.57 $\pm$ 0.18 \\ [0.75ex]
				
\cite{Sullivan00}    		& $0<z<0.3$   			&  $\sim$10   			&	UV--selected, FOCA					& 216 			 	  & 0.55	& ...		  						& -1.86 $\pm$ 0.06 \\ [0.75ex]
				
\cite{Tresse02}       		& $0.5<z<1.1$   		&  0.12 				&	$I$--selected, CFRS					& 33 				 	  & 0.88	& -1.37$^{+ 0.07}_{-0.08}$	 		& -1.06$^{+ 0.07}_{-0.08}$ \\[0.75ex] 
				
\cite{Fujita03}        		& $0.234<z<0.252$   	&  0.2				&	NBF H$\alpha$ selected				& 348 			 	  & 1.00	&  -1.90$^{+ 0.08}_{-0.17}$	 		& -1.50$^{+ 0.08}_{-0.17}$ \\ [0.75ex]
					
\cite{Perez03}        		& $0<z<0.05$   		&  ...					&	SLS H$\alpha$ selected, UCM		         & 79				 	  & 1.00	& ...		  				 		& -1.61$^{+ 0.11}_{-0.08}$ \\ [0.75ex]
				
\cite{Nakamura03}        	& $0<z<0.12$   		&  229.7				&	SDSS DR1\footnotemark[6]			& 1482(SF=665)	 	  & 1.00	& ...		 				 		& -1.94$^{+ 0.11}_{-0.082}$ \\ [0.75ex]
				
\cite{Hippelein03}        	& $0.238<z<0.252$   	&  0.1				&	NBF H$\alpha$ selected, CADIS		& 92				 	  & 1.00	& ...		  				 		& -1.83$^{+ 0.10}_{-0.13}$ \\ [0.75ex]
				
\cite{Brinchmann04}     	& $0.09<z<0.11$   		& ...					&	SDSS							& ...				 & 1.00				& ...		 			& -1.54 $\pm$ 0.07 \\ [0.75ex]
				
\cite{Westra08}     		& $0.229<z<0.261$   	& 0.262				&	NBF H$\alpha$ selected, CDFS		& 371			 & 1.00				& ...		  			& -1.77$^{+ 0.08}_{-0.10}$ \\ [0.75ex] 

					& 				   	& 					&									&				&	 	 			 & 	 				 &  (-1.93$^{+ 0.08}_{-0.10}$) \\ [0.75ex] 
				
					& $0.229<z<0.261$   	& 0.23				&	NBF H$\alpha$ selected, S11			& 335			 	  & 1.00	& ...		 				 	& -2.12$^{+ 0.09}_{-0.12}$ \\ [0.75ex] 
					
					& 				   	& 					&									&				&	 	  & 			 				 	&  (-2.24$^{+ 0.11}_{-0.14}$) \\ [0.75ex] 
					
{\cite{Westra10}}& $0<z<0.1$   			& 4					&	$R$--selected, SHELS								& 322			 	  & 1.00	& ...		 				 		& -2.19 $\pm$ 0.17 \\ [0.75ex]
				
					& $0.1<z<0.2$   		& "					&	"								& 1127			 	  & 1.00	& ...		 				 		& -1.92 $\pm$ 0.12 \\ [0.75ex]
												   
					& $0.2<z<0.3$   		& "					&	"								& 1268			 	  & 1.00	& ...		 				 		& -1.81 $\pm$ 0.10 \\ [0.75ex]
							   
					& $0.3<z<0.4$   		& "					&	"								& 848			 	  & 1.00	& ...		 				 		& -1.82 $\pm$ 0.08 \\ [0.75ex]
							   
{\cite{Westra10}}& $0.233<z<0.251$  	& 4					&	$R$--selected, SHELS				& -			 	  & 1.00	& ...		 				 		& -1.86 $\pm$ 0.13 \\ [0.75ex] 
								
+ \cite{Shioya08}		&					& + 1.5				&       + NBF H$\alpha$ selected			& -				  &  		& 		 				  		& 			        \\ [0.75ex] 
				
\cite{Shioya08}	   		& $0.233<z<0.249$   	& 1.54			 	&	NBF H$\alpha$ selected				& 980			 	  & 1.00	& ...		 				 		& -1.74$^{+ 0.17}_{-0.097}$ \\ [0.75ex] 
								
\cite{Ly07}	   		& $0.065<z<0.095$   	& 0.24				 &	NBF H$\alpha$ selected				& 318			 	  & 1.00	&								&-1.87 $\pm$ 0.29 \\ [0.75ex]
				
					& $0.239<z<0.251$   	& 0.24				 &	"								& 259			 	  & 1.00	&  -2.37 							& -2.11 $\pm$ 0.24 \\ [0.75ex]
							  
					& $0.382<z<0.418$   	& 0.24				 &	"								& 391			 	  & 1.00	& -2.10							& -1.79 $\pm$ 0.20 \\ [0.75ex]
							  
\cite{Hanish06}	   		& $0<z<0.12$   		& ... 					&	H\textsc{i} selected, SINGG			& 110			 	  & 1.00	& ...		 						& -1.80$^{+ 0.13}_{-0.07}$ \\ [0.75ex] 
				
\cite{Geach08}	   		& $2.214<z<2.246$   	& 0.60 				&	NBF H$\alpha$ selected				& 55				 	  & 1.00	& ...		 						& -1.00\footnotemark[8] \\ [0.75ex]
				
\cite{Morioka08}  		& $0.233<z<0.251$   	& 0.24+SDSS			& 	NBF H$\alpha$ selected + SDSS 		& 575			 	  & 1.00	& ...								& -1.456$^{+ 0.3}_{-0.174}$ \\ [0.75ex]
				
\cite{Villar08} 	 		& $0.831<z<0.849$   	& 0.17				&    NBF H$\alpha$, NIR selected                   	& 165			 	  & 1.00	& -1.009 [$\alpha$ constrained]		& -0.77 $\pm$ 0.077 \\ [0.75ex]
				
\cite{Shim09} 	 		& $0.7<z<1.4$   		& $\sim$0.03			&     SLS H$\alpha$ selected, HST--NICMOS    & 35				 	  & 1.00	& 					& {-1.056$^{+0.21}_{-0.44}$} \\ [0.75ex]
				
			 		& $1.4<z<1.9$   		& "					&        "               							& 45				 	  & 1.00	& 					& {-0.577$^{+0.22}_{-0.46}$} \\ [0.75ex]
					
					& $0.7<z<1.9$   		& "					&        "               							& 80				 	  & 1.00	& 					& {-0.86$^{+0.15}_{-0.24}$} \\ [0.75ex]
							
\cite{Sobral09} 	 		& $0.829<z<0.851$   	& 1.3					&      NBF H$\alpha$ selected, HiZELS  	         & 743			 	  & 1.00	& ...								& -0.96\footnotemark[7] \\ [0.75ex]

\cite{Dale10} 	 		& $0.14<z<0.18$   		& 4.19			  	 &    NBF H$\alpha$ selected, WySH                 	& 214			 	  & 1.00	& ...								& {-2.002$^{+0.041}_{-0.046}$} \\ [0.75ex]
				
					&  $0.22<z<0.26$ 	   	& 4.03				  &           "            						& 424			 	  & 1.00	& ...								&  {-1.89$^{+0.032}_{-0.034}$}\\ [0.75ex]
							
					& $0.30<z<0.34$   		& 4.13				  &            "         							& 438			 	  & 1.00	& ...								&  {-1.7$^{+0.022}_{-0.021}$}\\ [0.75ex]
							
					& $0.38<z<0.42$	  	& 1.11				  &             "          						& 91				 	  & 1.00	& ...								&  {-1.66$^{+0.04}_{-0.04}$}\\ [0.75ex]
				
\cite{Ly11} 	 		& $0.801<z<0.817$   	& 0.82				  &   NBF 1.18$\mu$m selected  			& 522 			 	  & 1.00	& ...								& -1.00 $\pm$ 0.18\footnotemark[8]\\ [0.75ex]
							
					& 				  	&	 				  &   NEWFIRM H$\alpha$ (total sample)  		&				 	 &  		&								&			\\ [0.75ex]
				
					& $0.801<z<0.817$   	& 0.82				  &   NBF 1.18$\mu$m selected 			 & 414 			 	 &  1.00	& ...								& -1.10 $\pm$ 0.09\\ [0.75ex]
							 
					& 				  	&	 				  &   NEWFIRM H$\alpha$ (L$\geqslant$ L$_{lim}$)  			 	 & 		&  								&	\\ [0.75ex]				
			
\cite{Hopkins00}		& $0.7<z<1.8$  		& 0.001				&	SLS H$\alpha$ selected, NICMOS		& 37				 	&   0.57	&  -0.74							& -0.588$\pm$ 0.064 \\ [0.75ex] 
				
\cite{Moorwood00}  		& $2.178<z<2.221$  	&  0.028			&	NBF 	H$\alpha$ selected				&  10			& 0.79			&  					&      -0.70	 \\ [0.75ex] 
				
\cite{Glazebrook99}           &   0.885$\pm$0.099	& 0.12			&   {Drawn from $I$--selected CFRS sample}		&  13			 & 0.89\footnotemark[9]				&  ... 	& -0.972$^{+ 0.15}_{-0.14}$\footnotemark[9]\\ [0.75ex] 
				
\cite{Glazebrook04}  	&   0.384$\pm$0.006	&  0.006			& NBF selected						& 	-			 	& 1		&   ...							& -1.7$^{+ 0.14}_{-0.21}$	\\ [0.75ex] 
				
\cite{Glazebrook04}[H$\beta$]  &   0.458$\pm$0.099& 	 "			&  "								&     -			 	&   1	&  	 ...							& -1.04$^{+ 0.17}_{-0.14}$\\ [0.75ex] 
				
\cite{Pascual01}  		&  $0.228<z<0.255$   	&  0.19			&	NBF 	H$\alpha$ selected				& 52				 	&  1.4		&  ...				& -1.3138$^{+ 0.08}_{-0.07}$\\ [0.75ex] 
				
\cite{Hayes10}  &  $2.214<z<2.246$ 	&  0.016		&	NBF 	H$\alpha$ selected				&  55			&  1 		&  	 ...			 	& -0.74$\pm$ 0.2\\ [0.75ex] 
+ \cite{Geach08} &  $2.214<z<2.246$     &  0.6		&	NBF 	H$\alpha$ selected   &  55
     & 				& 							& 				\\ [0.75ex]
					
\cite{James08}  		&   $0<z<0.01$   		 & ...			&	H$\alpha$ imaging					&  $\sim330$		&  1				&	...			& -1.72 $\pm$ 0.08 \\ [0.75ex] 
				
\cite{Karachentsev10}  	&   within 10 Mpc &  0.002 		   	&	H$\alpha$ imaging						& 52				 	&   1		&	 ...			& -1.72 $\pm$ 0.06 \\ [0.75ex] 

\cite{Tadaki11}  		& $2.214<z<2.246$   	&  $\sim0.016$	&  NBF H$\alpha$ selected				& 66 				 	&    1    	&  ...			& -0.51	  \\ [0.75ex] 
				
\cite{Pascual05}\footnotemark[10]  &   $\sim 0.24$	& 					&									& 				 	&      		&  								&-1.39$^{+ 0.3}_{-0.3}$ \\ [0.75ex]

					&   $\sim 0.4$		  	& 					&									& 				 	&      		&  						 		&-1.26$^{+ 0.4}_{-0.2}$ \\ [0.75ex] 
				
\cite{Doherty06}  	&  $0.77<z<1.0$ 			 &  0.026	& H$\alpha$ survey using CIRPASS\footnotemark[11]	& 38 			&  1     		&  ... 		 		& -1.13$\pm$ 0.1 \\ [0.75ex] 		
										
\hline		 
			{\footnotesize
			\footnotetext[1]{H$\alpha$ selected surveys use either slitless spectroscopy (SLS), or narrow--band filters (NBF).}
		         \footnotetext[2]{Number of galaxies.}
			\footnotetext[3]{The factor used in converting $\dot{\rho}_*$ from the cosmology assumed in the original reference to the cosmology assumed here.  Eq.\,1 in \cite{Hopkins04} is used to obtain the conversion factor.}
			\footnotetext[4]{The original reference only reports an observed $\dot{\rho}_*$, the value given here is cosmology/IMF corrected. IMF used is Salpeter.}
			\footnotetext[5]{ The final value converted to our assumed cosmology, and (Salpeter) IMF. Even though we assumed a \cite{BG03} IMF for the analysis presented in the main paper, the SFR density measurements presented in this table and in Table\,B2 are based on a \cite{Salpeter55} IMF as many of the SFR density measurements in the literature are based on \cite{Kennicutt98} relation that assume a \cite{Salpeter55} IMF. To change the IMFs from \cite{Salpeter55} (used to estimate the SFR densities in Tables B1 and B2) to \cite{BG03} (used in the main paper) simply add $-0.43$ to $\log\,\dot{\rho}_{*, int}$ measurements. 
			
			If the quoted SFR density measurement in the original reference is uncorrected for dust obscurations, a corrected based on the assumption of a one magnitude extinction in H$\alpha$ \citep{HB06} is applied to $\log\,\dot{\rho}_{*, int}$ measurements presented above. 
			
			 All SFR density values reported here are $\log \dot{\rho}_* (L>0)$. If the original reference reports a SFR density above a limiting flux, then it is indicated here within the brackets underneath.}
			\footnotetext[6]{Optically selected and morphology--classified bright galaxies from the SDSS northern stripe}
			\footnotetext[7]{The SFRD value reported here is from \cite{Ly11}}
			\footnotetext[8]{Estimates of cosmic (sample) variance is included in the uncertainties}
			\footnotetext[9]{Measurements from \cite{Hopkins04}}
			\footnotetext[10]{measurement taken from \cite{Villar08}}
			\footnotetext[11]{Near IR multi-object spectograph Cambridge Infrared Panoramic Survey Spectograph on the William Herschel Telescope.}}
			
			\vspace{-0.01\skip\footins}
			\renewcommand{\footnoterule}{}
			\end{longtable}\par
\end{landscape}

\onecolumn
{\small
\begin{landscape}
\setlength\LTleft{0pt}
\setlength\LTright{0pt}
\begin{longtable}{lllllllllll}
			
\caption[]{Compilation of [\ion{O}{ii}]~$\lambda$3727 and [\ion{O}{iii}]~$\lambda$5007 emission--line SFR density measurements} \label{table1} \\

\hline Reference  &Indicator & Redshift  ($z$)  & Selection & N   &  C$_1$     &  $\log\,\dot{\rho}_{*, obs}$    &$\log\,\dot{\rho}_{*, int}$  \\ [0.15ex]
 					&	&     ($z$)                 & (deg$^2$)  &     &     &     (M$_{\odot}$ yr$^{-1}$ Mpc$^{-3}$)        & (M$_{\odot}$ yr$^{-1}$ Mpc$^{-3}$)\\  [0.15ex]
\endfirsthead

\hline \hline

\hline

\cite{Sullivan00}     		&[\ion{O}{ii}]~$\lambda$3727	& $0<z<0.3$ 						&	UV--selected, FOCA 			& $\sim$216  				   	  & 0.55     	& ...		  					&  -1.64 $\pm$ 0.06  \\[0.15ex]

\cite{Hippelein03}     		&[\ion{O}{ii}]~$\lambda$3727	& $0.866<z<0.894$ 		 		&	NBF H$\alpha$--selected 		& $\sim$222(total detected)  	   	  & 1.00     	& ...		  				&  -1.00$^{+ 0.12}_{-0.17}$  \\[0.15ex]
		
					&[\ion{O}{ii}]~$\lambda$3727	& $1.175<z<1.211$ 			 				&	"							& "		 				   	  & 1.00     	& ...		  					&  -0.68$^{+ 0.09}_{-0.12}$  \\[0.15ex]

					&[\ion{O}{iii}]~$\lambda$5007	& $0.39<z<0.412$ 						&	"							& $\sim$124(total detected)	   	  & 1.00     	& ...		  				&  -1.33$^{+ 0.13}_{-0.18}$  \\[0.15ex]
					
					&[\ion{O}{iii}]~$\lambda$5007	& $0.626<z<0.646$ 						&	"							& "		 				   	  & 1.00     	& ...		  				&  -1.61$^{+ 0.09}_{-0.11}$  \\[0.15ex]

\cite{Gallego02}     		&[\ion{O}{ii}]~$\lambda$3727	& $0<z<0.05$ 					 		&	SLS H$\alpha$--selected, UCM 	& 191				  	   	  & 0.67     	& -3.02$\pm$ 0.15				&  -2.03$\pm$ 0.11  \\[0.15ex]

\cite{Ly07}     			&[\ion{O}{ii}]~$\lambda$3727	& $0.877<z<0.905$ 					&	NBF H$\alpha$--selected 		& 673				 	   	  & 1.00     	& ...		  				&  -1.26  \\[0.15ex]

					&[\ion{O}{ii}]~$\lambda$3727	& $0.902<z<0.922$ 			 			&	"					 		& 818				 	   	  & 1.00     	& ...		  					&  -0.97  \\[0.15ex]
					
					&[\ion{O}{ii}]~$\lambda$3727	& $1.171<z<1.203$ 						&	"					 		& 894					    	  & 1.00     	& ...		  					&  -0.82  \\[0.15ex]
					
					&[\ion{O}{ii}]~$\lambda$3727	& $1.450<z<1.485$ 				 			&	"					 		& 951					    	  & 1.00     	& ...		  					&  -0.55  \\[0.15ex]
					
					&[\ion{O}{iii}]~$\lambda$5007	& $0.391<z<0.431$ 							&	"					 		& 351					    	  & 1.00     	& ...		  					&  -1.87  \\[0.15ex]
					
					&[\ion{O}{iii}]~$\lambda$5007	& $0.416<z<0.444$ 						&	"							& 209				 	   	  & 1.00     	& ...		  				&  -2.03  \\[0.15ex]
					
					&[\ion{O}{iii}]~$\lambda$5007	& $0.616<z<0.656$ 						&	"							& 293				 	   	  & 1.00     	& ...		  				&  -1.66  \\[0.15ex]
					
					&[\ion{O}{iii}]~$\lambda$5007	& $0.823<z<0.868$ 				 			&	"					 		& 662				 	   	  & 1.00     	& ...		  				&  -1.30  \\[0.15ex]

\cite{Hogg98}    &[\ion{O}{ii}]~$\lambda$3727	& $0.1<z<0.3$ 						 &	$R$--selected, CFGRS 			& 375(total)				    	  & 0.625   & ...		  					& -1.865$^{+ 0.101}_{-0.094}$\\[0.15ex]

					 &[\ion{O}{ii}]~$\lambda$3727	& $0.2<z<0.4$ 						 	&	"		 					& "						    	  & 0.601   & ...		  					& -1.925$^{+ 0.163}_{-0.118}$\\[0.15ex]
				
					 &[\ion{O}{ii}]~$\lambda$3727	& $0.3<z<0.5$ 							 	&	"		 					& "						    	  & 0.583   & ...		  					& -1.271$^{+ 0.105}_{-0.085}$\\[0.15ex]
					 
					 &[\ion{O}{ii}]~$\lambda$3727	& $0.4<z<0.6$ 							 	&	"		 					& "						    	  & 0.570   & ...		  				& -1.020$^{+ 0.075}_{-0.067}$\\[0.15ex]
					 
					 &[\ion{O}{ii}]~$\lambda$3727	& $0.5<z<0.7$ 						 	&	"		 					& "						    	  & 0.559   & ...		  					& -1.188$^{+ 0.072}_{-0.062}$\\[0.15ex]
					 
					 &[\ion{O}{ii}]~$\lambda$3727	& $0.6<z<0.8$ 							 	&	"		 					& "						    	  & 0.552   & ...		  					& -1.272$^{+ 0.088}_{-0.073}$\\[0.15ex]
					 
					 &[\ion{O}{ii}]~$\lambda$3727	& $0.7<z<0.9$ 						 	&	"		 					& "						    	  & 0.547   & ...		  					& -1.247$^{+ 0.081}_{-0.071}$\\[0.15ex]
					 
					 &[\ion{O}{ii}]~$\lambda$3727	& $0.8<z<1.00$ 						 	&	"		 					& "						    	  & 0.543   & ...		  				& -1.146$^{+ 0.090}_{-0.074}$\\[0.15ex]
					 
					 &[\ion{O}{ii}]~$\lambda$3727	& $0.9<z<1.10$ 						 	&	"		 					& "						    	  & 0.540   & ...		  				& -0.941$^{+ 0.107}_{-0.087}$\\[0.15ex]

					 &[\ion{O}{ii}]~$\lambda$3727	& $1.00<z<1.20$ 						 	&	"		 					& "						    	  & 0.539   & ...		  					& -1.046$^{+ 0.199}_{-0.136}$\\[0.15ex]

					 &[\ion{O}{ii}]~$\lambda$3727	& $1.10<z<1.30$ 						 	&	"		 					& "						    	  & 0.538   & ...		  					& -1.066$^{+ 0.301}_{-0.176}$\\[0.15ex]

\cite{Takahashi07}    	&[\ion{O}{ii}]~$\lambda$3727	& $1.17<z<1.20$ 			 	&	NBF, [\ion{O}{ii}]~$\lambda$3727--selected, HST COSMOS & 3176					    	  & 1.00	& ...		  					& -0.495$^{+ 0.075}_{-0.058}$\\[0.15ex]

					&[\ion{O}{ii}]~$\lambda$3727	& $1.17<z<1.20$ 			 	&	NBF, [\ion{O}{ii}]~$\lambda$3727--selected, SDF			 & 294					    	  & 1.00	& ...		  					& -0.854$^{+ 0.216}_{-0.105}$\\[0.15ex]

\cite{Teplitz03}		    	&[\ion{O}{ii}]~$\lambda$3727	& $0.46<z<1.415$ 				 	&	SLS, [\ion{O}{ii}]~$\lambda$3727--selected, HST STIS		 & 71						    	  & 1.00	& -1.55$\pm$ 0.06				& -1.005$\pm$ 0.11\\[0.15ex]

\cite{Hammer97}		&[\ion{O}{ii}]~$\lambda$3727	& $0.25<z<0.5$ 					 	&	$I$--selected, CFRS				 & $\sim212$(total sample)	    	  & 1.04	& -2.20$^{+ 0.070}_{-0.080}$	& -1.705$^{+ 0.070}_{-0.080}$\\[0.15ex]

					&[\ion{O}{ii}]~$\lambda$3727	& $0.5<z<0.55$ 					 	&	"							 &"						    	  & 0.95	& -1.72$^{+ 0.11}_{-0.15}$		& -1.226$^{+ 0.11}_{-0.15}$\\[0.15ex]
					
					&[\ion{O}{ii}]~$\lambda$3727	& $0.55<z<1.00$ 						 	&	"							 &"						    	  & 0.892	& -1.35$^{+ 0.20}_{-0.38}$	& -0.855$^{+ 0.20}_{-0.38}$\\[0.15ex]
								
\cite{Bayliss11}			&[\ion{O}{ii}]~$\lambda$3727	& $1.822<z<1.878$ 				&  NBF, [\ion{O}{ii}]~$\lambda$3727--selected, HAWK--I VLT 	& 26						    	  & 1.00	& 					 	& -0.42$^{+ 0.064}_{-0.075}$\\[0.1ex]

					&		& 			 	& 											 		& 						 & 	   	  & 		 					 		&(assuming A$_{H\alpha}=1\,mag$)\\[0.15ex]
					
							& 			 	& 											 		& 						 & 	   	  & 							 &		& -0.62$^{+ 0.097}_{-0.12}$\\[0.1ex]
					
						& 			 	& 											 		& 						 & 	   	  & 		 						&	& (Independent estimate)\\[0.15ex]
					
\hline

			\end{longtable}\par
						
			\vspace{-0.75\skip\footins}
			\renewcommand{\footnoterule}{}

\end{landscape}

\twocolumn

\appendix

\bsp

\label{lastpage}

\end{document}